\documentclass[iop,revtex4]{emulateapj}
\usepackage{amsmath}

\newcommand{\uvwone}{\textit{uvw1}}
\newcommand{\uvwtwo}{\textit{uvw2}}
\newcommand{\uvmtwo}{\textit{uvm2}}
\newcommand{\Galex}{\textit{GALEX}}
\newcommand{\Swift}{\textit{Swift}}

\defcitealias{menard10}{MSFR10}
\defcitealias{weingartner01}{WD01}

\begin{document}

\slugcomment{submitted to ApJ}
\shorttitle{ULTRAVIOLET GALAXY HALOS}
\shortauthors{HODGES-KLUCK \& BREGMAN}

\title{Detection of Ultraviolet Halos around Highly Inclined Galaxies}

\author{Edmund Hodges-Kluck$^{1}$ \& Joel N. Bregman$^{1}$}
\altaffiltext{1}{Department of Astronomy, University of Michigan, Ann
  Arbor, MI 48109}
\email{hodgeskl@umich.edu}

\begin{abstract}
We report the discovery of diffuse ultraviolet light around late-type
galaxies out to 5--20\,kpc from the midplane using \Swift{} and \Galex{}
images.  The emission is consistent with the stellar outskirts in the early-type
galaxies but not in the late-type galaxies, where the emission is quite blue
and consistent with a reflection nebula powered by light escaping from the
galaxy and scattering off dust in the halo.  Our results agree with
expectations from halo dust discovered in extinction by Menard et
al. (2010) to within a few kpc of the disk and imply a comparable
amount of hot and cold gas in galaxy halos (a few$\times 10^8
M_{\odot}$ within 20\,kpc) if the dust resides primarily in Mg~II
absorbers.  The results also highlight the potential of UV photometry
to study individual galaxy halos.
\end{abstract}


\section{Introduction}

The composition and masses of galaxy halos are important predictions of 
galaxy formation and evolution models.  Material can be ejected from the
disk into the halo through galactic winds (stellar feedback) or winds and jets
driven by active galactic nuclei (AGN), and fresh material can fall into the
halo from the circumgalactic medium (CGM).  Both of these processes are thought
to occur in the lives of normal galaxies, but the relative contribution of
each process over cosmic time is unclear as they both produce hot and cold
components that occupy roughly the same space.

One of the most striking examples of similar looking halos produced by very
different processes concerns the ``missing baryon problem'' \citep{bregman07}.
Galaxies (and even groups and some clusters of galaxies) are ``missing'' 
a substantial fraction of the baryons they are expected to have from the cosmic
baryon fraction if the baryons all cooled onto the disk.  There are two basic
possibilities for \citet{schechter76} $L_*$ galaxies: either the baryons 
might all have cooled onto the galactic disk before most of them were expelled by 
stellar feedback, or most baryons have been prevented from accreting onto the
disk, having been heated to the virial temperature (a few million degrees) on
infall.  Both scenarios produce extended hot halos, and the cooling rate is
slow enough for the gas to reach hydrostatic equilibrium.  Thus, neither gas
temperature nor the surface brightness profile distinguishes the scenarios.
However, the gas metallicity is dispositive, as gas that has cycles through the
disk is enriched with metals from stellar mass loss and supernovae (SNe).

\begin{deluxetable*}{llccccc}
\tablenum{1}
\tabletypesize{\scriptsize}
\tablecaption{Properties of \textit{Galex} and \textit{Swift} UVOT Filters Used}
\tablewidth{0pt}
\tablehead{
\colhead{Instrument} & \colhead{Filter} & \colhead{$\lambda_{\text{eff}}$} & \colhead{$\Delta\lambda_{\text{FWHM}}$} & \colhead{Pixel Scale} & \colhead{FWHM psf} & \colhead{Flux Conversion} \\
                     &                  & \colhead{\AA}                    & \colhead{\AA} 		      & \colhead{(arcsec\,pixel$^{-1}$)}	  & \colhead{(arcsec)} & \colhead{(erg\,cm$^{-2}$\,Hz)}
} 
\startdata
UVOT	& UVW1	& 2600 & 693	& 0.502	& 2.37\tablenotemark{a} & $9.524\times10^{-28}$\tablenotemark{a} \\
Galex	& NUV	& 2267 & 616	& 1.5	& 5.3\tablenotemark{c}  & $3.529\times10^{-28}$\tablenotemark{b} \\
UVOT    & UVM2  & 2246 & 498	& 0.502 & 2.45\tablenotemark{a} & $1.396\times10^{-27}$\tablenotemark{a} \\
UVOT    & UVW2	& 1928 & 657	& 0.502 & 2.92\tablenotemark{a} & $8.225\times10^{-28}$\tablenotemark{a} \\
Galex 	& FUV	& 1516 & 269	& 1.5	& 4.2\tablenotemark{c}  & $1.073\times10^{-28}$\tablenotemark{b} 
\enddata
\tablecomments{Note that the \uvwone{} and \uvwtwo{} filters have red leaks, making their
effective wavelengths source dependent.}
\tablerefs{\label{table.filters} (a) \citet{breeveld10} (b) \citet{poole08} (c) \citet{morrissey07} 
}
\end{deluxetable*}

The situation is similar for the cool components.  In $L_*$ and smaller halos,
accretion of fresh material from the CGM can occur through cool flows,
but tidal interactions with other galaxies (especially satellites) can also 
produce cool streams in the halo \citep[for a review, see][]{putman12},
and cool clouds such as the high velocity clouds may be 
produced by a ``galactic fountain'' powered by SNe in the disk \citep{bregman80}.
It is not generally possible to determine the origin of cool halo gas through
where it is detected, and again the metallicity is a good way to distinguish
between different scenarios.  In some galaxies, it is also possible to map the
spatial and velocity structure of the cool gas, but this technique requires very
deep \ion{H}{1} exposures that are only feasible for nearby large systems
\citep[e.g.,][]{oosterloo07,westmeier07}.  

While metallicity is an important and unambiguous indicator of halo gas history,
it is hard to measure because the halo gas is tenuous and therefore faint.  There
are only a few ``normal'' galaxies with X-ray halos that are sufficiently bright
to make a measurement in the hot halo \citep[of which the brightest is NGC~891;][]{hodges-kluck13},
and measurements in the cooler gas rely on absorption lines in the continua of
background quasars \citep[e.g., the COS-Halos survey;][]{tumlinson13}  
The quasar absorption-line method has been very productive, but it suffers from
important limitations.  First, there are few galaxies with more than one bright
quasar behind the halo, so we are limited to a statistical view.  Second, the
cross-section to background quasars is very small close to the galaxy, so our
view of the halo between 5--20\,kpc is limited, and this is the region where we
expect interaction between stellar feedback and accreted components in the halo.
Third, making these measurements requires long exposures to get a high
signal-to-noise ratio ($S/N$) in the spectrum.  

Recently, \citet[][MSFR10]{menard10} discovered that quasars are reddened by
dust extinction in galaxy halos out to 1\,Mpc from the galaxy 
\citep[evidence for dust extinction out to 200\,kpc towards two background galaxies was first
reported by][]{zaritsky94}.  This dust is evidently a
distinct component from the well known extraplanar dusty clouds near the disk,
which form filamentary structures that do not extend beyond a few kpc from the
disk and are likely related to the disk--halo cycle \citep[e.g.,][]{rossa04,howk09}.
The discovery of dust at large radii 
suggests that halo gas may be observable through its dust content apart from
the extinction towards background quasars, since the dust will also emit in
the mid-infrared band and scatter ambient photons.  Diffuse intergalactic
dust has also been discovered in galaxy clusters using the same method
\citep{chelouche07,mcgee10}.

Diffuse dust emission is only visible within several kpc of the disk because of the 
high IR background \citep[e.g.,][]{howk09}, although a few polycyclic aromatic
hydrocarbon features were detected to above 6\,kpc around NGC~5907
\citep{irwin06}, and \citet{burgdorf07} find continuum dust emission up to a similar
height.  Also, \citet{mccormick13} found extraplanar PAH emission to 6\,kpc
in galactic winds.  
However, dust extinction is most efficient in the ultraviolet where the sky is also
quite dark.  In the UV band, the scattering albedo (the proportion of scattering
to the total extinction, which comprises scattering and absorption) is around
0.5 \citep[e.g.,][]{draine03}, so we might expect to detect this dust as reflection
nebulas around highly inclined, star-forming galaxies.  This idea is similar to
seeing (in the Earth's atmosphere) the beam of a searchlight that is pointed
away from the observer.

The reflection nebula luminosity depends on the UV luminosity of the
galaxy, escape fraction of UV photons from the disk, and the type and
quantity of the dust.  The reflection nebula spectrum depends on the
grain composition and size, which in turn is tied to the gas
metallicity, as many of the metals in the interstellar medium are
depleted onto grains \citet{jenkins04,draine04}.  If the metallicity
of the dust-bearing gas can be measured, one can place a lower bound
on the metallicity of the halo gas with some estimate of the gas mass
($N_H$, hot gas mass, or a known dust-to-gas ratio).  Motivated by
this possibility, we searched for reflection nebulas in the NUV and
FUV bands around highly inclined, star-forming galaxies to measure a
photometric SED.

In this paper, we present initial results for a sample of
these galaxies with both \Swift{} Ultraviolet and Optical Telescope (UVOT) and
\textit{Galaxy Evolution Explorer} (\textit{GALEX}) archival data where we have
detected UV halos out to 5--20\,kpc around late-type galaxies.  Our main result is the detection of the
halos, which we attribute to dust-scattered light.  These halos are seen around
every galaxy in the sample (and many more that are not included in the present
work).  With the existing data, we cannot constrain the metallicity of the gas,
but with better spectral coverage (especially in the far UV), dust
models using different proportions of silicate and carbonaceous grains
can be distinguished \citep[e.g., the models in][]{nozawa13}.  The
detection of the halos is important in its own right because if they
trace cold, dust-bearing gas, the comparatively high resolution of UV
imaging can isolate halo gas with much less integration time than is
required for high resolution 21-cm interferometry.

The remainder of this paper is organized as follows: in
Section~\ref{section.sample} we describe the sample of galaxies and
data processing, then we discuss instrumental scattered-light
artifacts in Section~\ref{section.scattered_light}.  In
Section~\ref{section.results} we present our results and describe some
basic trends.  In Section~\ref{section.sed_fits} we use the spectral
information to argue that the light comes from a reflection nebula
around late-type galaxies, and in Section~\ref{section.msfr10} we find
that the amount of light is consistent with extrapolating the \citetalias{menard10}
extinction profile to near the disk.  In
Section~\ref{section.discussion} we briefly discuss the implications
of our work.  Throughout this paper, we use the dust models of
\citet{weingartner01} (herafter, WD01), but note that others exist.

\begin{deluxetable*}{llllccccc}
\tablenum{2}
\tabletypesize{\scriptsize}
\tablecaption{Basic Parameters of Galaxies in this Work}
\tablewidth{0pt}
\tablehead{
\colhead{Name} & \colhead{Type} & \colhead{$i$} & \colhead{$d$} &
\colhead{$M_K$} & \colhead{$E(B-V)$} & \colhead{$B-V$} & \colhead{$M_*$} & \colhead{SFR$_{\text{IR}}$} \\
               &                & \colhead{(deg)} & \colhead{(Mpc)} &
\colhead{(mag)} & \colhead{mag} & \colhead{(mag)} & \colhead{($10^{10} M_{\odot}$)} & \colhead{($M_{\odot}$\,yr$^{-1}$)}\\
			   & \colhead{(1)}  & \colhead{(2)} &
\colhead{(3)} & \colhead{(4)} & \colhead{(5)} & \colhead{(6)} &
\colhead{(7)} & \colhead{(8)}
} 
\startdata
ESO 243-049  & S0    & 90   & 91    & -24.1 & 0.011 & -     & -     & -     \\
IC 5249      & SBcd  & 90   & 29.5  & -19.8 & 0.030 & 0.95  & 0.15  & -     \\
NGC 24	     & Sc    & 70.1 & 9.1   & -20.6 & 0.017 & 0.48	& 0.15	& 0.11	\\
NGC 527	     & S0a   & 90   & 76.3  & -24.4 & 0.024 & -	& -	& -	\\
NGC 891	     & SBb   & 88   & 10.0  & -24.0 & 0.057 & 0.70	& 5.0	& 2.4	\\
NGC 1426     & E4    & 90   & 22.7  & -23.0 & 0.014 & 0.87  & 2.6   & -     \\
NGC 2738     & Sbc   & 66.1 & 45.5  & -23.8 & 0.029 & -     & -     &       \\
NGC 2765     & S0    & 90   & 50    & -24.3 & 0.029 & -     & -     & -     \\
NGC 2841     & Sb    & 68   & 14.1  & -24.6 & 0.013 & 0.79	& 9.6	& $<$0.48\\
NGC 2974     & E4    & 90   & 24.7  & -25.7 & 0.048 & 0.93  & 32.   & $<$0.45\\
NGC 3079     & SBcd  & 82.5 & 16.5  & -23.7 & 0.010 & 0.53	& 2.8	& 6.1	\\
NGC 3384     & E-S0  & 90   & 11.8  & -23.5 & 0.024 & 0.88	& 4.1	& -     \\
NGC 3613     & E6    & 90   & 29.5  & -24.3 & 0.011 & 0.90  & 8.7   & -     \\
NGC 3623     & SABa  & 90   & 12.6  & -24.4 & 0.022 & 0.92  & 10.   & - \\
NGC 3628     & Sb    & 79.3 & 11.3  & -24.4 & 0.024 & 0.68	& 7.0	& 4.8	\\
NGC 3818     & E5    & 90   & 36.1  & -23.9 & 0.031 & 0.91  & 6.0   & -     \\
NGC 4036     & S0    & 90   & 20.8  & -24.0 & 0.021 & 0.85  & 6.2   & $<$0.32\\
NGC 4088     & SABc  & 71   & 16.2  & -23.6 & 0.017 & 0.69  & 2.8   & 2.8   \\
NGC 4173     & SBcd  & 90   & 7.8   & -18.2\tablenotemark{a} & 0.018 & -     & -     & -     \\
NGC 4388     & Sb    & 82   & 17.1  & -23.0 & 0.029 & 0.57	& 1.4	& 2.2	\\
NGC 4594     & Sa    & 78.5 & 9.8   & -24.9 & 0.046 & 0.88	& 15.	& 0.28	\\
NGC 5301     & SBc   & 90.0 & 23.9  & -22.7 & 0.015 & 0.55  & 1.1   & 1.0   \\
NGC 5775     & SBc   & 83.2 & 26.7  & -24.4 & 0.037 & 0.66	& 6.8	& 6.9	\\
NGC 5866     & S0a   & 90   & 15.3  & -24.0 & 0.011 & 0.78	& 5.3	& 0.81	\\
NGC 5907     & SABc  & 90   & 16.6  & -24.0 & 0.009 & 0.62	& 4.4	& 1.9	\\
NGC 6503     & Sc    & 73.5 & 5.3   & -21.2 & 0.028 & 0.58	& 0.31	& 0.15	\\
NGC 6925     & Sbc   & 84.1 & 30.7  & -24.4 & 0.052 & 0.57  & 5.6   & 2.8   \\
NGC 7090     & SBc   & 90	& 6.3	& -20.6	& 0.020 & 0.45	& 0.14	& 0.14	\\
NGC 7582     & SBab  & 68.2 & 23.0  & -24.4 & 0.012 & 0.66	& 6.7	& 13	\\
UGC 6697     & Sm    & 90	& 96.6	& -24.1	& 0.019 & 0.25	& 2.5	& $<$5.6 \\
UGC 11794    & Sab   & 78.0 & 75.4  & -23.7 & 0.092 & -	    & -	    & $<$5.2
\enddata
\tablenotetext{a}{NGC~4173 is not in any 2MASS catalogs, so we
measured this from the 2MASS $K$-band image.}
\tablecomments{\label{table.sample} Cols. (1) Morphological type and
(2) Disk Inclination angle (taken from the HyperLeda galaxy database available
at http://leda.univ-lyon1.fr) 
(3) Distance in Mpc taken from NED (http://nedwww.ipac.caltech.edu), either using the
average redshift-independent distance or the redshift distance in WMAP cosmology 
(4) 2MASS Extended Source Catalog \citep{skrutskie06} absolute K-band magnitude 
(5) Foreground Galactic extinction from NED \citep{schlafly11}
(6) The extinction- and redshift-corrected $B-V$ color from HyperLeda 
(7) Stellar masses computed from $M_K$ and color corrections from \citet{bell01}
(8) Star formation rate estimated from the \citet{kennicutt98}
relation $\text{SFR} = 4.5\times 10^{-44}
L_{\text{IR}}$\,$M_{\odot}$\,yr$^{-1}$.  $L_{\text{IR}}$ we measure as
defined by \citet{rice88} $L_{\text{IR}} = 5.67\times 10^5 d_{\text{Mpc}}^2
(13.48f_{12} +5.16f_{25}+2.58f_{60}+f_{100})L_{\odot}$, where the
fluxes at 12, 25, 60, and 100\,$\mu$m are in Jy from the IRAS catalog
(http://irsa.ipac.caltech.edu/Missions/iras.html)
}
\end{deluxetable*}

\section{Sample, Instruments, and Data Processing}
\label{section.sample}

Although the scattering cross-section is highest in the UV, the column 
densities are low, so we expect to see a small fraction of the UV light
that leaks into the halo.  Our sample therefore
consists of nearby, highly inclined, star-forming galaxies that are luminous
in the $B$-band in regions of low UV background.  

Our choice of instruments is motivated by the UV filters aboard \Swift{} and
\Galex{} and the abundance of archival data for each.  
If UV halos exist, they are evidently faint (as they have not
previously been reported), so systematic instrumental issues (such as
scattered light artifacts and filter defects) are
important.  The UV foreground is also variable, so comparing multiple
exposures from different epochs is useful.  Thus, we also 
restricted our sample to galaxies with both UVOT and \Galex{} data to compare
any halos detected in either instrument (and to give us as many points
as possible for an SED).

In this section, we describe the UVOT and
\Galex{} filters, our sample criteria, and our data processing methods.  
Part of our processing
includes removing instrumental artifacts from the UVOT exposures, which is
described in the following section.  

\subsection{\Galex{} and \Swift{}-UVOT Filters}

\Galex{} is a 50\,cm
telescope that obtains near UV (NUV) and far UV (FUV) images simultaneously in
a wide (1.2$^{\circ}$ diameter) field of view \citep{martin05,morrissey07}.  The NUV filter
is a wide-band filter covering 1700--3000\AA\ with a peak effective area at around
2200\AA\ and an angular resolution of 5.3\,arcsec.  The FUV filter covers 1400--1800\AA\
and has a peak effective area at 1500\AA, with an angular resolution of 4.3\,arcsec.
Additional properties are listed in Table~\ref{table.filters}.

The UVOT \citep{roming05} is one of three instruments aboard the \Swift{} 
observatory \citep{gehrels04} and collects data simultaneously with the other
instruments.  The UVOT has a much smaller ($17\times17$\,arcmin$^{2}$) 
field of view but a higher ($\sim$1\,arcsec) angular resolution and filters
extending from 1600--8000\AA.  In this paper, we use only the UV filters
(\uvwone{}, \uvmtwo{}, and \uvwtwo{}).  A summary of their properties is 
given in Table~\ref{table.filters}.  

\begin{figure}
\begin{center}
\hspace{-0.5cm}\includegraphics[width=0.5\textwidth]{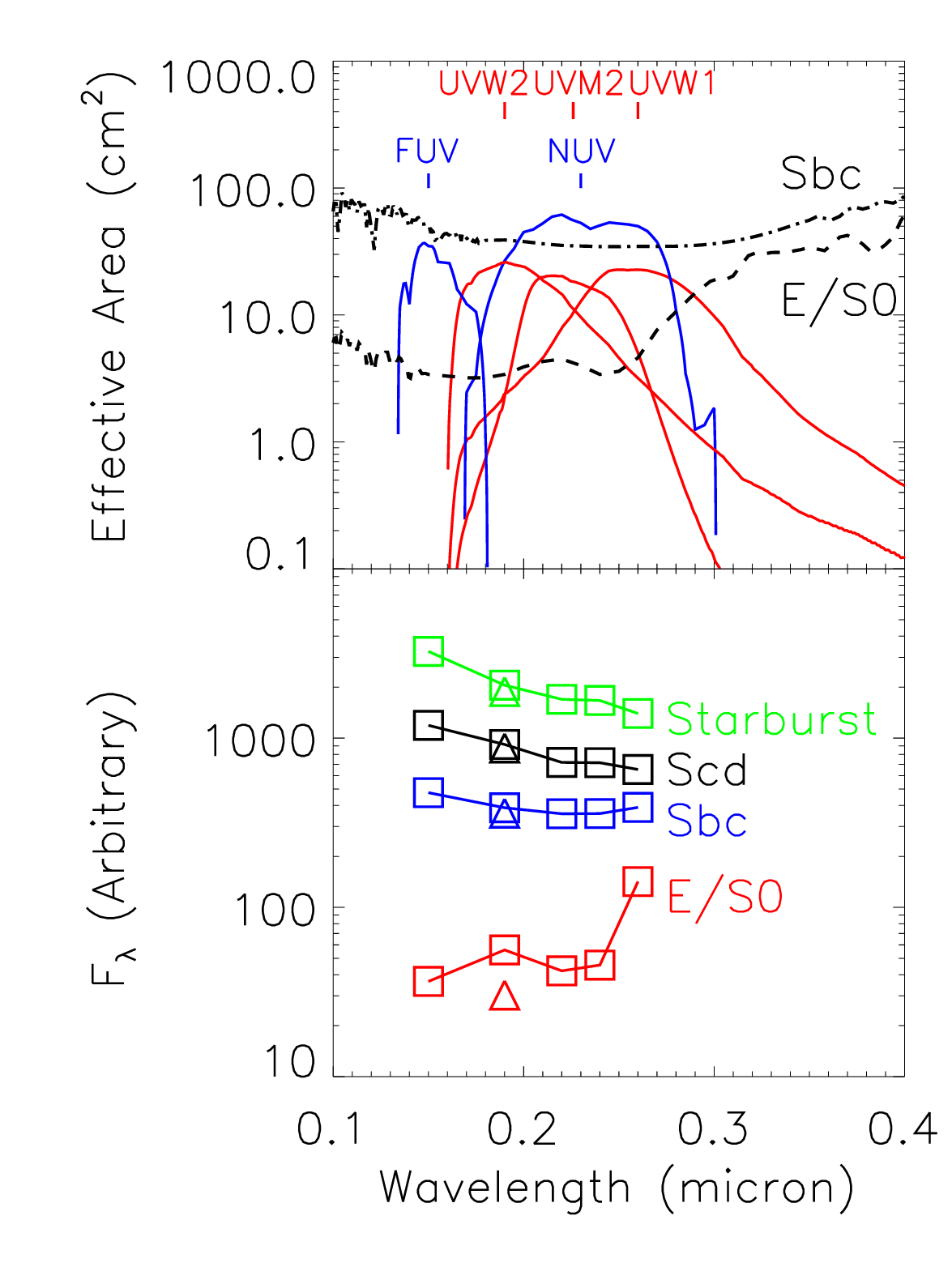}
\caption{\scriptsize \textit{Top}: \Galex{} and UVOT effective area curves for each
filter with Sbc and E/SO templates from \citet{bolzonella00} overplotted with
arbitrary units.  The red tails of the \uvwtwo{} and \uvwone{} move the
effective wavelength of the filters to longer wavelengths, depending on the
spectrum.  \textit{Bottom}: SEDs for four template galaxies in the \Galex{}
and UVOT filters.  The triangles denote the corrected \uvwtwo{} flux when the
filter response is clipped at 0.24\,micron.}
\label{figure.uv_responses}
\vspace{-0.5cm}
\end{center}
\end{figure}

The \uvwone{} and \uvwtwo{} filters have ``red tails'' in their effective
area curves\footnote{see http://heasarc.gsfc.nasa.gov/docs/\\swift/analysis/uvot\_digest/redleak.html},
meaning that a portion of the measured flux comes from the optical band.  Thus,
the effective wavelengths in these filters are higher than their nominal
wavelengths by an amount that depends on the source spectrum.  This can
be seen in Figure~\ref{figure.uv_responses}, where we show the effective area
curves for the UVOT and \Galex{} filters in the top panel along with arbitrarily
scaled UV spectra of template E/S0 and Sbc galaxies from \citet{coleman80}.
It is clear that the effective wavelength for early-type galaxies is higher
than for late-type systems.  Applying a correction to measure only the true
UV fluxes requires knowledge of the intrinsic spectrum (in this case the
galaxy halo), which has not previously been measured.  The nature of this
correction can be used to help determine the underlying spectrum in combination
with the \Galex{} and \uvmtwo{} filters, which we discuss in 
Section~\ref{section.sed_fits}.

Although we do not know the underlying halo spectrum, the
red tail is much smaller in the \uvwtwo{} than \uvwone{} filter, and the FUV
and \uvmtwo{} fluxes bracket the \uvwtwo{} and provide a useful constraint
on the real UV flux.  There are two basic
possibilities for the halo emission: a stellar halo or reflection nebula
(discussed in detail in Section~\ref{section.sed_fits}), but in either case
the correction to the \uvwtwo{} flux, constrained by the other filters, is 
constrained enough to apply a correction to the reported
fluxes.  We do not correct the \uvwone{} fluxes, where the effective wavelength is a
much stronger function of the underlying spectrum (but see Section~\ref{section.sed_fits}).

To correct the \uvwtwo{} flux we use a template-based method similar to 
those in \citet{brown10} and \citet{tzanavaris10}.  We use the template
galaxy spectra from \citet{coleman80} \citep[as updated by][]{bolzonella00} and \citet{kinney96}
and the filter effective area curves, and measure the \uvwtwo{} flux below
2400\AA. The \uvwtwo{} flux density is 
\begin{equation}
F_{uvw2} = \frac{\int F_{\lambda} A_{\text{eff}}(\lambda) d\lambda}{\int A_{\text{eff}}(\lambda) d\lambda},
\end{equation}
so the correction factor $c_{uvw2}$ is the ratio of $F_{uvw2}$ computed using
a \uvwtwo{} response cut off at 2400\AA\ (rest-frame) and all wavelengths.  
The lower panel of Figure~\ref{figure.uv_responses} shows the \Galex{}$+$UVOT
SEDs for several templates with the corrected \uvwtwo{} fluxes shown as
triangles.  $c_{uvw2}$ ranges from 0.6--1.0 depending on the galaxy template.
E/S0 galaxies have $c_{uvw2} = 0.55-0.65$, whereas spirals of type Sb or later
have $c_{uvw2} = 0.90-0.97$.  We adopt correction factors of
$c_{uvw2} = 0.60$ for elliptical galaxies, 0.65 for S0 galaxies, 0.80
for type Sa galaxies, and 0.93 for all late-type galaxies.  We apply the same
corrections to the halo fluxes.  

\subsection{Sample Criteria}

We used the HyperLeda database\footnote{http://leda.univ-lyon1.fr/} to define a
sample of candidate galaxies that are within 50\,Mpc, have an inclination of
$i \gtrsim 65^{\circ}$, and $M_B < -19$\,mag.  These criteria were guided by
the need to spatially resolve the halo in galaxies with a large intrinsic UV
brightness.  
The ``inclination'' of the early-type galaxies refers to
the projected elongation.

The working sample consists of galaxies on this list with good \Swift{} and
\Galex{} data.  For most galaxies, we require at least 1000\,s of exposure 
time in each UV filter except \uvwone{}.  All of the data we use are archival,
so there is a wide range in exposure times.  We also included a few galaxies
at larger distances with deep \Swift{} data to determine the limits of our
ability to detect halos.  

The sample was further culled by excluding interacting galaxies with tidal
tails and those that contain or are near very bright point sources (or bright
starbursts) that might produce instrumental scattered-light rings (Section~\ref{section.scattered_light}).
We also excluded galaxies such as M82 with galactic winds that are clearly
visible in the halo, galaxies that are too large (a major
axis larger than $\sim 10$\,arcmin, such as for NGC~4945) to define meaningful
background regions on the UVOT chip, and galaxies in regions of bright or
visibly structured ``cirrus'' (scattered UV emission from dust in our Galaxy)
on scales relevant to measuring a flat background.  In a few cases, we are
able to use emission from one side of a galaxy but not the other.  Cirrus
emission is all-sky, but for most of the galaxies in
the working sample the surrounding region has approximately uniform foreground
emission.  
Finally, we omit some galaxies for idiosyncratic reasons such as 
instrumental artifacts, proximity of the target to another extended object, or 
very crowded fields.  

\begin{deluxetable}{lrrrrr}
\tablenum{3}
\tabletypesize{\scriptsize}
\tablecaption{Total Useful UV Exposure Times}
\tablewidth{0pt}
\tablehead{
\colhead{    } & \multicolumn{3}{c}{UVOT}                          & \multicolumn{2}{c}{Galex} \\
\colhead{Name} & \colhead{\uvwone{}} & \colhead{\uvmtwo{}} & \colhead{\uvwtwo{}} & \colhead{NUV} & \colhead{FUV} \\
               & \colhead{(s)}   & \colhead{(s)}   & \colhead{(s)}   & \colhead{(s)}  & \colhead{(s)}
} 
\startdata
ESO 243-049  & 5459   & 8304  & 28463 & 13214 & 7966 \\
IC 5249      & -      & 8784  & 7716  & 4467  & 1705 \\
NGC 24	     & 8208   & 9298  & 9247  & 1577  & 1577 \\
NGC 527	     & 817    & 2052  & 2738  & 1532  & 1632 \\
NGC 891	     & 9787   & 15353 & 15200 & 6283  & 6047 \\
NGC 1426     & 10640  & 5904  & 10736 & 1696  & 1696 \\
NGC 2738     & 253    & 453   & 563   & 3203  & 3203 \\
NGC 2765     & 3958   & 17204 & 11843 & 1693  & 1693 \\
NGC 2841     & 873    & 1128  & 1744  & 14387 & 10723 \\
NGC 2974     & -      & 6767  & 16543 & 2694  & 2694 \\
NGC 3079     & -      & 8512  & 1185  & 16108 & 16108 \\
NGC 3384     & 560    & 768   & 1118  & 1667  & 1667 \\
NGC 3613     & 4851   & 12637 & 9105  & 1680  & 1680 \\
NGC 3623     & 3193   & 7122  & 5300  & 1656  & 1656 \\
NGC 3628     & 1579   & 1728  & 3059  & 17076 & 5812 \\
NGC 3818     & 2853   & 8497  & 10616 & 1166  & 1166 \\
NGC 4036     & 1643   & 1713  & 2512  & 2445  & 2438 \\
NGC 4088     & -      & 17781 & 8031  & 3493  & 3493 \\
NGC 4173     & 2161   & 1908  & 2353  & 1648  & 1648 \\
NGC 4388     & 6234   & 7242  & 10792 & 4994  & 2538 \\
NGC 4594     & 1027   & 1354  & 2055  & 1917  & 1917 \\
NGC 5301     & 578    & 880   & 1129  & 4657  & 1549 \\
NGC 5775     & 4181   & 23678 & 14979 & 5346  & 2776 \\
NGC 5866     & 1052   & 1428  & 2109  & 1526  & 1526 \\
NGC 5907     & 9722   & 2695  & 14484 & 5423  & 1544 \\
NGC 6503     & -      & 12598 & 6546  & 6502  & 2431 \\
NGC 6925     & 1259   & 18355 & 13360 & 1835  & 1677 \\
NGC 7090     & -      & 19697 & 2646  & -     & - \\
NGC 7582     & 5060   & 406   & 544   & 4817  & 4817 \\
UGC 6697     & 2451   & 2552  & 1972  & 2915  & 2915 \\
UGC 11794    & 821    & 639   & 3228  & 4551  & 4551 
\enddata
\tablecomments{\label{table.observations} The total exposure time is the total amount of good
  time in the combined images we used for each filter. }
\end{deluxetable}

These criteria leave us with 30 galaxies with sufficient spectral coverage to
measure a 4-point SED, and we also include NGC~7090 because of its very deep
\uvmtwo{} data, but do not use it for most of the analysis.  
There are many other galaxies with either \Swift{} or \Galex{} data, but
not both; as our major goal is to search for halo emission, we require both
instruments to identify instrumental issues.  

The basic properties of the sample are listed in Table~\ref{table.sample}, where
we have used values from HyperLeda and the NASA/IPAC Extragalactic Database\footnote{http://ned.ipac.caltech.edu/}
along with the mass-to-light ratio of \citet{bell01} and IR star-formation law
of \citet{kennicutt98} to estimate the stellar masses and star formation rate (SFR).
The $M_K$ values were obtained from the 2MASS
archive\footnote{http://irsa.ipac.caltech.edu/applications/2MASS/PubGalPS/}, whereas the SFR was computed using IRAS
fluxes\footnote{http://irsa.ipac.caltech.edu/Missions/iras.html}.  
The inclinations for some galaxies are suspect, and inclinations reported for
early-type galaxies may not be meaningful.  The total good exposure times in
each UV filter are given in Table~\ref{table.observations}.

Since we are using archival data from a variety of programs, the sample is not
complete in any metric, and includes a variety of distances,
inclinations, SFR, and galaxy morphologies.  Still, most of the galaxies have 
$M_K \sim -24$\,mag and are roughly Milky Way sized.
Many of the galaxies are in
the \Galex{} Ultraviolet Atlas of Nearby Galaxies \citep{gildepaz07}, but
the authors did not examine the halo emission. 

\subsection{Data Processing}

The archival \Galex{} data were
retrieved\footnote{http://galex.stsci.edu/GR6/} as fully reduced
datasets, and for most galaxies we did no
further processing.  The \Galex{} data reduction pipeline is described
in \citep{morrissey07}.  For our analysis, we use the reduced
intensity maps but not the background-subtracted versions.  The background
files provided by the \Galex{} pipeline are not adequate for detecting diffuse
emission \citep{sujatha10}.  

Even in the intensity maps that include the background, the background is not flat because the images are flat fielded
based on forcing observations of the standard white dwarf LDS 749B to be 
uniform everywhere on the chip.  However, the main source of NUV background is
the zodiacal light, which is redder than the standard star\footnote{http://www.galex.caltech.edu/researcher/techdoc-ch4.html}.
It is difficult to correct this in post-processing because the \Galex{} field
of view virtually always contains some Galactic cirrus with spatial variations
on similar scales.  Because the UVOT field of view is much smaller than the
\Galex{} field, as long as the galaxy is within a region where the background
varies slowly we do not need to correct the entire field.  

The \Swift{}
data\footnote{http://heasarc.nasa.gov/docs/swift/archive/} were
reprocessed from the most basic data product (a list of events and
telemetry information or raw images) so that we could combine
exposures from within a single observation and between observations to
achieve a high $S/N$.  This is required for almost every \Swift{}
dataset because in order to study the evolution of gamma-ray burst
afterglow SEDs, a single \Swift{}/UVOT observation typically consists
of a large number of shorter exposures in several of the available
filters.  

\begin{figure*}
\begin{center}
\includegraphics[width=0.95\textwidth]{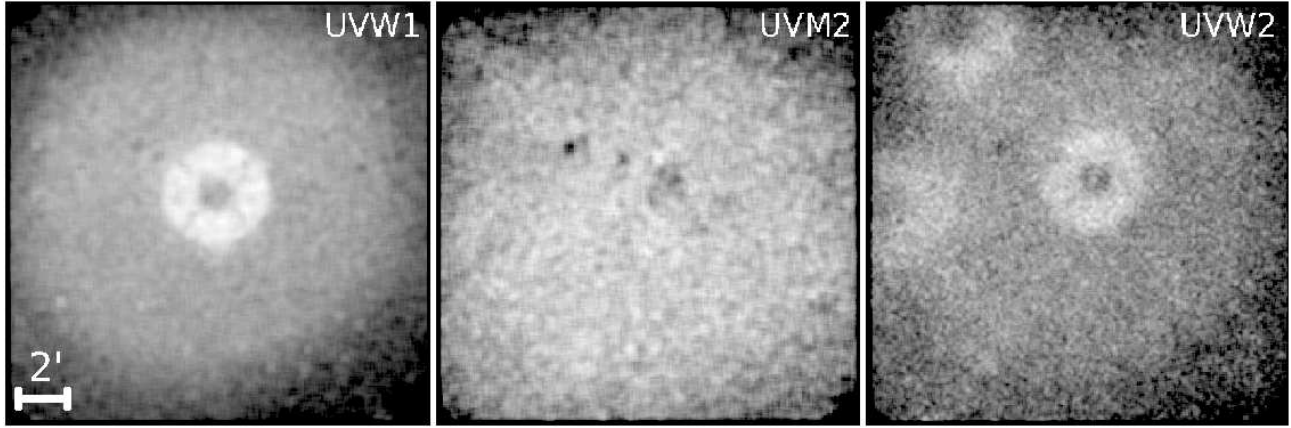}
\caption{\scriptsize Images of the persistent scattered-light artifacts in the UVOT
filters.  The images were generated by stacking many deep exposures after
removing point sources and processing up through flat-fielding and exposure
correction, and are in detector coordinates.  The final images have been lightly
smoothed.  Each of the visible rings has an outer radius of 2\,arcmin.}
\label{figure.uvot_artifacts}
\end{center}
\end{figure*}

To create the each analysis image, we first reduced and calibrated each
exposure in the observations.  We mostly
followed the processing flowchart in the UVOT Software User's
Guide\footnote{see Figure~2.1 in
http://archive.stsci.edu/swiftuvot/\\UVOT\_swguide\_v2\_2.pdf}, and
applied the latest calibrations described in
\citet{breeveld10,breeveld11}\footnote{see also
http://heasarc.gsfc.nasa.gov/docs/heasarc/\\caldb/swift/docs/uvot/}.
This includes the correction for large-scale sensitivity variations.
\citet{breeveld10} note that for point source photometry, the
large-scale sensitivity correction should not be applied directly to
the image.  However, for extended diffuse emission it is necessary to
do so.  We also added a few steps that are described in
Section~\ref{section.scattered_light} to remove persistent scattered
light artifacts.  Finally, we summed and exposure corrected all the
good data.  It is important to note that because the background varies
between exposures, exposure correction does not produce a uniform
sensitivity where there is imperfect overlap, and we reject exposures with
backgrounds much higher than the average.  We used the USNO-B1.0
star catalog \citep{monet03} to compute the astrometry for each
exposure individually and to verify the final image. 
 
We masked point sources outside the galaxies using the tool
{\sc uvotdetect}, which is based on the SExtractor code \citep{bertin96}.  Many
point sources are bright at longer wavelengths but fade in the far UV, so we used
a very conservative mask generated from the NUV to minimize contamination from
undetected point sources in the FUV.  We excised detected point sources out to 90\% of
encircled energy based on the latest psf calibration
available\footnote{http://heasarc.gsfc.nasa.gov/docs/heasarc/\\caldb/swift/docs/uvot/}
and used the nominal psf FWHM in each filter to exclude undetected
point sources seen in other filters.  For very bright sources, we used a larger
mask.

\section{UV Background and Scattered Light Artifacts}
\label{section.scattered_light}

In most cases, our measured fluxes (Section~\ref{section.results}) are 2--50\%
of the background.  Detecting halo emission requires averaging over many
pixels to improve the $S/N$ and measuring the systematic error in the background,
which is dominated by real spatial variation.  This
variation can be due to Galactic cirrus, instrumental scattered light, or, 
as in the \Galex{} NUV, due to the flat-fielding issue when the background
is dominated by zodiacal light (zodiacal light itself is not spatially 
variable in the field of view).  

The diffuse scattered light artifacts are the most significant obstacle to 
detecting halo flux with the UVOT because they occur in every exposure, cover
most of the chip, vary with position, and have a comparable or greater
``surface brightness'' than the halo flux.  However, some of these artifacts
have a persistent and predictable nature and can be subtracted.

In this section, we describe the background and foreground components,
the scattered light artifacts and our methods to mitigate them, and
the remaining systematic uncertainty in the background, which
dominates the uncertainty in our flux measurements.  Finally, we
discuss the potential for contamination by diffuse scattered light
artifacts associated with individual stars (and, by extension,
extended sources).

\begin{figure}
\begin{center}
\includegraphics[width=0.45\textwidth]{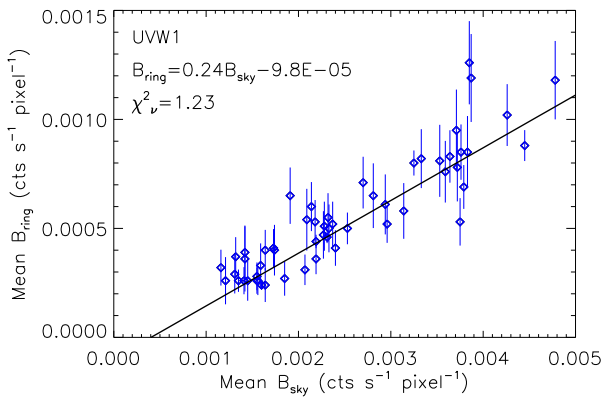}
\caption{\scriptsize The background-subtracted surface brightness the central ring of the \uvwone{} filter
pattern $B_{\text{ring}}$ as a function of the sky background measured near the chip edges 
$B_{\text{sky}}$.  The relation appears to be linear in each UVOT filter.  Most observations used
in our sample have a \uvwone{} $B_{\text{sky}} \sim 0.002$\,counts\,s$^{-1}$\,pixel$^{-1}$.}
\label{figure.uvw1_artifact_brightness}
\end{center}
\end{figure}

\subsection{Background and Foreground Components}

The detector background consists of foreground and background
components.  The foreground components include the instrumental dark
current (which is negligible for both the UVOT and \Galex{}),
instrumental scattered light (produced by off-axis light reflected
from the detector housing and filter windows onto the chip), airglow,
zodiacal light (which is much more important in the near UV bands),
and the Galactic cirrus.  Although the contribution of cirrus
decreases with increasing Galactic latitude, it is important at all latitudes.
Finally, there is the
extragalactic UV background that likely originates in galaxies.

The airglow, zodiacal light and amount of instrumental scattered light
vary with the orbital position of the satellite and pointing
direction, so the total on-field background varies between
observations of a given object.  This is most relevant for \Swift{},
where we combine multiple exposures to create a final image.  
In the rest of this paper, we use the term ``background'' to refer to
the sum of the foreground and background components including residual
instrumental scattered light unless otherwise specified.

\subsection{Diffuse Scattered Light Artifacts}

Both \Galex{} and the UVOT have diffuse scattered light artifacts that can
cover large portions of the chip.  There are both persistent and transient
artifacts, where persistent artifacts are those that are produced predictably
under given circumstances.  

The persistent artifacts include source-specific patterns that are most obvious
around bright stars as well as filter-specific patterns in the UVOT that cover
the entire chip.  The most common source-specific patterns are out-of-focus ghost
images that occur near the primary image and take the form of rings owing to the
shape and geometry of the telescope mirrors (often called ``smoke'' or ``halo''
rings).  These rings have a characteristic
size and morphology based on the optics (in the UVOT, for example, the dimmer
``halo rings'' have an outer radius of 2\,arcmin).  The ghost images result from
light that internally reflects in the detector window \citep{breeveld10}, so they
actually occur around every source for which the ghost image falls onto the chip;
whether a ring is visible depends on the source brightness.  The filter
patterns (Figure~\ref{figure.uvot_artifacts}) affect every UVOT exposure and
are produced by off-axis light that reflects onto the chip.   

Transient artifacts are most commonly large streaks of scattered light across
the chip, but some exposures also have a scattered light gradient that covers
most of the chip.  The brightnesses and widths of the streaks vary, and other
artifacts may appear if there is a very bright source nearby but outside the
field of view.

\begin{figure}[t]
\begin{center}
\includegraphics[width=0.5\textwidth]{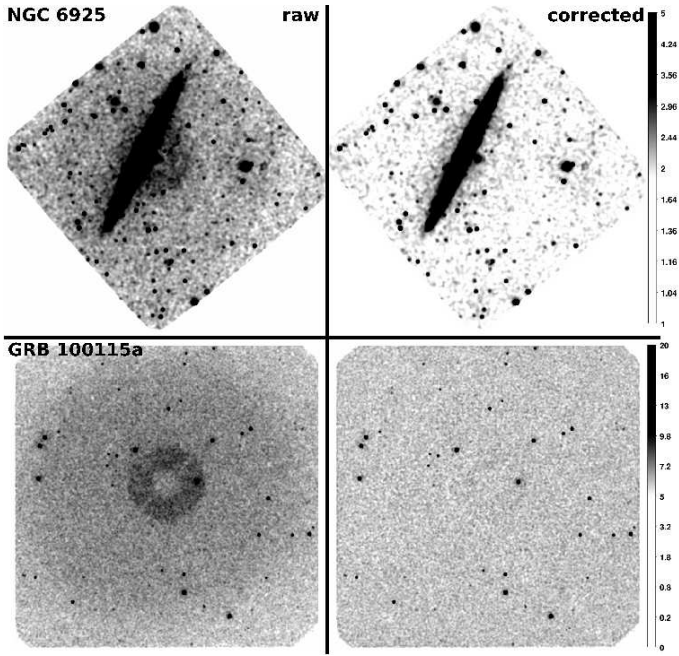}
\caption{\scriptsize Raw and corrected $\sim$1.5\,ks images in two fields in the \uvwone{} filter 
(where the central halo ring is a useful reference).  The scale and color map
are the same between raw and corrected images, and in each field we have clipped
the scale at a low level to highlight the background.}
\label{figure.before_after}
\end{center}
\end{figure}

With \Galex{}, both persistent and transient artifacts are an issue because
there are typically just a few deep exposures of a given target.  \Galex{} 
counts events and reconstructs images based on knowledge of the
spacecraft attitude, so in principle it is possible to filter the event list
by time.  However, the transient artifacts are dim (so they cannot be easily
identified in a light curve) and have an unknown time dependence, so we discard
exposures where the target galaxy is affected by large-scale transient artifacts.
Only a few galaxies in the sample are near bright stars that produce smoke
rings, and in these cases we simply mask the artifact.

\Swift{} observations consist of many short exposures with small
pointing offsets, so the persistent artifacts are more important than
in \Galex{}.  Thus, we developed a way to subtract the filter patterns.
 These patterns
consist of a large disk of scattered light (the large feature that
fills most of each panel in Figure~\ref{figure.uvot_artifacts}) whose 
brightness declines radially outwards.  The variation is significant, as is the 
absolute contribution to the background (up to 30\%).  In the \uvwone{} and \uvwtwo{} filters,
these patterns include 2\,arcmin halo rings that are fixed in position in 
chip coordinates.  The \uvmtwo{} filter does not appear to have halo rings,
possibly because it is a narrower filter \citep[for a more in-depth discussion
of these patterns, see][]{breeveld10}.  The patterns are produced by the optics,
so if only a portion of the chip is exposed (a ``windowed'' image), that section 
receives the same part of the pattern it would if the entire chip were exposed.
However, the pattern does shift slightly (a few pixels) between exposures in
detector coordinates because the optical axis is not always centered
on the chip center.

\begin{figure}[t]
\begin{center}
\hspace{-0.5cm}\includegraphics[width=0.5\textwidth]{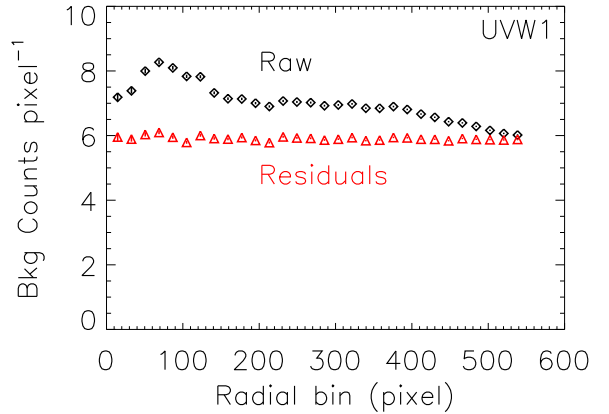}
\caption{\scriptsize Radial profiles for the images shown in the bottom two panels of
Figure~\ref{figure.before_after} (GRB~100115a).  We have subtracted point
sources and measure the radial profile from the center of the halo ring (which
is not perfectly centered on the chip).  The error bars are smaller than the 
plot symbols.  Cleaning removes over 95\% of the scattered light.}
\label{figure.grb_radial_profile}
\end{center}
\end{figure}

There is presently no correction in the UVOT calibration software for this 
artifact, so we created templates in each filter that we scale and subtract
to leave a flat background.  This is possible because the scattered light is
always ``extra'' light that is a linear function of the true sky background
and the patterns are fixed in detector coordinates.  Assuming the true sky
background is flat across the field of view, the background in pixel $i$ is
\begin{equation}
B_i = B_{\text{sky}} + m B_{\text{sky}} f(x_i,y_i) 
\end{equation}
where $m$ is the scaling factor and $f(x_i,y_i)$ is the fixed normalized
distribution of the filter pattern.  The appropriateness of adopting $m$ can
be seen in Figure~\ref{figure.uvw1_artifact_brightness},
which shows the dependence of the background-subtracted ``surface brightness'' in
the brighter halo ring that is part of the filter pattern in the
\uvwone{} filter (Figure~\ref{figure.uvot_artifacts}) on the sky
background measured at the chip edges.  The slopes are 
different in the other filters (probably because the feature we
normalize to is different), but the relation remains linear.  

The filter pattern $f(x,y)$ was generated in a similar way to the ``blank sky''
images in Figure~\ref{figure.uvot_artifacts}. We used full-frame exposures of 1\,ks or
more in sparse fields with no bright sources, extended emission, or background
gradients.  The source exposures were chosen from the large number of gamma-ray
burst follow-up observations in the archive.  Each exposure was processed
using the \Swift{} pipeline up through flat-fielding,
whereupon we masked all sources, subtracted the sky background
measured at the mostly unaffected edges of the chip, and corrected for
the large-scale sensitivity variations on the chip.  The
exposure-corrected images were stacked in detector coordinates,
smoothed by a Gaussian kernel with $\sigma=3$\,pixels, 
and normalized.  

We subtract the filter pattern after flat-fielding but before the transformation from detector to
sky coordinates.  The position of the template is first matched to the image
binning and window, then we mask point sources and smooth a copy of the image.
The ``true'' sky background is measured at the chip edges where the chip is
fully exposed and the template surface brightness is less than 5\% of its peak.
Because of the scatter in the relationship between $B_{\text{sky}}$ and $m$
(e.g., Figure~\ref{figure.uvw1_artifact_brightness}), we use least-squares
fitting to minimize the residuals against what we assume to be a flat
background.  Since the background may not be perfectly flat, 
the flatness of the residuals is verified by
binning the image by 16\,pixels and measuring the variance and azimuthally
averaged radial profile over the fully exposed portion of the chip.  When the
results are not flat or the variance is high, we inspect the images visually
and adjust the scaling factor (in such cases, the pattern or its inverse is
typically visible in a smoothed image). 

\begin{figure*}
\begin{center}
\hspace{-0.5cm}\includegraphics[width=0.85\textwidth]{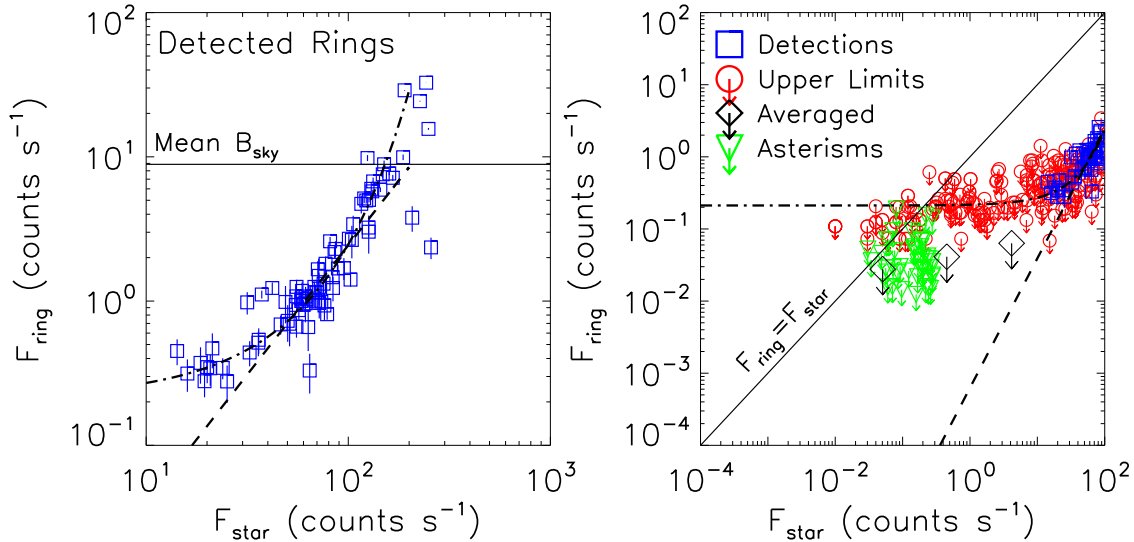}
\caption{\scriptsize \label{figure.uvm2_halo_ring_flux}
\textit{Left:} Instrumental scattered light flux ($F_{\text{ring}}$) as a
function of the incident flux of the associated star ($F_{\text{star}}$) for the
\Swift{} \uvmtwo{} filter.  The best exponential and log-linear models have
been overplotted.  \textit{Right:} $F_{\text{ring}}$ detections and
limits at lower $F_{\text{star}}$, including limits from averaged data. 
See Section~\ref{section.scattered_light} for discussion.}
\end{center}
\end{figure*}

This method is demonstrated in the \uvwone{} filter in Figure~\ref{figure.before_after}.
The top two panels show the method as applied to a 1.4\,ks exposure of NGC~6925
and the bottom two panels show a 1.6\,ks exposure of a sparse field with no 
extended emission (GRB~10015a, obsID~20126003).  For both objects, we have
clipped the maximum pixel values to show the background more clearly
and lightly smoothed the field, but the scaling and color map are the
same between the images.  In the gamma-ray burst field one can
see the filter pattern very clearly along with the telescope struts in
the center halo ring.  

Subtraction typically accounts for more than 95\% of the filter pattern
light, making the residual smaller and more spatially uniform than other background sources.  
Figure~\ref{figure.grb_radial_profile} shows the azimuthally averaged
radial profile for the two gamma-ray burst field images in 
Figure~\ref{figure.before_after}.  

After subtracting the filter pattern, we transform to sky coordinates
and proceed with the remainder of the UVOT pipeline to produce
reduced, calibrated exposures.  However, we must insert one more step
into the pipeline because the UVOT pixel scale is slightly non-uniform
\citep{breeveld10}, but during the sky transformation (using the UVOT
tool {\sc swiftxform}) the plate scale is made uniform in a way
that conserves flux.  In other words, a truly uniform sky background
would \textit{not} appear perfectly uniform in the raw image, but our
template subtraction assumes that it is.  Thus, we apply a distortion
correction to account for this assumption analogous to
{\sc swiftxform}.  The result is a flat background in the sky image.

Halo rings produced by bright stars are not removed in the pipeline process.
The scatter in the relationship between the source count
rate and the ring ``surface brightness'' is too high to make automatic
removal reliable, especially for short exposures and dimmer sources.  It is
possible to detect and subtract these rings because they are anchored to their
progenitor star, so we can use a ring template even for faint rings.  However,
if the star is within the halo detection zone, the residuals can have a large
variance and it is better to mask the region or discard the exposure.  

\subsection{Residual Background Variability and $\sigma_{\text{sky}}$}

The sensitivity in the \Swift{} and \Galex{} data is limited by background
spatial variation (photon counting noise can be
reduced well below 1\% of background by using deep exposures and averaging
the flux over large bins).  

Many \Swift{} exposures have faint transient instrumental artifacts, such as
broad, faint streaks parallel to a chip edge or background gradients.  When
these are bright enough to detect (which still requires smoothing and clipping
the image), we discard the exposure, but often the artifacts are undetectable
(when combining exposures, they are smoothed to some extent because of the
different roll angles).
Likewise, there are small variations in the astrophysical scattered light from
the Galactic cirrus across many fields, so the ``true'' mean background light
differs across the chip.  These differences are generally larger than the 
standard deviation measured in a small region of uniform background and
dominate the uncertainty in our flux measurements.

To characterize the uncertainty, we measure the variance in the background
count rate measured in many regions across the fully exposed portion of the
chip.  For \Galex{} we restrict the region of
interest to the \Swift{} UVOT field of view for consistency.  The
background fields are then defined in the UVOT images using
source-free regions more than 2\,arcmin
away from sources whose count rates might produce a bright halo ring
(typically we use a cutoff of a total count rate less than
50\,counts\,s$^{-1}$).  The scattered light background count
rate appears to follow a Poisson distribution when examining
``event''-mode data and the number of background counts is quite high, so
we adopt the standard deviation of the background
$\sigma_{\text{sky}}$ as our estimate for the uncertainty in
background subtraction.  For most of the galaxies in our sample,
$\sigma_{\text{sky}} \sim 0.1-4$\% of the mean sky background.

\subsection{Stellar Halo Rings}

In both \Swift{} and \Galex{} a small fraction of the incident photons
from a given source are scattered by the optics into a region
around the source, producing ghost images with an instrument-specific
pattern as described above.  For example, in the UVOT, which produces
visible ghost images for fainter stars than \Galex{}, the
characteristic outer radius of the larger ghost images (the ``halo
rings'') is 2\,arcmin.  The rings are only visible for the brightest
sources, but dimmer sources still scatter incident flux; they simply
do not fill in the pattern.  We denote this scattered light
$F_{\text{ring}}$.  This behavior
can be generalized to extended sources such as galaxies, where the
total instrumental scattered light $F_{\text{scat}}$ is the sum
of $F_{\text{ring}}$ over all the sources in the galaxy.

Considering the angular sizes of galaxy halos,
the contribution of
$F_{\text{scat}}$ to halo light may be important.
However, the dependence of $F_{\text{ring}}$ 
on the count rate of the associated
star, $F_{\text{star}}$, is unknown.  Because of the faintness of the
rings it is also not clear whether $F_{\text{ring}}$ is a continuous
function of $F_{\text{star}}$ or 
of there is a threshold $F_{\text{star}}$ below which
reflected light is not transmitted to the chip.  

In this subsection, we examine the dependence of $F_{\text{ring}}$
on $F_{\text{star}}$ at low count rates to assess what
$F_{\text{scat}}$ we might expect from the galaxies in our sample.
These galaxies have total count rates between $1-200$\,counts\,s$^{-1}$,
but the 
compact clumps that characterize the emission typically have count
rates of $1-10\times 10^{-4}$\,counts\,s$^{-1}$.
Clearly,
$F_{\text{ring}}$ must be less than this, but if it is not much less then
$F_{\text{scat}}$ will be a significant fraction of the background.

To address this issue we measured $F_{\text{ring}}$ around a sample of
stars from the \Swift{} gamma-ray burst observations between 2008--2013,
which randomly sample the sky.  We only included exposures greater than 800\,s
for good background statistics and rejected exposures with strong background
variation.  The data were processed in an analogous way to the analysis sample.

For stars with visible halo rings we measured the
background-subtracted count rate.  Stars without immediately apparent halo rings
may still have \textit{detectable} halo rings, but there are two challenges.  First,
halo rings have large offsets from the primary images when the source
has a large offset from 
the optical axis, so we restricted our sample to sources within 3--4\,arcmin
of the chip center where this is not a significant issue.  Unfortunately, the
\uvwone{} and \uvwtwo{} filter patterns have bright halo ring features near
the chip center (Figure~\ref{figure.uvot_artifacts}), so we restrict the analysis
to the \uvmtwo{} filter.  Second, if there is no visible halo ring we require
the star to be by far the brightest in its vicinity so we do not measure
competing halo rings.  We then measure $F_{\text{ring}}$ 
within the 2\,arcmin around the star while also avoiding the
primary image of the star and any other nearby sources to 95\% of the encircled
energy in the point-spread function.  The values are scaled by the
area in a complete ring.  We measured $F_{\text{ring}}$ for all stars
above $F_{\text{star}} = 10$\,counts\,s$^{-1}$ that meet these criteria.
The sample was extended to $F_{\text{star}} > 0.01$\,counts\,s$^{-1}$,
but isolated stars are less common, and we limited the lower flux
sample to fields with lower, more uniform backgrounds.

\begin{figure*}
\begin{center}
\mbox{}
\vspace*{-0.65cm}\includegraphics[width=0.85\textwidth]{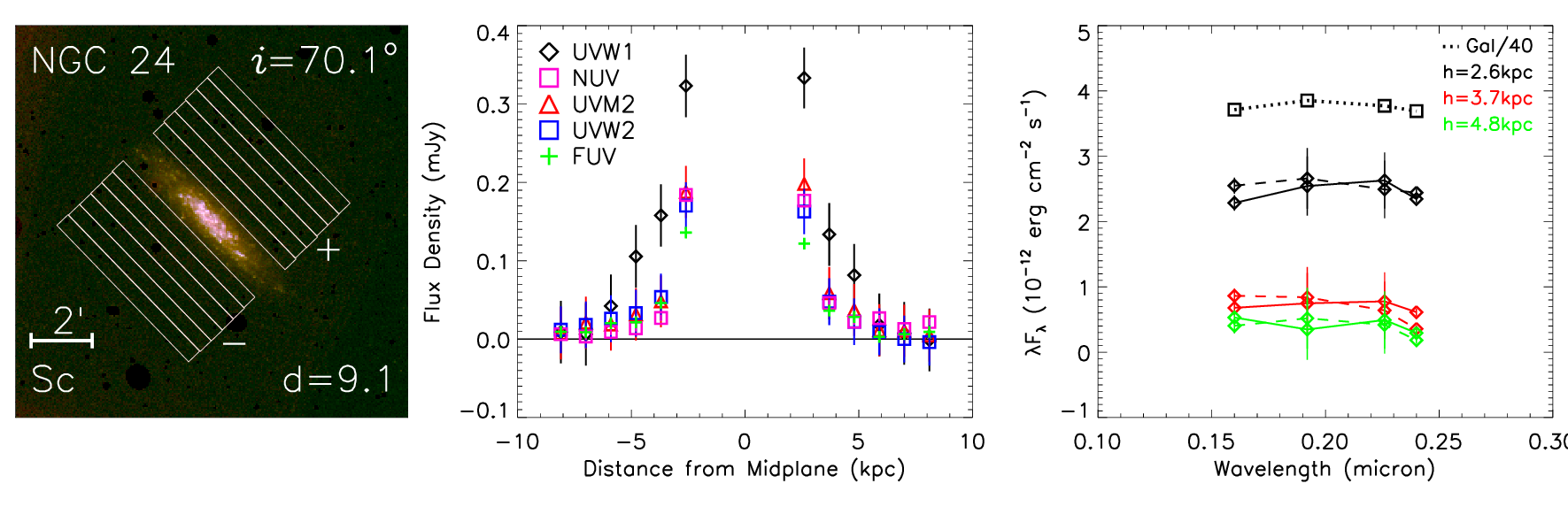}
\vspace{-0.65cm}\includegraphics[width=0.85\textwidth]{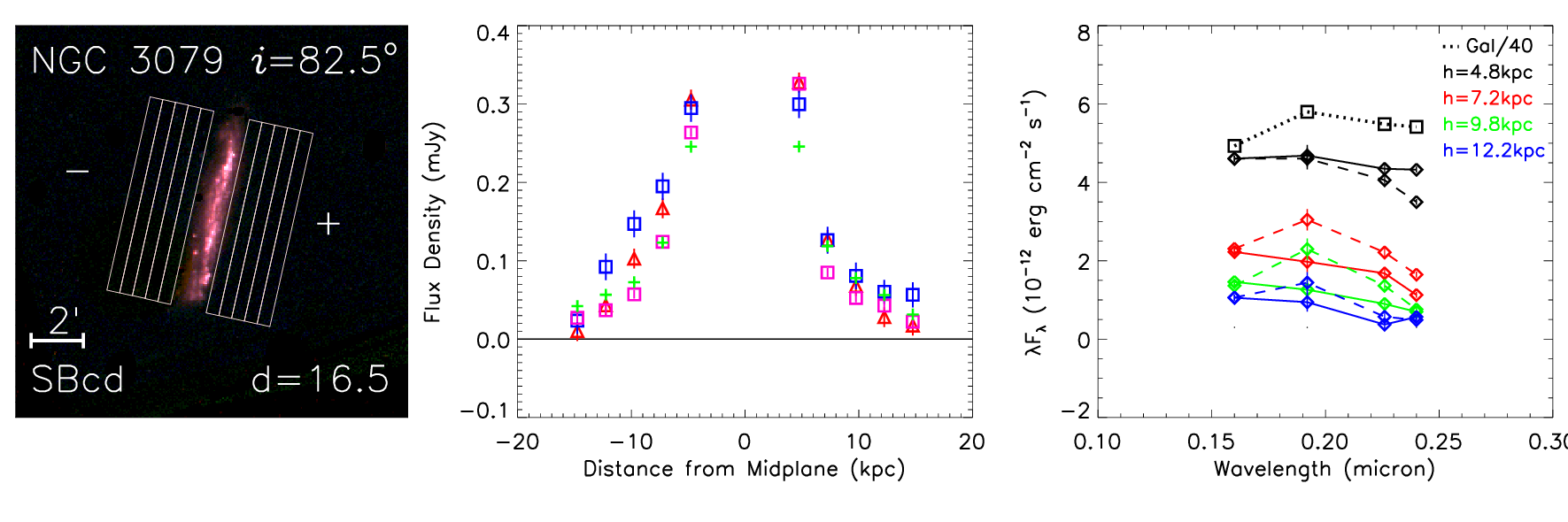}
\vspace{-0.65cm}\includegraphics[width=0.85\textwidth]{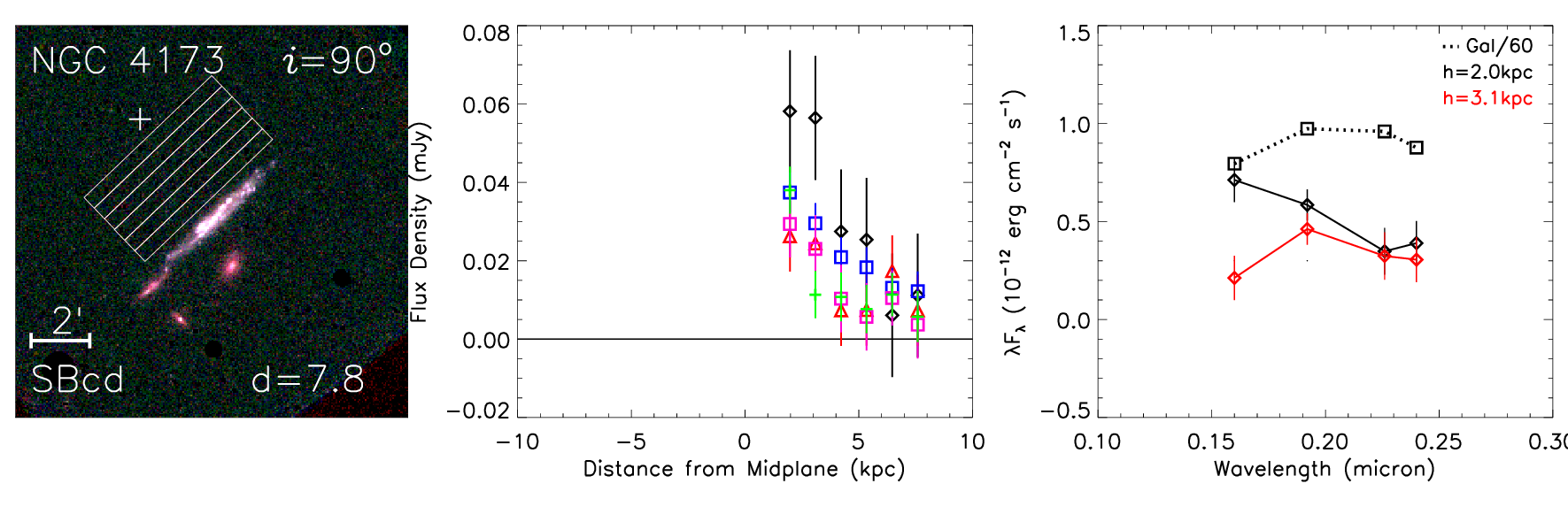}
\vspace{-0.65cm}\includegraphics[width=0.85\textwidth]{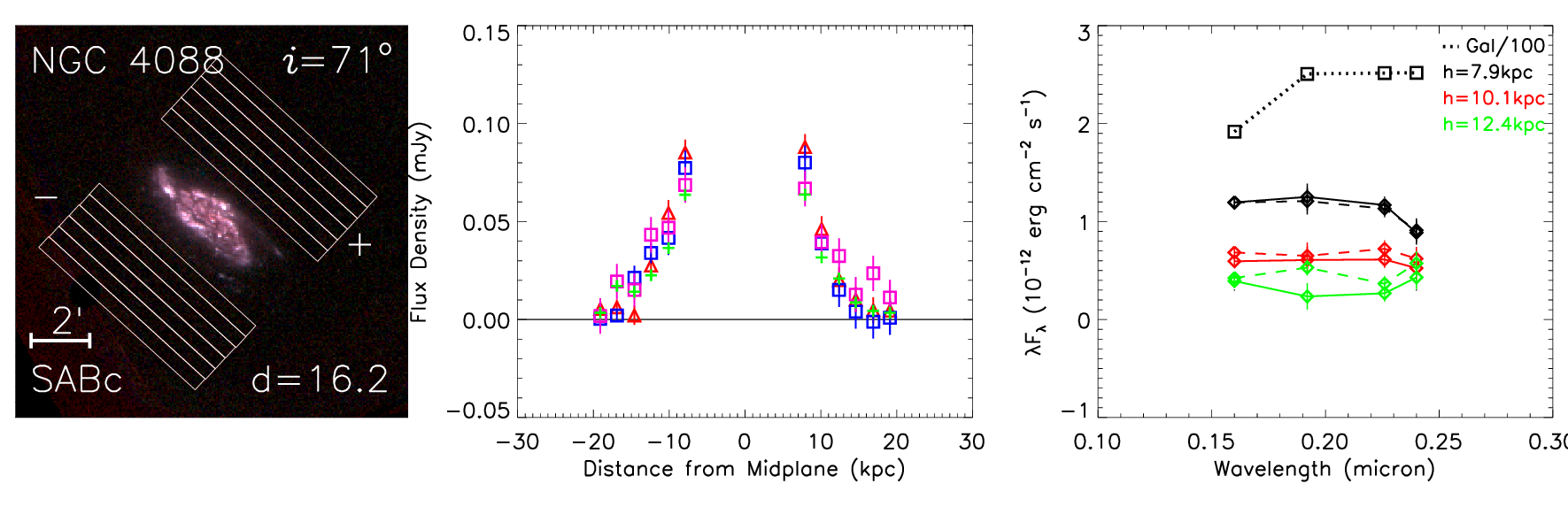}
\includegraphics[width=0.85\textwidth]{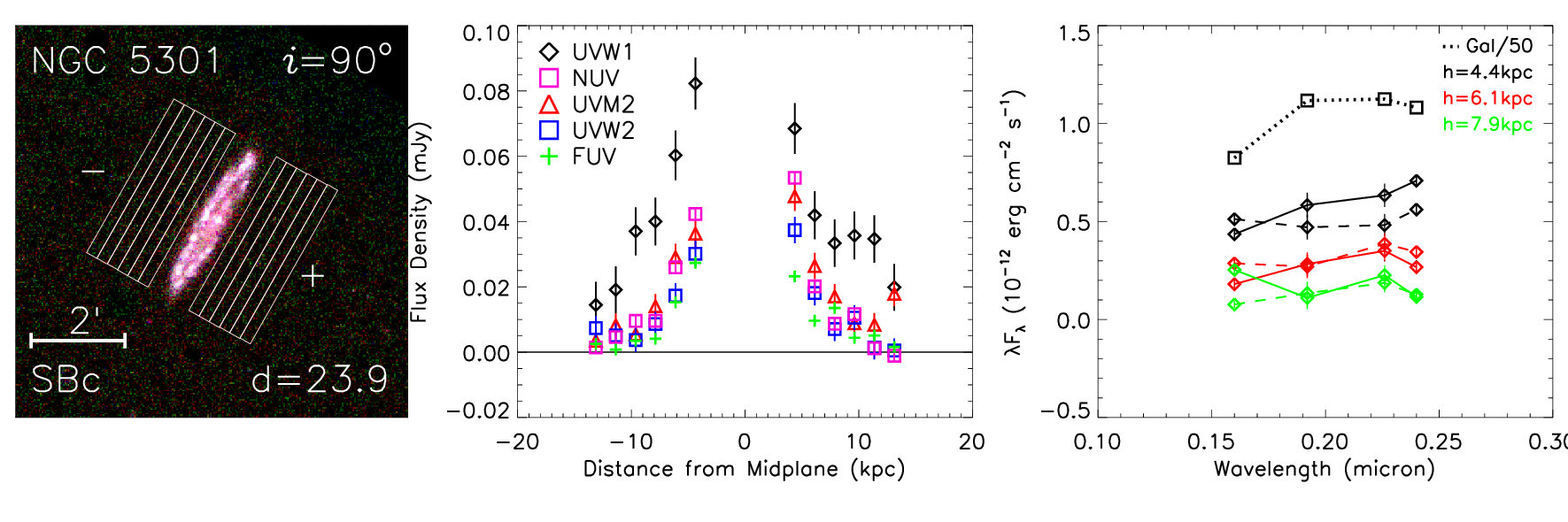}
\caption{\scriptsize
\label{figure.results_sc}
Halo emission fluxes and SEDs for Sc through Sd morphology galaxies in our sample.
\textit{Left panels}: RGB image made from the \uvwone{}, \uvmtwo{}, and \uvwtwo{}
images, annotated with the galaxy name, inclination, morphological type, and
distance in Mpc from Table~\ref{table.sample}.  Flux extraction bins are overlaid
in white, and extragalactic point sources have been masked.  \textit{Central panels}:
$F_{\nu}$ as a function of height above the midplane in each of the UV bands
(\uvwone{}: black diamonds, NUV: pink squares, \uvmtwo{}: red triangles,
\uvwtwo{}: blue squares, and FUV: green crosses).    
\textit{Right panels}: 4-point SED as a function of height above the midplane
using (from left to right) the FUV, \uvwtwo{}, \uvmtwo{}, and NUV fluxes. 
The SEDs are colored as a function of height, and solid/dashed lines at each
height represent the fluxes on the ``positive'' and ``negative'' sides of the
midplane respectively.}
\end{center}
\end{figure*}

\setcounter{figure}{6}
\begin{figure*}
\begin{center}
\mbox{}
\vspace*{-0.65cm}\includegraphics[width=0.85\textwidth]{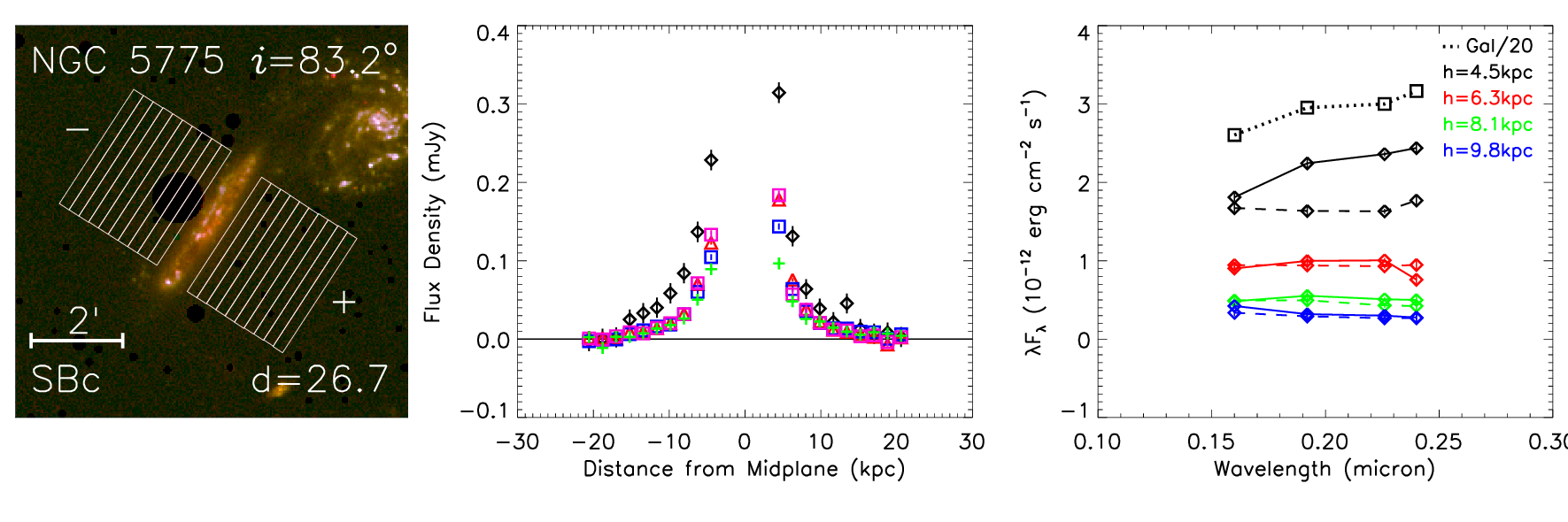}
\vspace{-0.65cm}\includegraphics[width=0.85\textwidth]{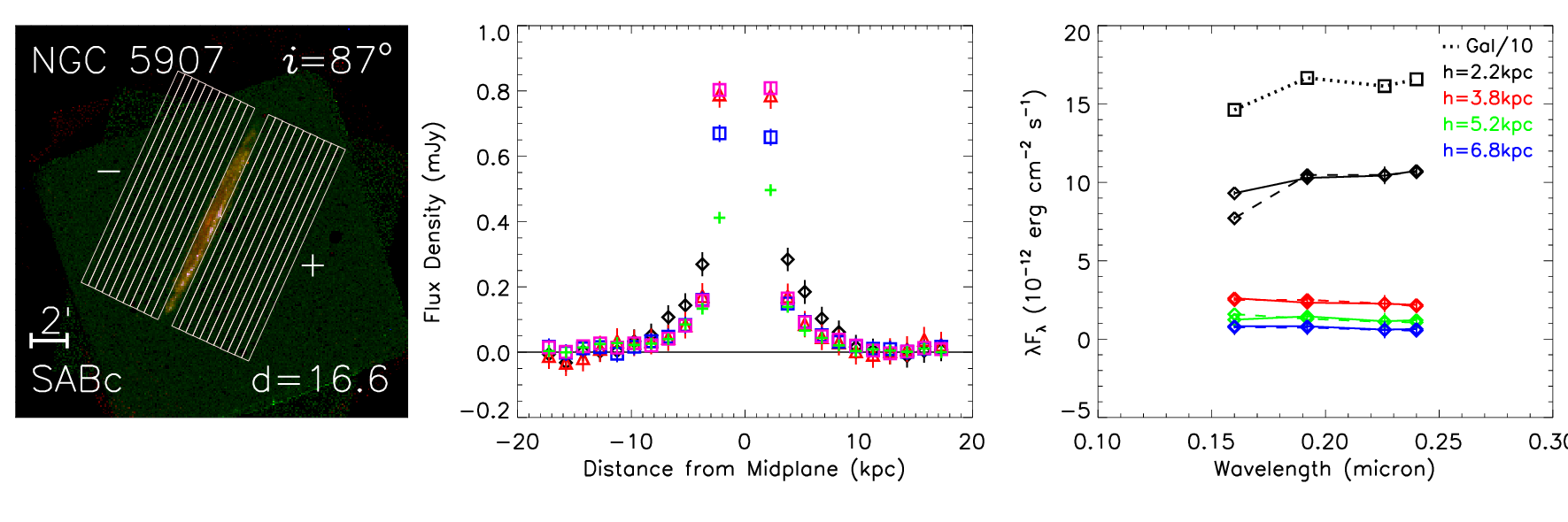}
\vspace{-0.65cm}\includegraphics[width=0.85\textwidth]{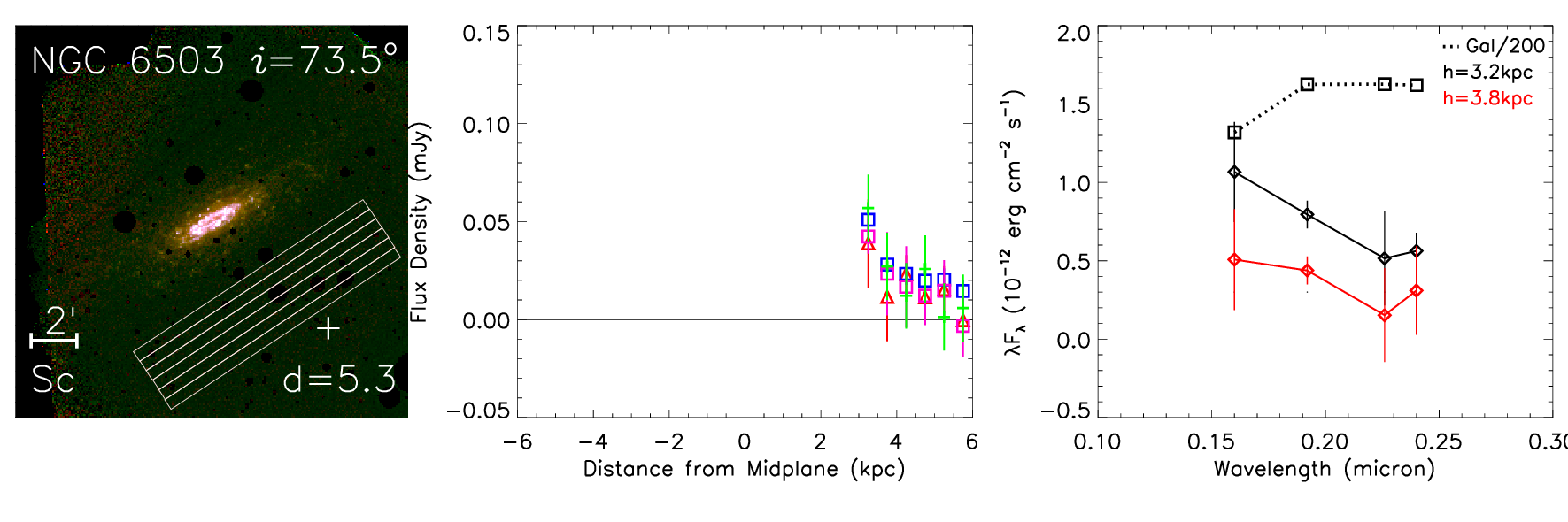}
\vspace{-0.65cm}\includegraphics[width=0.85\textwidth]{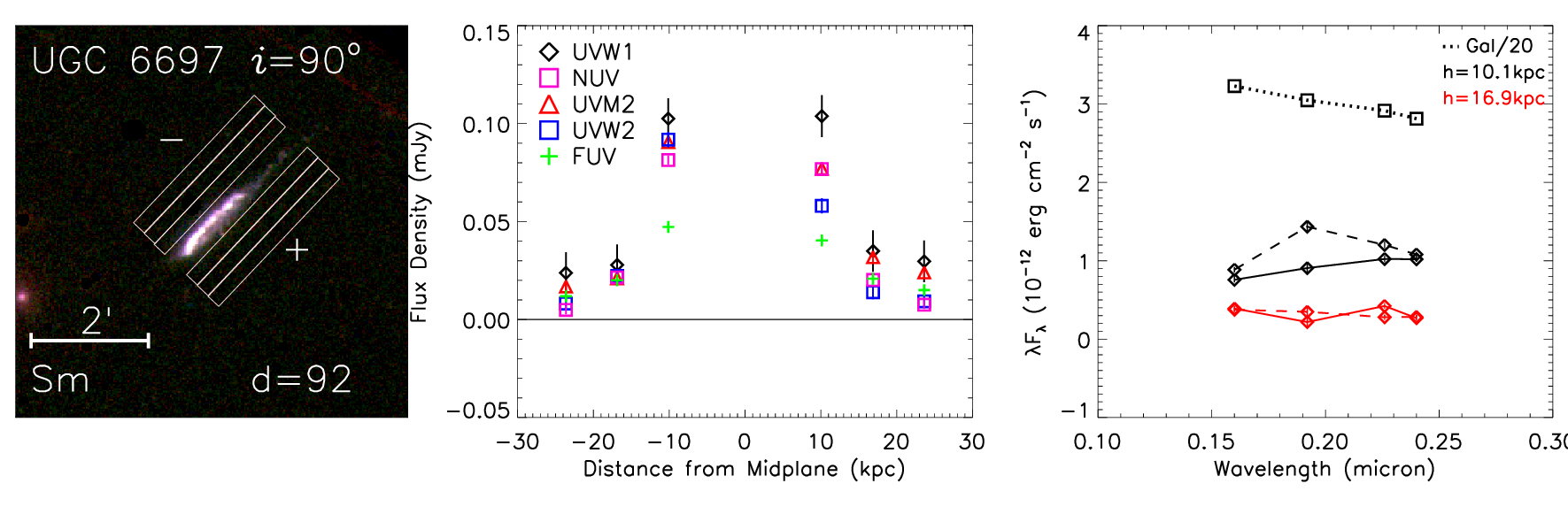}
\includegraphics[width=0.85\textwidth]{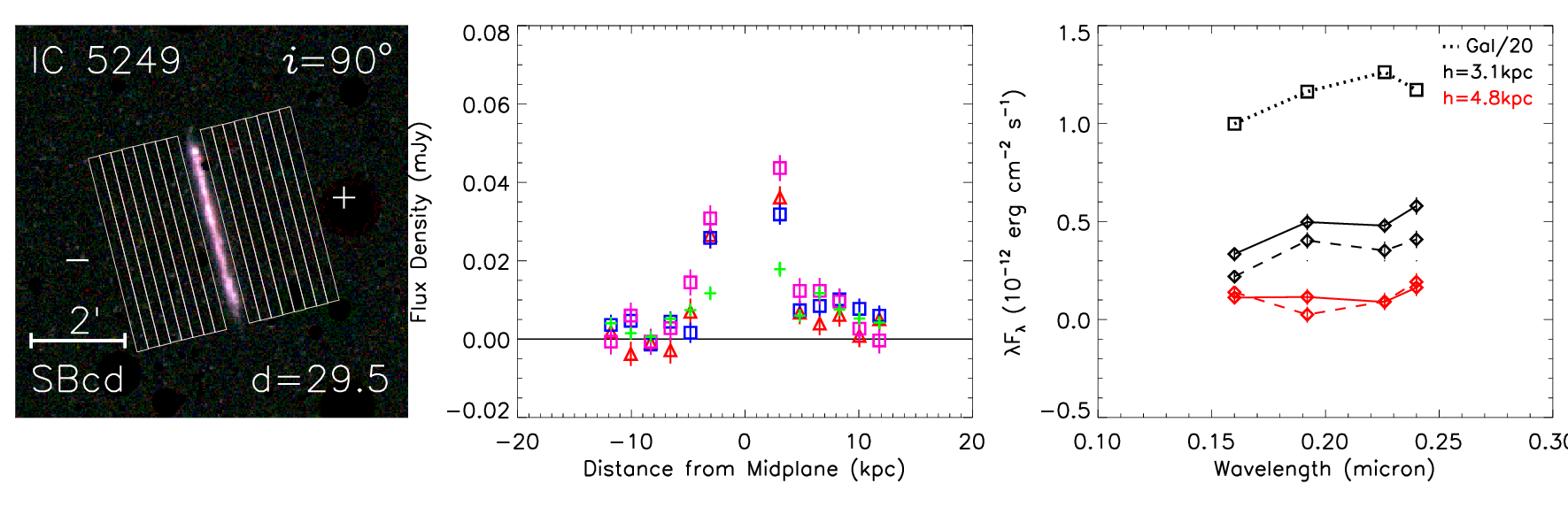}
\caption{Halo emission fluxes and SEDs for Scd spiral galaxies in our
  sample \textit{continued}.}
\end{center}
\end{figure*}

\begin{figure*}
\begin{center}
\mbox{}
\vspace*{-0.65cm}\includegraphics[width=0.85\textwidth]{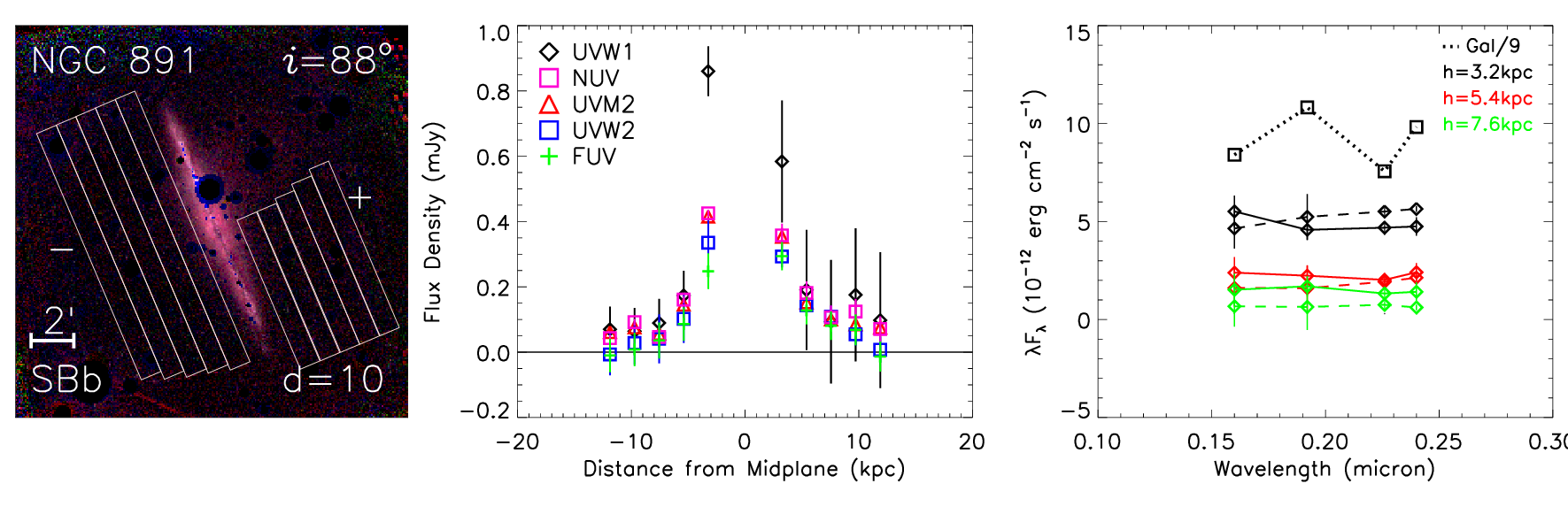}
\vspace{-0.65cm}\includegraphics[width=0.85\textwidth]{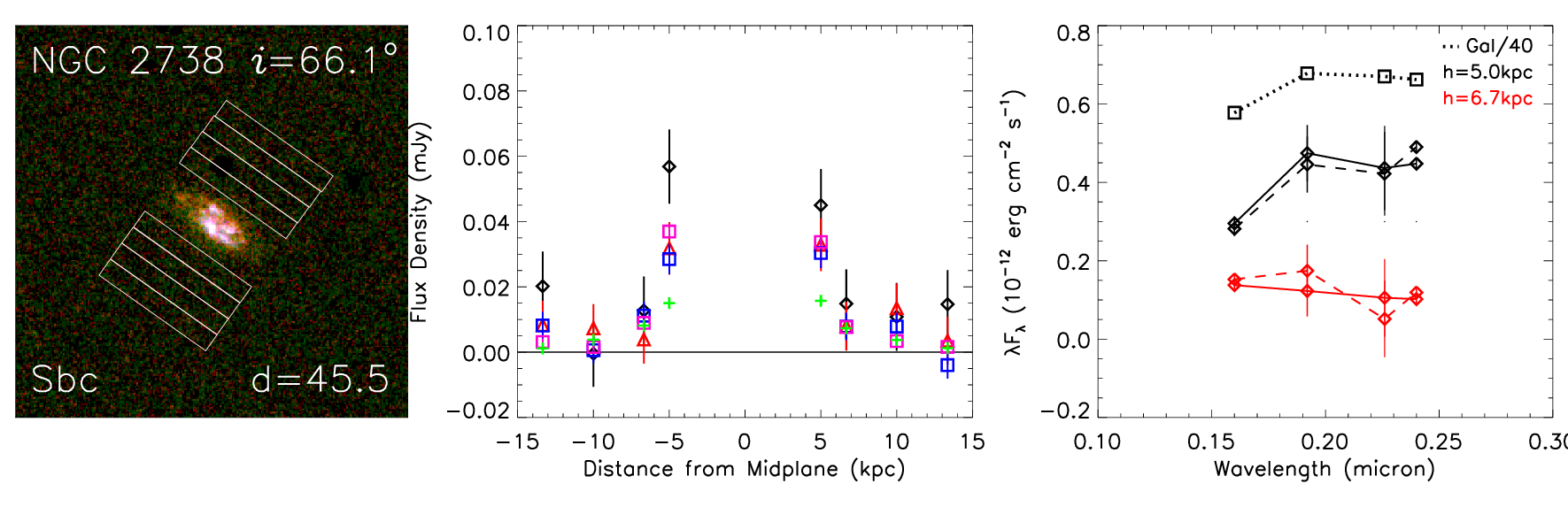}
\vspace{-0.65cm}\includegraphics[width=0.85\textwidth]{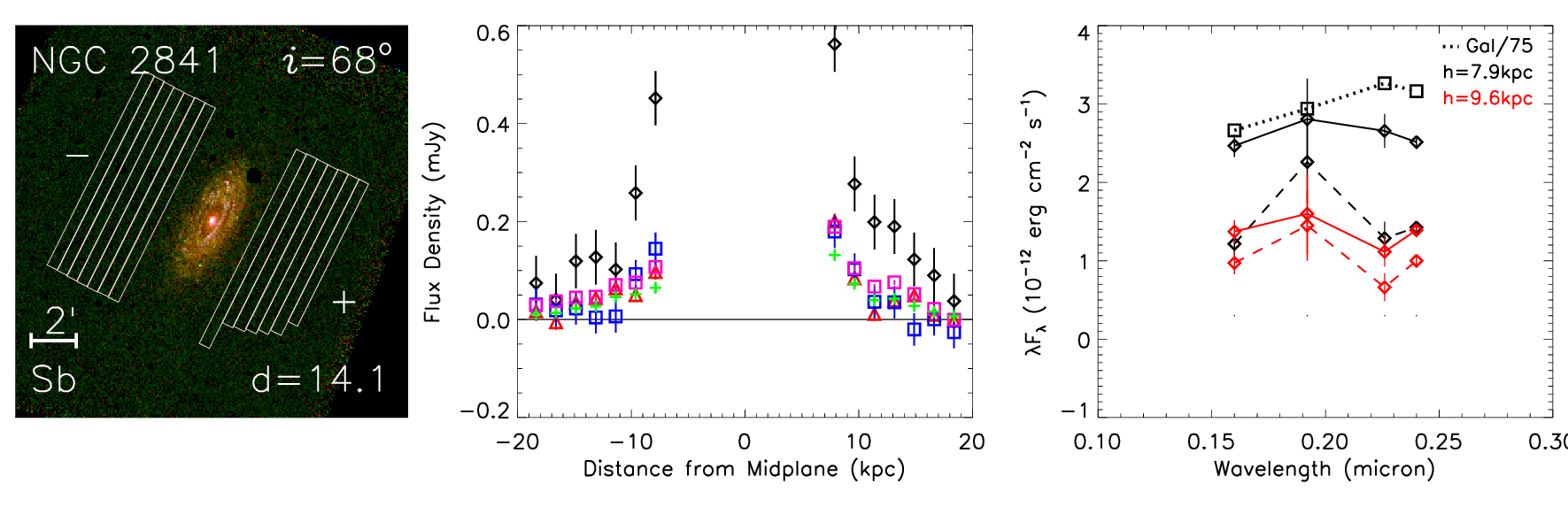}
\vspace{-0.65cm}\includegraphics[width=0.85\textwidth]{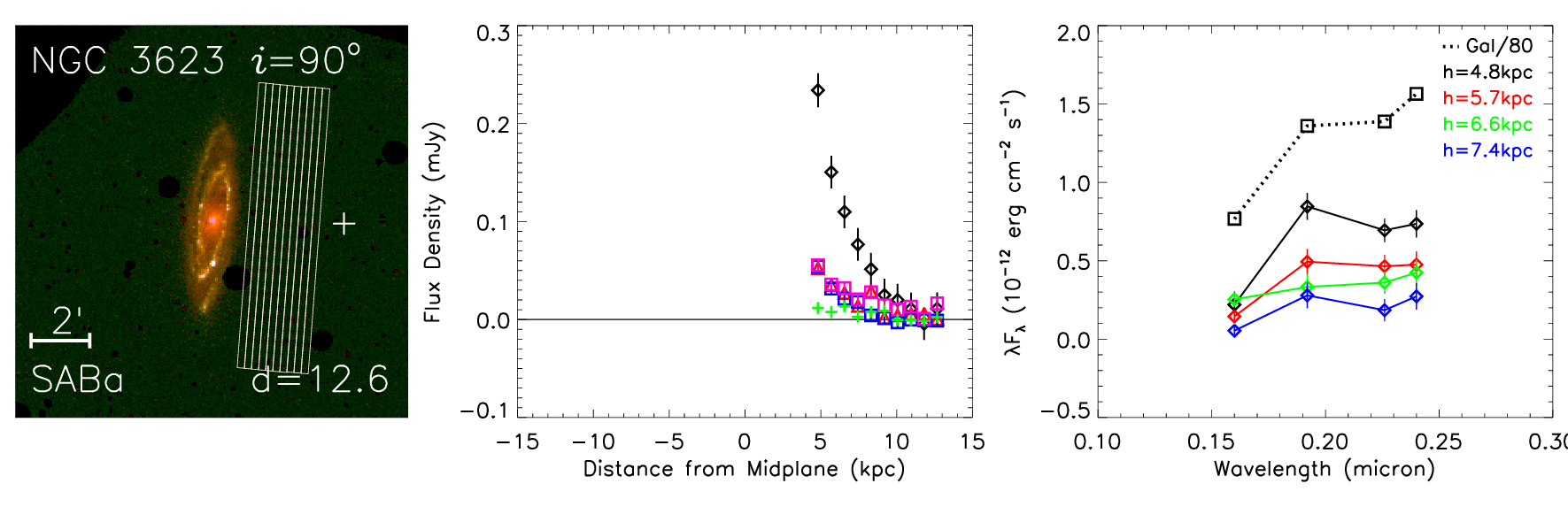}
\includegraphics[width=0.85\textwidth]{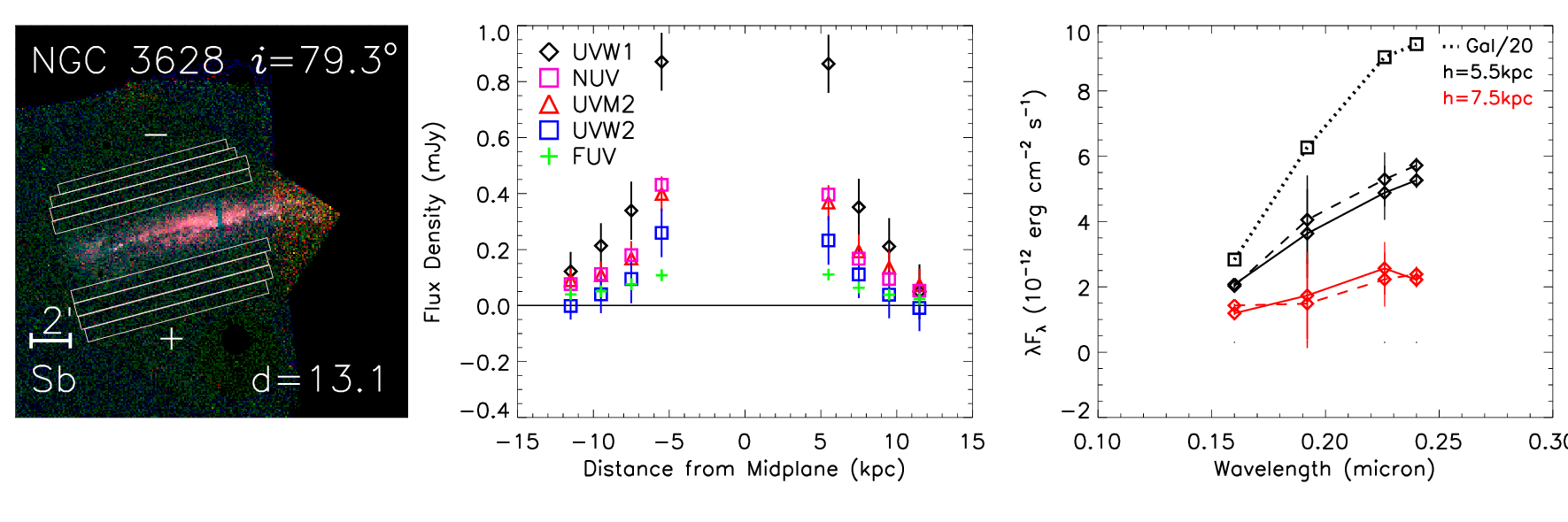}
\caption{\scriptsize \label{figure.results_sb}
Halo emission fluxes and SEDs for Sa through Sbc morphology galaxies
in our sample.  The plot format and symbols are the same as in
Figure~\ref{figure.results_sc}.}
\end{center}
\end{figure*}

\setcounter{figure}{7}
\begin{figure*}
\begin{center}
\mbox{}
\vspace*{-0.65cm}\includegraphics[width=0.85\textwidth]{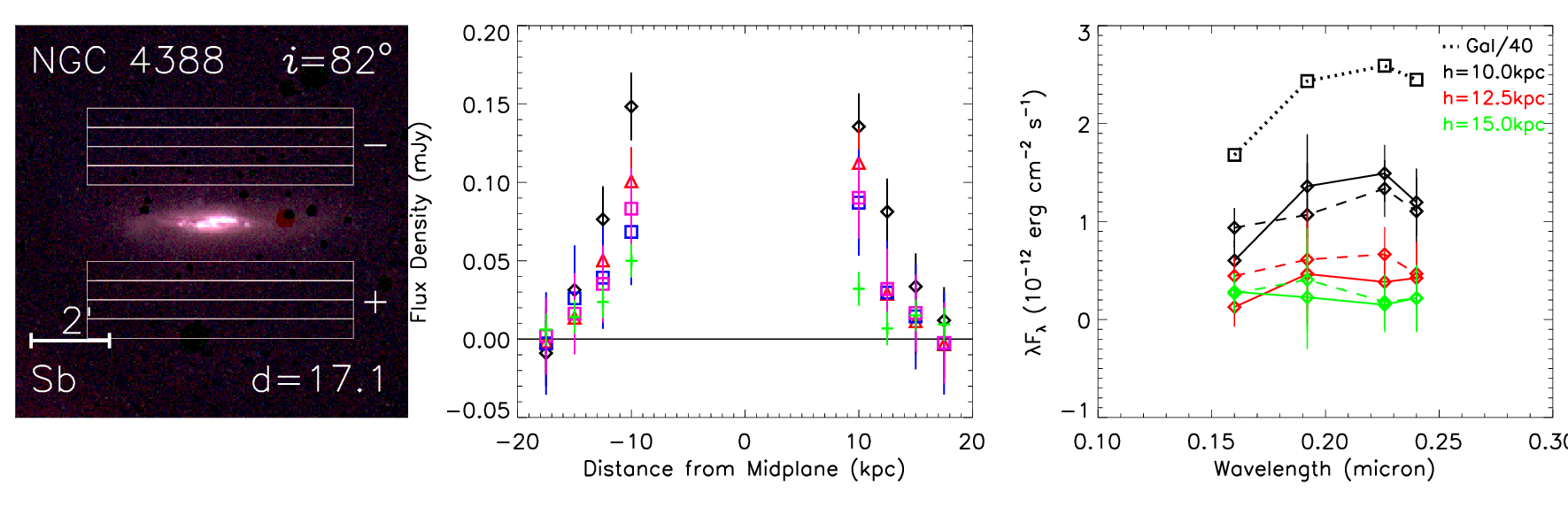}
\vspace{-0.65cm}\includegraphics[width=0.85\textwidth]{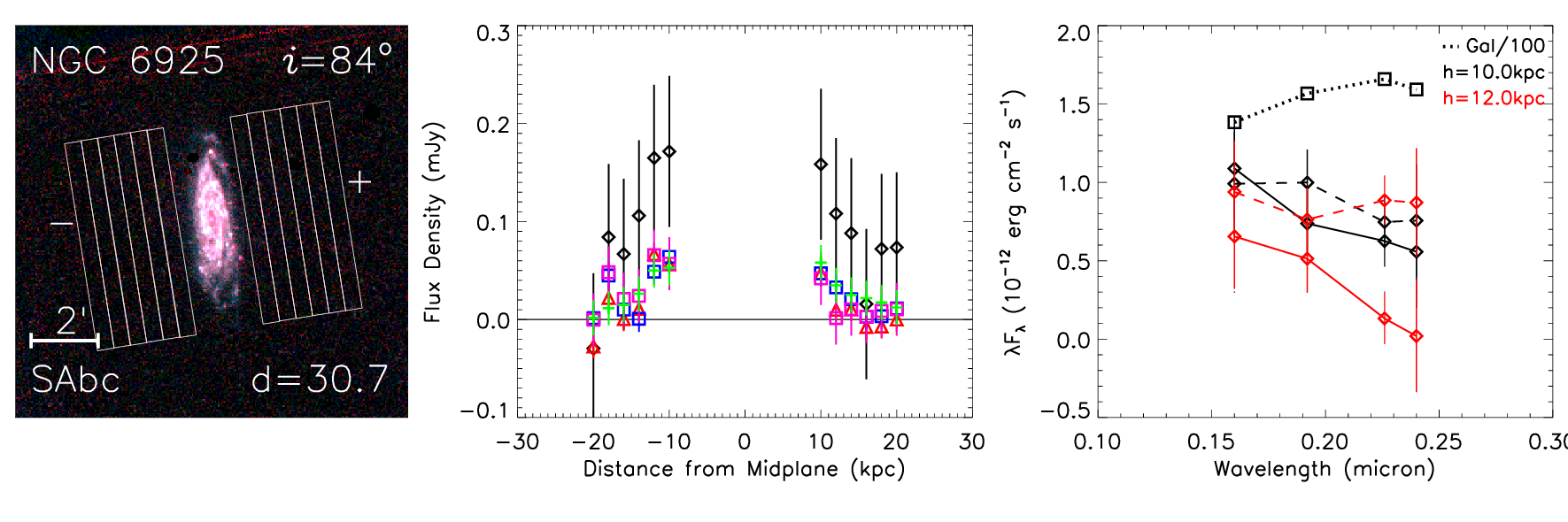}
\vspace{-0.65cm}\includegraphics[width=0.85\textwidth]{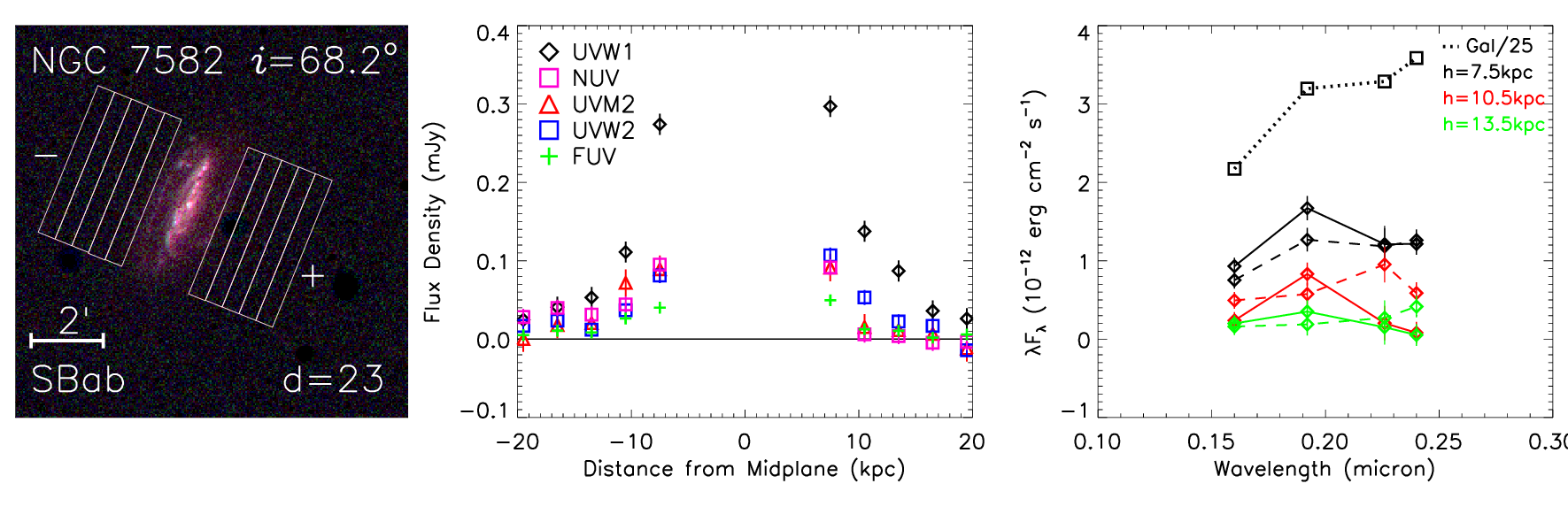}
\includegraphics[width=0.85\textwidth]{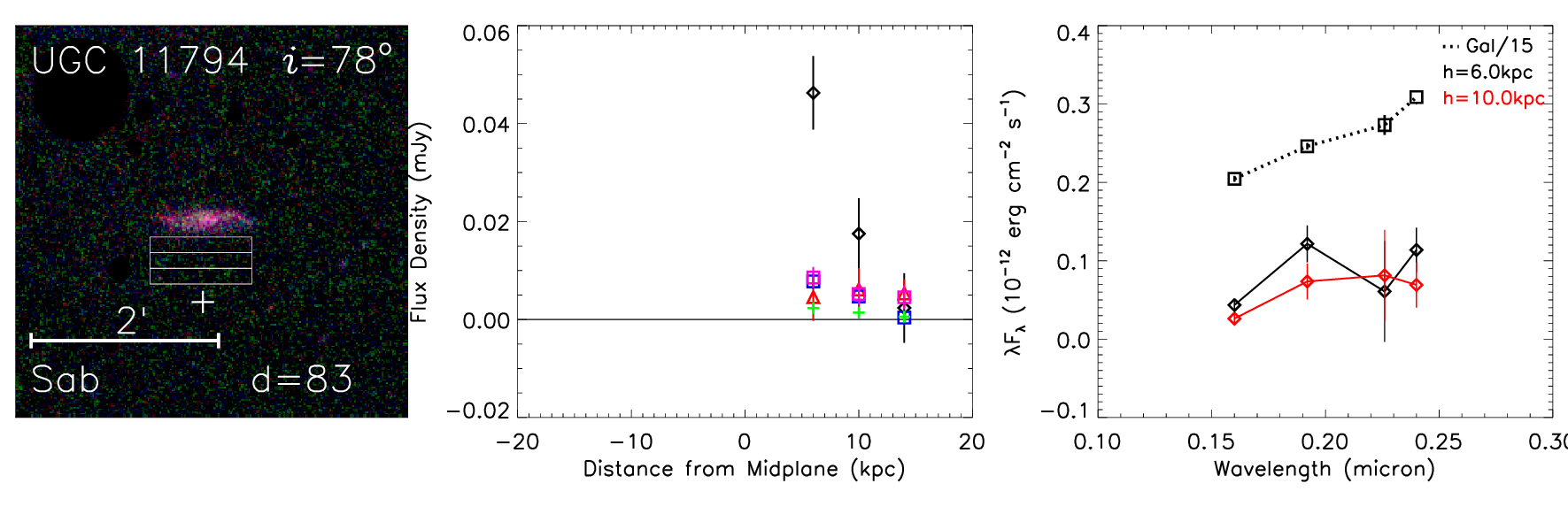}
\caption{\scriptsize
Halo emission fluxes and SEDs for Sa through Sbc morphology galaxies
in our sample \textit{continued}.}  
\end{center}
\end{figure*}

\begin{figure*}
\begin{center}
\mbox{}
\vspace*{-0.65cm}\includegraphics[width=0.85\textwidth]{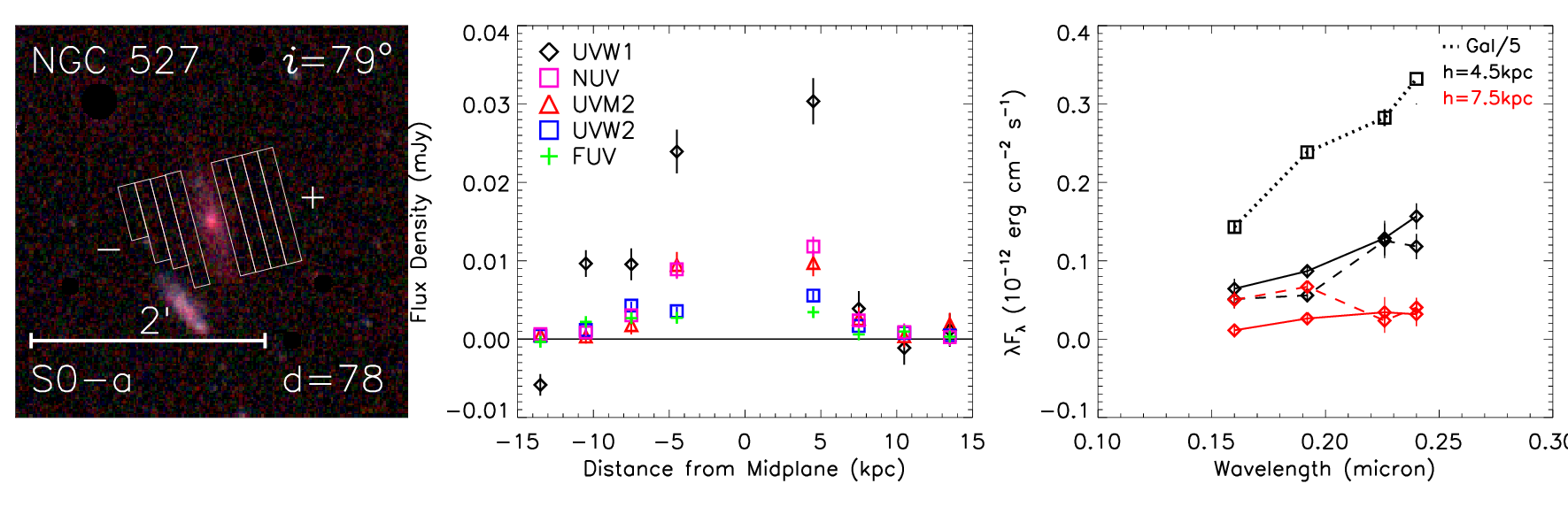}
\vspace{-0.65cm}\includegraphics[width=0.85\textwidth]{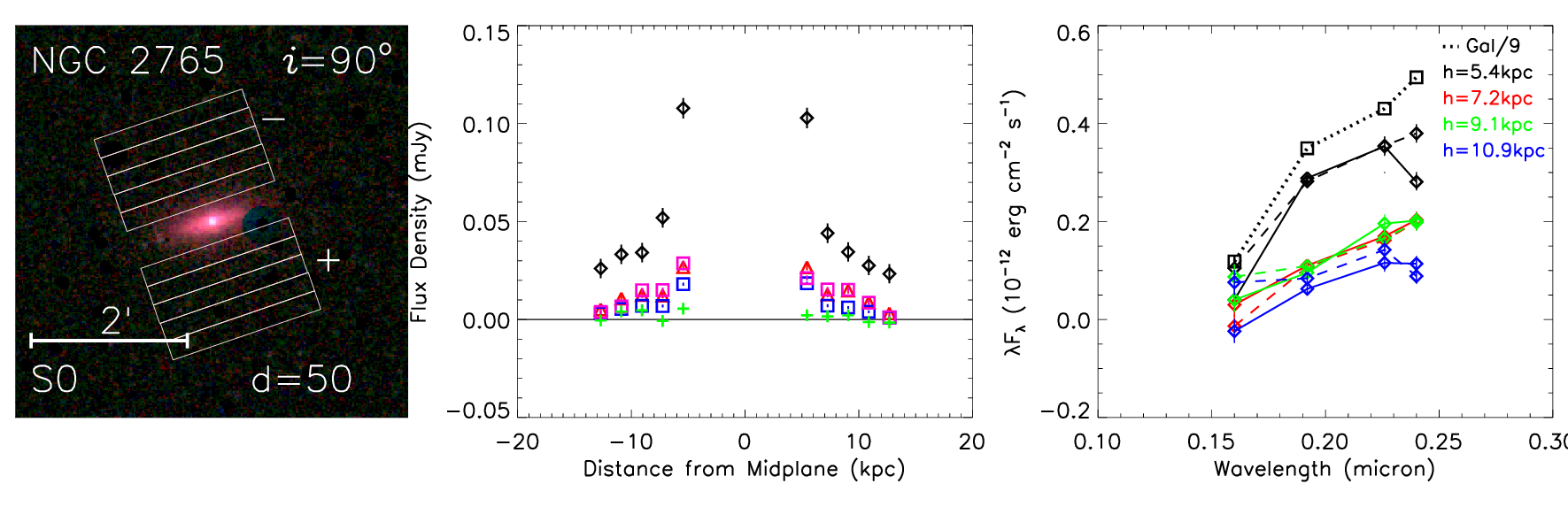}
\vspace{-0.65cm}\includegraphics[width=0.85\textwidth]{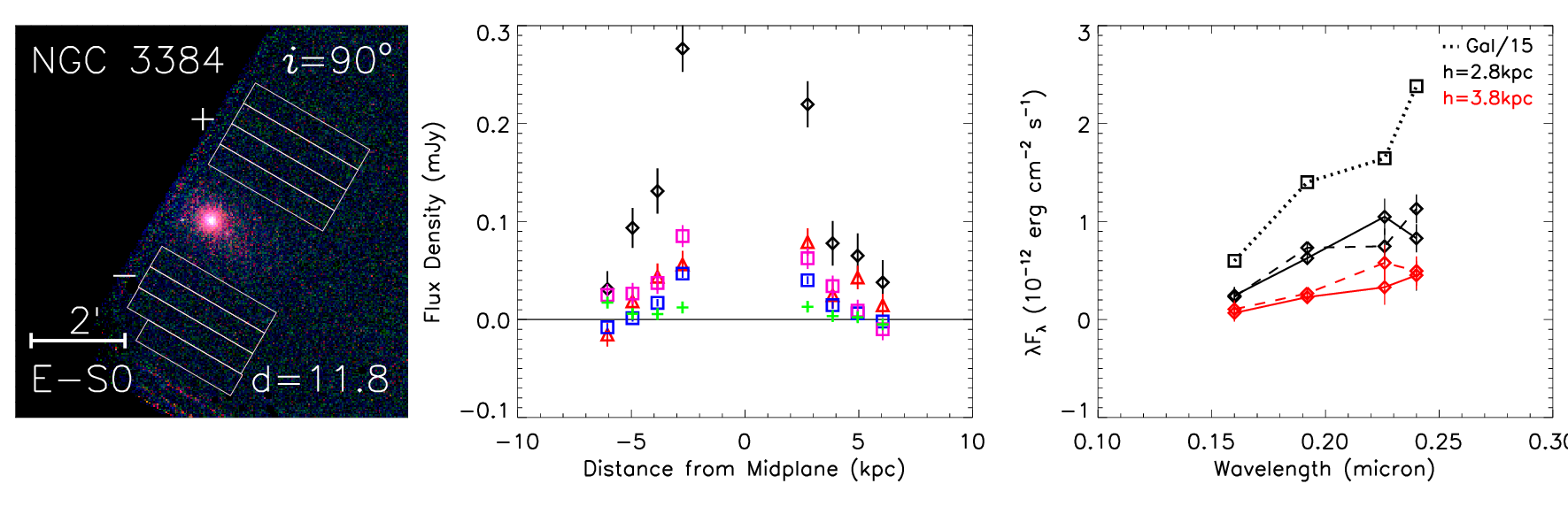}
\vspace{-0.65cm}\includegraphics[width=0.85\textwidth]{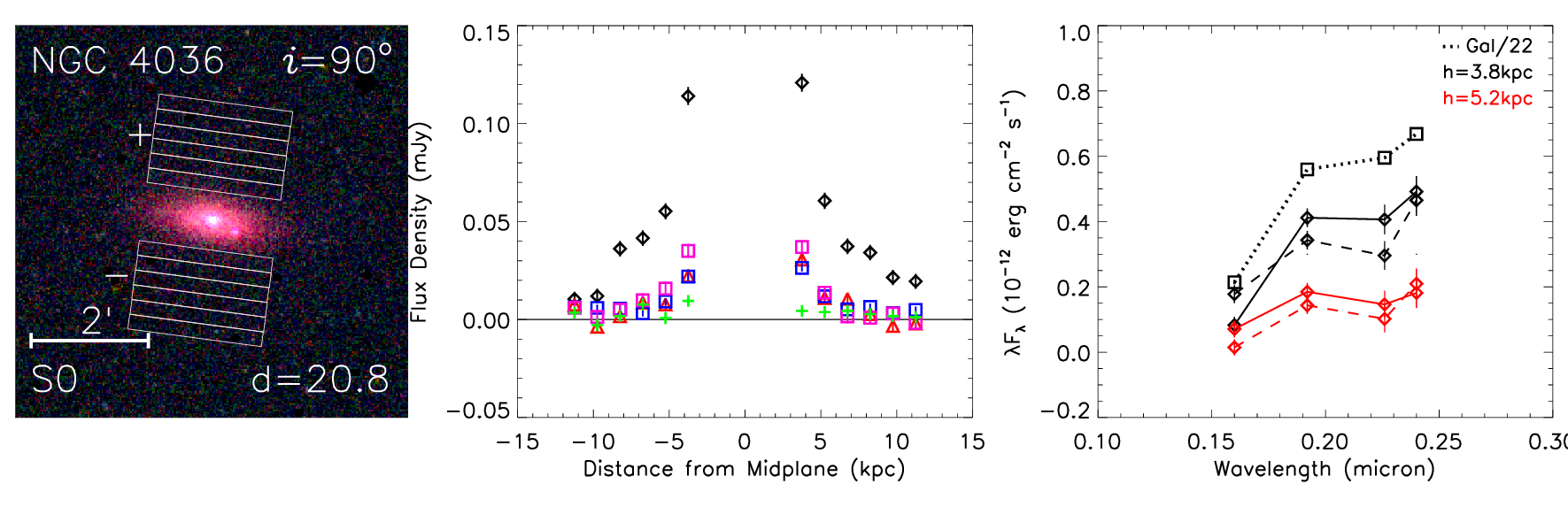}
\includegraphics[width=0.85\textwidth]{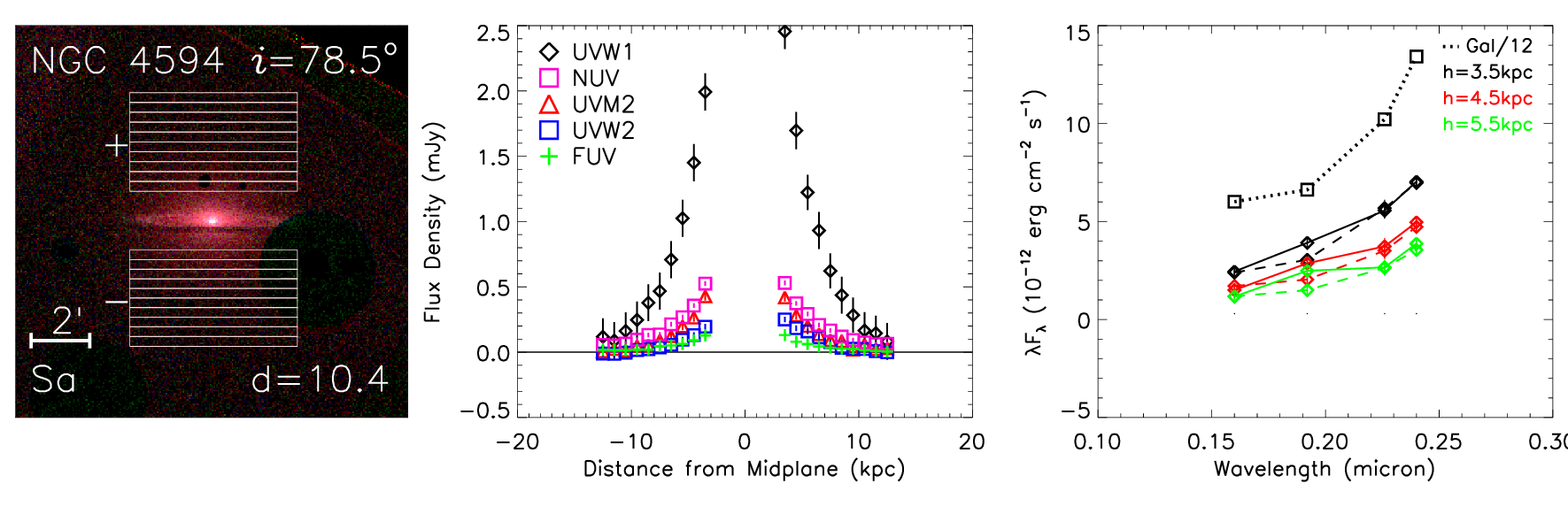}
\caption{\scriptsize \label{figure.results_s0}
Halo emission fluxes and SEDs for highly inclined S0 galaxies in our sample
(NGC~4594 is also included).  The plot format and symbols are the same as in
Figure~\ref{figure.results_sc}.}
\end{center}
\end{figure*}

\setcounter{figure}{8}
\begin{figure*}
\begin{center}
\mbox{}
\vspace*{-0.65cm}\includegraphics[width=0.85\textwidth]{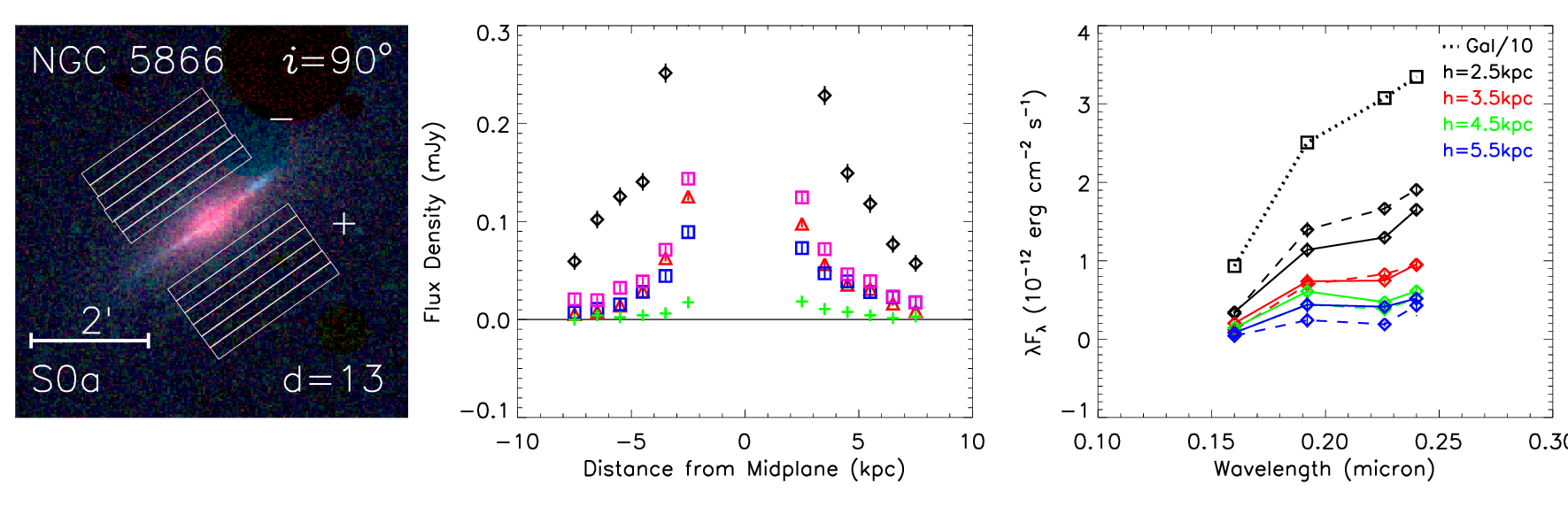}
\includegraphics[width=0.85\textwidth]{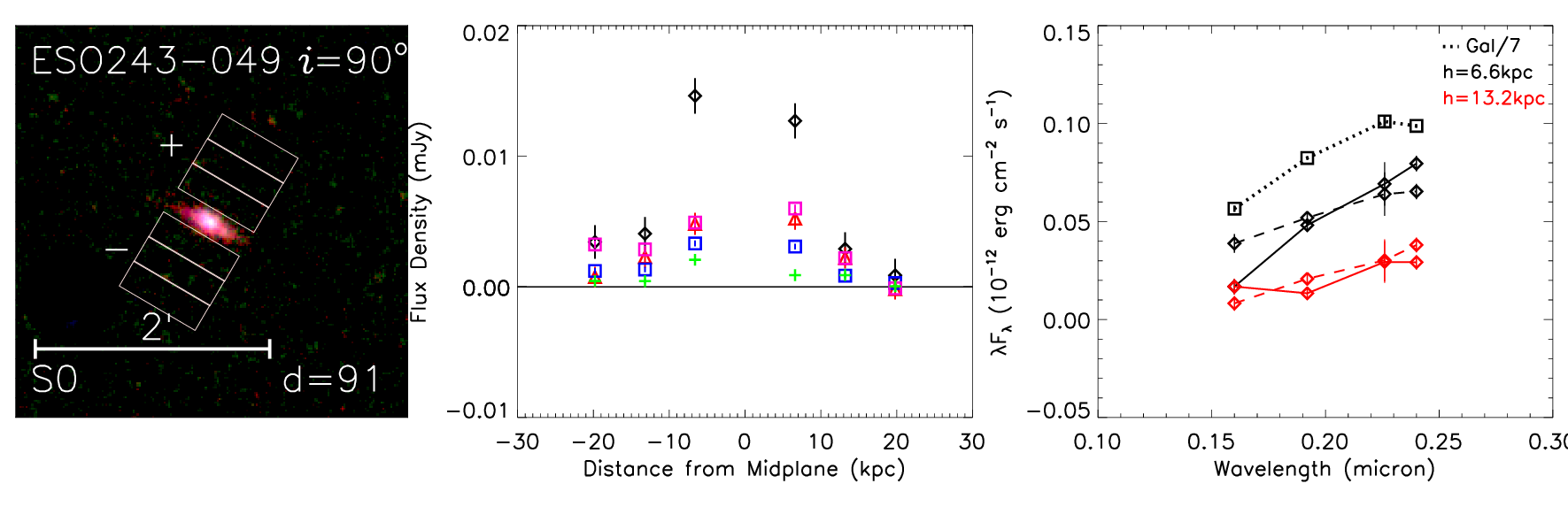}
\caption{\scriptsize Halo emission fluxes and SEDs for highly inclined S0 galaxies in our
  sample \textit{continued}}
\end{center}
\end{figure*}

\begin{figure*}
\begin{center}
\mbox{}
\vspace*{-0.65cm}\includegraphics[width=0.85\textwidth]{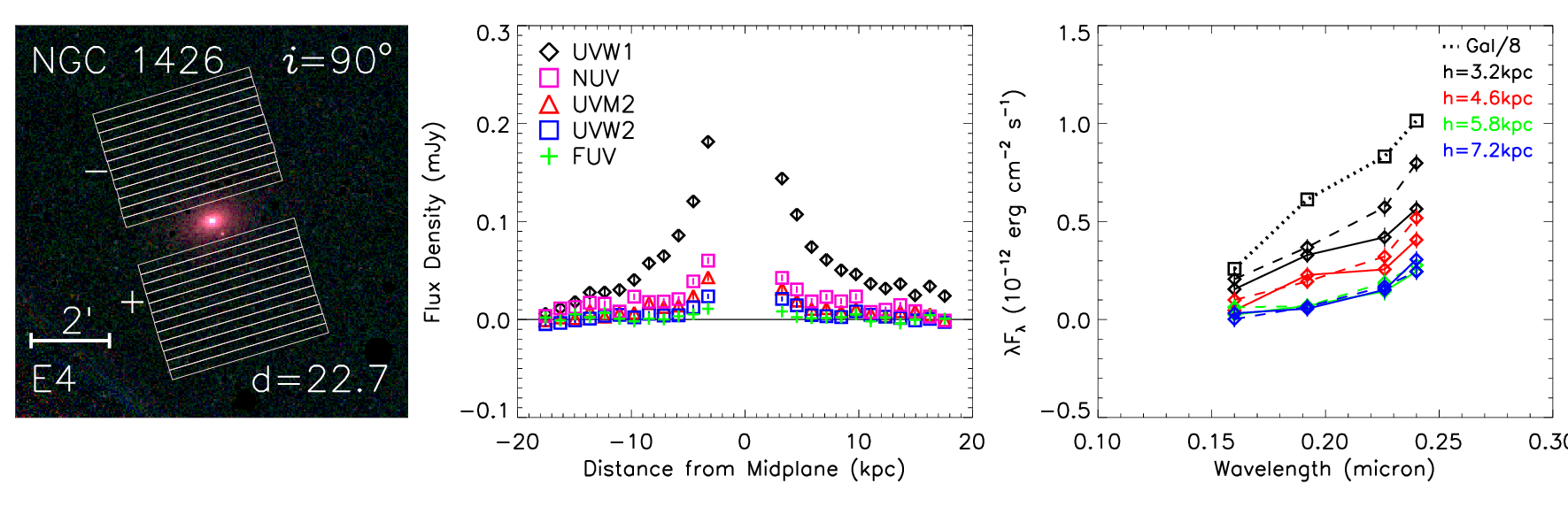}
\vspace{-0.65cm}\includegraphics[width=0.85\textwidth]{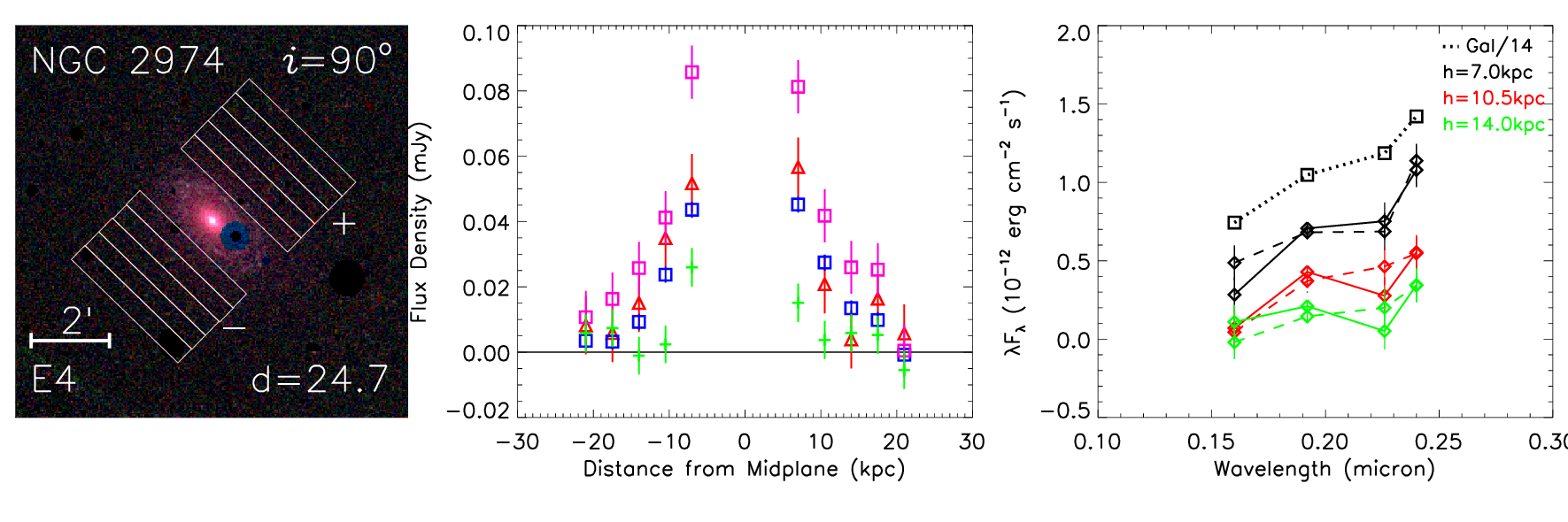}
\vspace{-0.65cm}\includegraphics[width=0.85\textwidth]{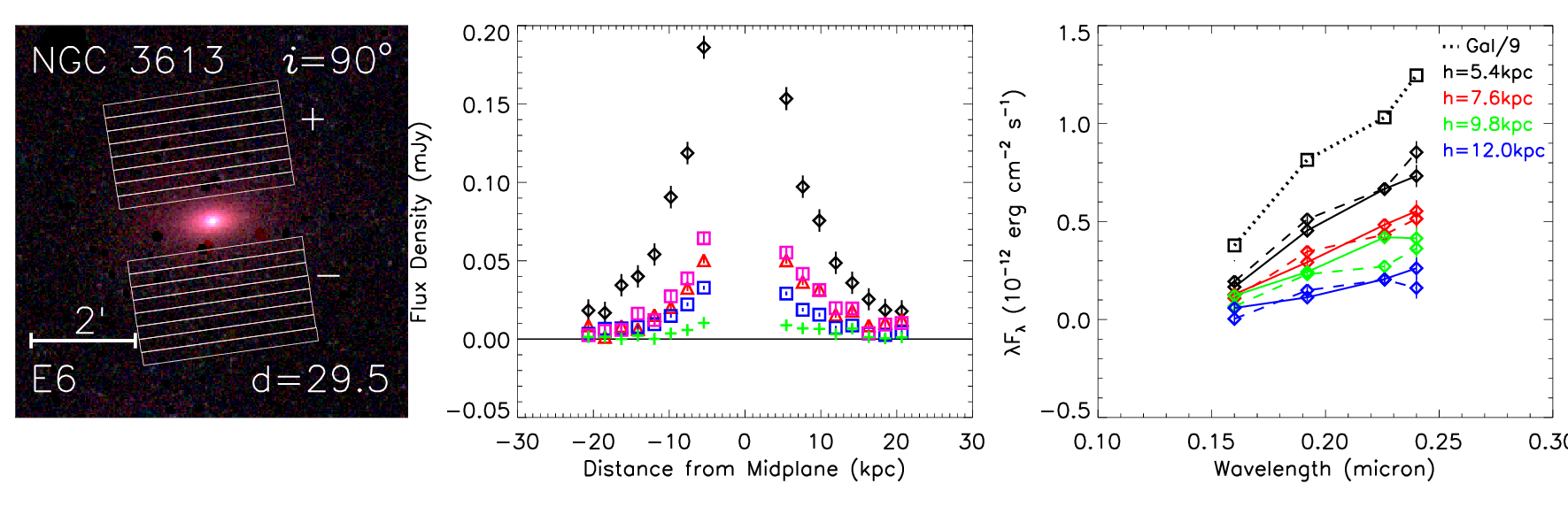}
\includegraphics[width=0.85\textwidth]{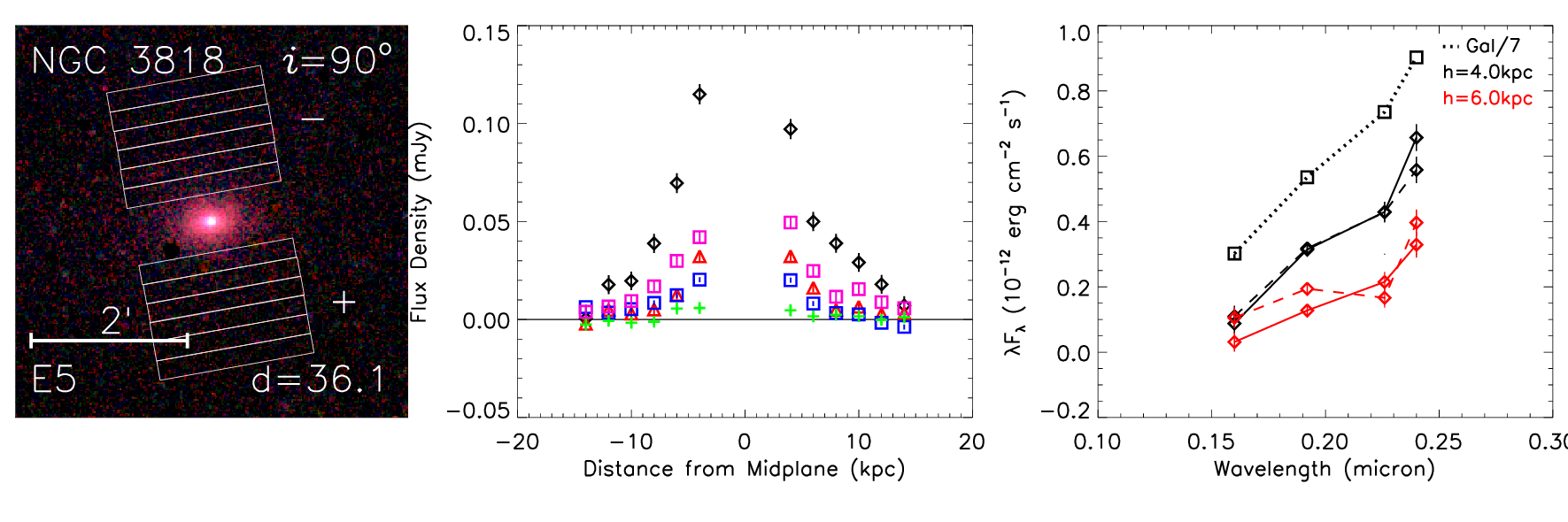}
\caption{\scriptsize \label{figure.results_e}
Halo emission fluxes and SEDs for elongated elliptical galaxies 
in our sample.  The plot format and symbols are the same as in
Figure~\ref{figure.results_sc}.}
\end{center}
\end{figure*}

Halo rings are only detected above $F_{\text{star}} = 10$\,counts\,s$^{-1}$.
The detections are shown in the left panel of Figure~\ref{figure.uvm2_halo_ring_flux}.
Nondetections are omitted for clarity, but are shown in the right panel.
Above $F_{\text{star}} \sim 130$\,counts\,s$^{-1}$ the scatter increases markedly
for unknown reasons.  One possibility is that we see a secondary
ring.  Very bright stars sometimes produce
multiple halo rings, where one is much brighter than the others.  If a bright
star's primary ring fell off the chip but its secondary did not, we might mistake
the secondary ring for the primary.  In any case, as a first step to determine
the dependence of $F_{\text{ring}}$ on $F_{\text{star}}$ we fit the 
detected $F_{\text{ring}}$ values below $F_{\text{star}} = 130$\,counts\,s$^{-1}$
with simple models.  The scatter precludes formally good fits for any model
we tried (linear, log-linear, quadratic, and exponential), but the exponential
model is the best fit.  The best exponential and linear models are overlaid in
Figure~\ref{figure.uvm2_halo_ring_flux}, where it is clear that the exponential
model does a better job at reproducing the curvature in the data.

However, it is clear from the right panel of Figure~\ref{figure.uvm2_halo_ring_flux}
that the exponential model becomes unphysical for stars with
$F_{\text{star}} \lesssim 0.2$\,counts\,s$^{-1}$ and is also disfavored by the 
upper limits we have measured.  
The constraints provided by individual stars
are limited by the \Swift{} background, so we averaged over
the $F_{\text{ring}}$ measuremeants below $F_{\text{star}} = 10$\,counts\,s$^{-1}$,
grouping stars in each decadal bin.  After averaging there were still no detections,
but the upper limits were reduced (diamonds in Figure~\ref{figure.uvm2_halo_ring_flux}).

Significantly tightening the constraints would require a much larger
sample, but at low count rates another way of averaging is to use small asterisms.
We measured $F_{\text{ring}}$ in the region around isolated groups of dim stars,
where the asterisms comprise 3--15 members contained within 1\,arcmin in the
inner portion of the chip.  Since these stars should produce overlapping halo
rings, we measured the average $F_{\text{ring}}$ per star and the mean count
rate $F_{\text{star}}$.  Again, there are no detections, but the upper limits
(shown as inverted triangles in Figure~\ref{figure.uvm2_halo_ring_flux}) improve
the constraint on $F_{\text{ring}}$ at low count rates. 

The functional form of $F_{\text{ring}}$ below $F_{\text{star}} \sim 10$\,counts\,s$^{-1}$
cannot be determined, but we can make some inferences from the limits and the
requirement that $F_{\text{ring}} < F_{\text{star}}$
(Figure~\ref{figure.uvm2_halo_ring_flux}).  The linear fit to detected
rings suggests a ratio of $F_{\text{ring}}/F_{\text{star}} \lesssim
0.01$, while the cluster of detected rings at $F_{\text{star}} \sim
10$\,counts\,s$^{-1}$ is consistent with a ratio of 0.02--0.04.  The
averaged limit between $F_{\text{star}} =
1-10$\,counts\,s$^{-1}$ rules out a significantly larger value.  
At lower count rates, the limits only constrain the ratio to be
$F_{\text{ring}}/F_{\text{star}} < 0.1$.

The total instrumental scattered light we expect from the galaxy is between
$F_{\text{scat}} = 0.01-0.1 F_{\text{gal}}$.  For $F_{\text{gal}} =
200$\,counts\,s$^{-1}$ (at the high end of our sample), we would
expect $F_{\text{scat}}$ between 2--20\,counts\,s$^{-1}$.  However,
galaxies with higher fluxes tend to be closer, and the angular size of
the galaxy is important because a given clump in the galaxy only
contaminates the surrounding 2\,arcmin with instrumental scattered
light. For a galaxy larger than 2\,arcmin, only some of the scattered
light contaminates the halo (the rest is coincident on the chip with
the galaxy itself), and the exact contribution at any point
in the halo depends on the region of the galaxy within 2\,arcmin of
that point on the chip.  In our sample, we find that the galaxy count
rates that could contribute to instrumental scattered light in the
halo are 0.1--50\,counts\,s$^{-1}$, with higher count rates nearer the
galaxy.  In the ``worst case scenario'' (bright galaxy, relatively
compact, measurements near the disk), we expect $F_{\text{scat}} \sim
0.1 B_{\text{sky}}$, whereas at the low end we expect $F_{\text{scat}}
< 0.001 B_{\text{sky}}$.  This is a wide range, but (based on count
rates, angular sizes, and Figure~\ref{figure.uvm2_halo_ring_flux}), we
suppose that $F_{\text{scat}}$ is 1\% or less of $B_{\text{sky}}$ for
most of our sample.  The typical $\sigma_{\text{sky}}$ is 0.1--4\% of
$B_{\text{sky}}$.

\begin{turnpage}
\begin{deluxetable*}{l|ccccc|ccccc|c}
\vspace{0.5cm}
\tablenum{4}
\tabletypesize{\scriptsize}
\tablecaption{Integrated UV Fluxes of Halos and Galaxies}
\tablewidth{0pt}
\tablehead{
\colhead{} & \multicolumn{5}{c}{Halo Flux} & \multicolumn{5}{c}{Galaxy Flux} & \\ 
\colhead{Galaxy} & \colhead{\uvwone{}} & \colhead{NUV} & \colhead{\uvmtwo{}} &
\colhead{\uvwtwo{}} & \colhead{FUV} & \colhead{\uvwone{}} & \colhead{NUV} & \colhead{\uvmtwo{}} &
\colhead{\uvwtwo{}} & \colhead{FUV} & \colhead{\uvwtwo{} Corr}  \\
\colhead{} & \colhead{(mJy)} & \colhead{(mJy)} & \colhead{(mJy)} &
\colhead{(mJy)} & \colhead{(mJy)} & \colhead{(mJy)} & \colhead{(mJy)} & \colhead{(mJy)} &
\colhead{(mJy)} & \colhead{(mJy)} & \colhead{} 
}
\startdata
ESO 243-049 & 0.039$\pm$0.003 & 0.019$\pm$0.001 & 0.015$\pm$0.002 & 0.010$\pm$0.001 & 0.005$\pm$0.001 & 0.175$\pm$0.002 & 0.052$\pm$0.001 & 0.053$\pm$0.002 & 0.037$\pm$0.001 & 0.021$\pm$0.001 & 0.65 \\
IC 5249     & 0.25$\pm$0.02   & 0.13$\pm$0.01   & 0.09$\pm$0.01   & 0.110$\pm$0.009 & 0.084$\pm$0.006 & 2.37$\pm$0.01 & 1.78$\pm$0.01 & 1.9$\pm$0.01 & 1.49$\pm$0.01 & 1.06$\pm$0.01 & 0.93 \\
NGC 24      & 1.220$\pm$0.014 & 0.55$\pm$0.01   & 0.6$\pm$0.1     & 0.55$\pm$0.09   & 0.45$\pm$0.02   & 14.3$\pm$0.05 & 11.1$\pm$0.02 & 11.3$\pm$0.04 & 9.86$\pm$0.04 & 7.92$\pm$0.03 & 0.93 \\
NGC 527     & 0.072$\pm$0.006 & 0.028$\pm$0.001 & 0.027$\pm$0.003 & 0.018$\pm$0.001 & 0.013$\pm$0.001 & 0.43$\pm$0.01 & 0.125$\pm$0.002 & 0.105$\pm$0.003 & 0.076$\pm$0.002 & 0.037$\pm$0.002 & 0.65 \\
NGC 891     & 2.4$\pm$0.5     & 1.6$\pm$0.1     & 1.52$\pm$0.07   & 1.1$\pm$0.2     & 0.9$\pm$0.1     & 13.0$\pm$0.1 & 6.5$\pm$0.1 & 5.2$\pm$0.1 & 6.21$\pm$0.05 & 4.13$\pm$0.05 & 0.93 \\
NGC 1426    & 1.34$\pm$0.02   & 0.45$\pm$0.02   & 0.26$\pm$0.02   & 0.11$\pm$0.01   & 0.06$\pm$0.01   & 1.91$\pm$0.01 & 0.612$\pm$0.005 & 0.502$\pm$0.005 & 0.314$\pm$0.005 & 0.11$\pm$0.004 & 0.60 \\
NGC 2738    & 0.17$\pm$0.03   & 0.097$\pm$0.003 & 0.11$\pm$0.02   & 0.09$\pm$0.01   & 0.056$\pm$0.002 & 2.94$\pm$0.05 & 2.00$\pm$0.01 & 2.02$\pm$0.04 & 1.84$\pm$0.02 & 1.233$\pm$0.007 & 0.93 \\
NGC 2765    & 0.49$\pm$0.02   & 0.131$\pm$0.004 & 0.133$\pm$0.004 & 0.077$\pm$0.003 & 0.016$\pm$0.004 & 1.01$\pm$0.01 & 0.335$\pm$0.003 & 0.292$\pm$0.002 & 0.205$\pm$0.003 & 0.057$\pm$0.004 & 0.65 \\
NGC 2841    & 2.6$\pm$0.2     & 0.92$\pm$0.03   & 0.69$\pm$0.01   & 0.3$\pm$0.1     & 0.58$\pm$0.03   & 35.6$\pm$0.1 & 17.87$\pm$0.03 & 18.4$\pm$0.05 & 14.11$\pm$0.06 & 10.65$\pm$0.02 & 0.93 \\
NGC 2974    & -               & 0.35$\pm$0.03   & 0.22$\pm$0.03   & 0.18$\pm$0.01   & 0.07$\pm$0.02   & - & 1.53$\pm$0.01 & 1.25$\pm$0.01 & 0.939$\pm$0.005 & 0.555$\pm$0.008 & 0.60 \\
NGC 3079    & -               & 1.04$\pm$0.02   & 1.19$\pm$0.04   & 1.38$\pm$0.05   & 1.07$\pm$0.01   & - & 16.32$\pm$0.01 & 16.53$\pm$0.02 & 14.87$\pm$0.04 & 10.52$\pm$0.01 & 0.93 \\
NGC 3384    & 0.93$\pm$0.06   & 0.27$\pm$0.03   & 0.26$\pm$0.04   & 0.12$\pm$0.02   & 0.055$\pm$0.01  & 7.56$\pm$0.04 & 2.69$\pm$0.01 & 1.47$\pm$0.04 & 1.35$\pm$0.03 & 0.48$\pm$0.01 & 0.65 \\
NGC 3613    & 1.03$\pm$0.03   & 0.36$\pm$0.02   & 0.327$\pm$0.007 & 0.194$\pm$0.008 & 0.060$\pm$0.005 & 2.85$\pm$0.01 & 0.845$\pm$0.006 & 0.699$\pm$0.003 & 0.471$\pm$0.003 & 0.181$\pm$0.004 & 0.60 \\
NGC 3623    & 0.69$\pm$0.05   & 0.23$\pm$0.02   & 0.18$\pm$0.02   & 0.13$\pm$0.02   & 0.052$\pm$0.006 & 22.2$\pm$0.2 & 9.45$\pm$0.07 & 8.36$\pm$0.06 & 7.05$\pm$0.06 & 3.27$\pm$0.03 & 0.85 \\
NGC 3628    & 3.0$\pm$0.3     & 1.51$\pm$0.04   & 1.53$\pm$0.2    & 0.8$\pm$0.2     & 0.511$\pm$0.007 & 33.0$\pm$0.1 & 14.21$\pm$0.02 & 14.0$\pm$0.1 & 8.03$\pm$0.05 & 3.02$\pm$0.01 & 0.93 \\
NGC 3818    & 0.50$\pm$0.02   & 0.22$\pm$0.01   & 0.122$\pm$0.008 & 0.085$\pm$0.005 & 0.017$\pm$0.005 & 1.44$\pm$0.01 & 0.48$\pm$0.01 & 0.074$\pm$0.004 & 0.242$\pm$0.003 & 0.113$\pm$0.004 & 0.60 \\
NGC 4036    & 0.56$\pm$0.01   & 0.13$\pm$0.01   & 0.09$\pm$0.01   & 0.11$\pm$0.01   & 0.038$\pm$0.004 & 4.85$\pm$0.02 & 1.11$\pm$0.01 & 0.99$\pm$0.01 & 0.792$\pm$0.003 & 0.25$\pm$0.04 & 0.65 \\
NGC 4088    & -               & 0.38$\pm$0.03   & 0.35$\pm$0.02   & 0.31$\pm$0.03   & 0.29$\pm$0.01   & - & 18.97$\pm$0.02 & 18.96$\pm$0.01 & 16.05$\pm$0.02 & 10.28$\pm$0.02 & 0.93 \\
NGC 4173    & 0.18$\pm$0.04   & 0.083$\pm$0.02  & 0.09$\pm$0.02   & 0.13$\pm$0.01   & 0.08$\pm$0.01   & 5.06$\pm$0.02 & 3.97$\pm$0.01 & 4.34$\pm$0.02 & 3.89$\pm$0.01 & 2.55$\pm$0.02 & 0.93 \\
NGC 4388    & 0.50$\pm$0.06   & 0.27$\pm$0.07   & 0.31$\pm$0.06   & 0.26$\pm$0.09   & 0.16$\pm$0.03   & 11.07$\pm$0.06 & 4.70$\pm$0.02 & 5.04$\pm$0.01 & 4.16$\pm$0.01 & 3.35$\pm$0.02 & 0.93  \\
NGC 4594    & 14.7$\pm$0.6    & 3.88$\pm$0.08   & 2.6$\pm$0.1     & 1.41$\pm$0.09   & 0.90$\pm$0.04   & 40.8$\pm$0.2 & 12.14$\pm$0.02 & 9.2$\pm$0.1 & 5.1$\pm$0.1 & 3.84$\pm$0.02 & 0.65 \\
NGC 5301    & 0.49$\pm$0.03   & 0.19$\pm$0.01   & 0.22$\pm$0.01   & 0.15$\pm$0.01   & 0.111$\pm$0.004 & 6.28$\pm$0.03 & 4.08$\pm$0.01 & 4.26$\pm$0.03 & 3.54$\pm$0.02 & 2.22$\pm$0.01 & 0.93 \\
NGC 5775    & 1.4$\pm$0.06    & 0.62$\pm$0.02   & 0.623$\pm$0.009 & 0.56$\pm$0.02   & 0.445$\pm$0.008 & 8.21$\pm$0.02 & 4.77$\pm$0.01 & 4.52$\pm$0.01 & 3.78$\pm$0.01 & 2.78$\pm$0.01 & 0.93 \\
NGC 5866    & 2.1$\pm$0.03    & 0.65$\pm$0.01   & 0.49$\pm$0.01   & 0.42$\pm$0.02   & 0.080$\pm$0.003 & 8.96$\pm$0.03 & 2.52$\pm$0.01 & 2.32$\pm$0.02 & 1.61$\pm$0.01 & 0.498$\pm$0.006 & 0.65 \\
NGC 5907    & 3.5$\pm$0.2     & 2.40$\pm$0.08   & 2.3$\pm$0.2     & 2.1$\pm$0.1     & 1.59$\pm$0.02   & 18.67$\pm$0.03 & 12.45$\pm$0.02 & 12.16$\pm$0.05 & 10.78$\pm$0.04 & 7.84$\pm$0.02 & 0.93 \\
NGC 6503    & 0.2$\pm$0.1     & 0.11$\pm$0.04   & 0.10$\pm$0.04   & 0.16$\pm$0.01   & 0.13$\pm$0.04   & 33.66$\pm$0.07 & 24.43$\pm$0.02 & 24.50$\pm$0.02 & 20.81$\pm$0.02 & 14.08$\pm$0.03 & 0.93 \\
NGC 6925    & 1.1$\pm$0.3     & 0.29$\pm$0.09   & 0.18$\pm$0.04   & 0.28$\pm$0.05   & 0.33$\pm$0.06   & 17.63$\pm$0.09 & 11.97$\pm$0.02 & 12.50$\pm$0.02 & 9.71$\pm$0.02 & 7.38$\pm$0.03 & 0.93 \\
NGC 7090    & -               & -               & 1.65$\pm$0.09   & 1.40$\pm$0.05   & -               & - & - & 11.05$\pm$0.02 & 9.49$\pm$0.02 & - & 0.93 \\
NGC 7582    & 1.09$\pm$0.04   & 0.33$\pm$0.03   & 0.31$\pm$0.05   & 0.36$\pm$0.04   & 0.17$\pm$0.02   & 12.27$\pm$0.03 & 6.76$\pm$0.02 & 6.19$\pm$0.06 & 5.11$\pm$0.04 & 2.92$\pm$0.01 & 0.85 \\
UGC 6697    & 0.32$\pm$0.02   & 0.212$\pm$0.005 & 0.261$\pm$0.006 & 0.203$\pm$0.008 & 0.155$\pm$0.004 & 4.65$\pm$0.03 & 4.23$\pm$0.01 & 4.39$\pm$0.02 & 3.90$\pm$0.01 & 3.44$\pm$0.01 &  0.93 \\ 
UGC 11794   & 0.07$\pm$0.01   & 0.018$\pm$0.004 & 0.016$\pm$0.007 & 0.012$\pm$0.004 & 0.004$\pm$0.001 & 0.65$\pm$0.01 & 0.349$\pm$0.003 & 0.33$\pm$0.01 & 0.26$\pm$0.01 & 0.164$\pm$0.003 & 0.85 
\enddata
\tablecomments{\label{table.halo_fluxes} We report the total halo flux measured in each filter by summing over
the extraction bins shown in Figures~\ref{figure.results_sc}, \ref{figure.results_sb},
\ref{figure.results_s0}, and \ref{figure.results_e}.  The galaxy flux is measured in a region roughly
spanning the space between the boxes, but the region is defined by SExtractor.  The error bar is dominated by uncertainty in
the background.  The \uvwtwo{} fluxes are corrected as described in the text, but the \uvwone{} fluxes
are not.}
\end{deluxetable*}
\end{turnpage}

\section{Results}
\label{section.results}

We measured halo fluxes above the midplanes of the galaxies in our sample
from the reduced, cleaned images in each filter.  The results are shown in
Figures~\ref{figure.results_sc}, \ref{figure.results_sb},
\ref{figure.results_s0}, and \ref{figure.results_e} which show,
respectively, spiral galaxies of type Sc through Sd, spiral galaxies
of type Sab through Sbc, S0 galaxies, and elliptical galaxies. 
For our purposes, we put the Sa galaxy NGC~4594 (M104) in the S0
category because of its giant stellar halo and bulge; as we shall see,
its spectral characteristics agree with this categorization. 

There are three panels per galaxy in these figures.  The left panel
is a false-color image made from the \uvwone{}, \uvmtwo{}, and
\uvwtwo{} filters with point sources outside the galaxy subtracted.  The
white boxes indicate the flux measurement boxes used for the central panel,
which shows the flux density at a projected height $z$ in
kpc based on the distance adopted for each galaxy (Table~\ref{table.sample}).  
We do \textit{not} correct the distances by $\sin i$ because in some cases the
inclination angle is suspect, but for the less inclined galaxies most
halo emission comes from regions closer to the galaxy than the nominal
projected distances.  The inclinations in the early-type
galaxies refer to the projected elongation.

Extended diffuse UV emission is detected around each galaxy in our
sample, with a maximum extent of 5--20\,kpc from the midplane.  The
flux profiles appear to decline exponentially, with some asymmetry 
across the midplane or profiles that differ between filters.
The halo fluxes range from 2--50\% of the background in bins with
3$\sigma$ detections.

The right-hand panel gives an orthogonal 
view of these fluxes by showing a four-point UV SED as a function of $z$
on both sides of the midplane (the solid and dashed lines of the same color).
We also plot the SED of the galaxy, arbitrarily scaled to fit on the plot,
as a dotted line.  The upper right-hand corner of the panel shows the scaling
factor and the heights of the halo SEDs.  We show the NUV point with an 
artificial offset in wavelength for clarity (it is nominally calibrated
to about the same wavelength as the \uvmtwo{} filter but covers most of the
\uvwone{} region of the spectrum; see Figure~\ref{figure.uv_responses}). 
The \uvwone{} flux is not included in the SEDs because the red tail can 
easily put the effective wavelength close to the optical band.

In the remainder of this section, we describe the flux
measurement, discuss the basic properties of the halo emission, and 
compare the different types of galaxies.  

\subsection{Flux Measurements}

Fluxes
are computed from the background-subtracted count rates in the extraction
boxes shown in the galaxy image panels.  The total halo fluxes in each filter
are given in Table~\ref{table.halo_fluxes}.

We sum the count rate in each box and scale it to a uniform box size
(accounting for point source masks and non-uniform box sizes).  
In
most cases, the correction is small (1--10\%), but in some bins can
exceed 50\% because of bright stars or chip edges.  The reported
fluxes in such cases may be slightly too high or low, but for most
galaxies this is not an issue.
The location of the innermost box
for each galaxy was chosen to be clearly above the disk for late-type galaxies
and outside a few scale heights for E/S0 galaxies, but we do not use a uniform
definition because of the different types and inclinations.  Nonetheless, the
boxes are defined conservatively to be at least a few kpc (projected)
above the midplane, outside of several optical scale heights as seen
around an edge-on system.
We then subtract a mean background from the count rates and convert them to
flux density using the conversion factors in Table~\ref{table.filters}, and
correct for Galactic extinction using the \citet{schlegel98} dust map
using the calibrations in \citet{wyder07} and \citet{roming09}.

Finally, we apply the correction for the red leak in the \uvwtwo{} filter
as described above.  As the true correction depends on the true spectrum,
we report both the corrected flux and the correction factor in Table~\ref{table.halo_fluxes}.
The \uvwone{} fluxes are not corrected and are not included in the SED plots
(but we return to these in Section~\ref{section.discussion}).  

In Section~\ref{section.scattered_light} we argued that the contribution of
instrumental scattered light from the galaxy to the measured light in the
halo is typically a few percent of the background.  
Given the agreement between the
\Galex{} and \Swift{} flux profiles in the bright late-type galaxies 
(e.g., Figure~\ref{figure.results_sc}) and the different systematic effects
in each instrument, we posit that the contamination from instrumental 
scattered light is no worse than the error on the background $\sigma_{\text{sky}}$,
which is dominated by spatial variation in the background.

\begin{figure}
\begin{center}
\hspace{-1.2cm}\includegraphics[width=0.4\textwidth,angle=-90]{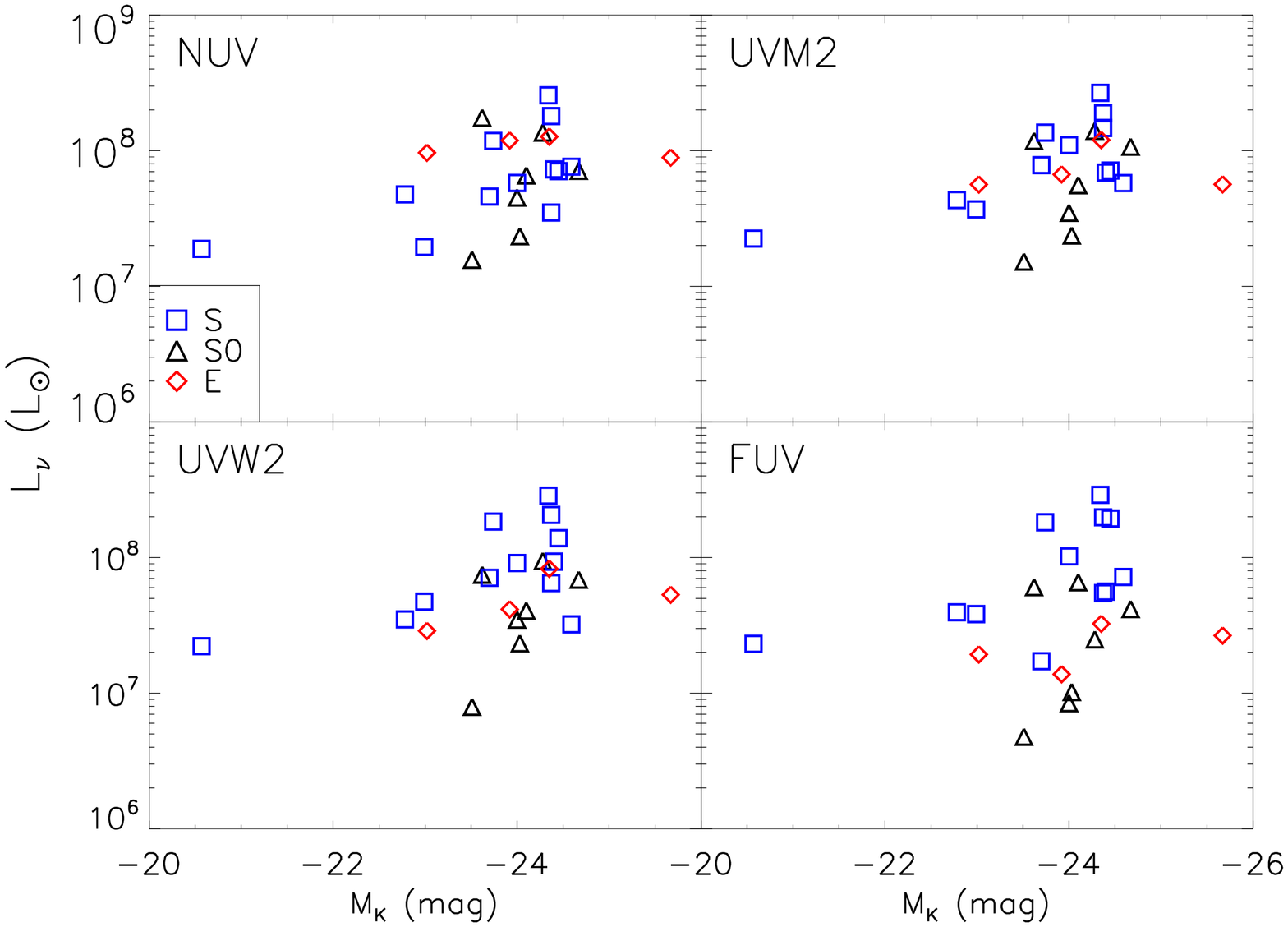}
\vspace{-0.5cm}
\caption{\scriptsize \label{figure.uv_lum}
UV halo luminosities in each filter as a function of the absolute $K$-band 
magnitude $M_K$ for spiral (blue squares), S0 (black triangles), and elliptical
(red diamonds) galaxies.  $L_{\nu}$ was computed in each filter based on the
distance and magnitudes in Table~\ref{table.sample} and UV fluxes in 
Table~\ref{table.halo_fluxes}.  See text for discussion.}
\end{center}
\end{figure}

\subsection{Properties of the Halo Emission}

From our flux measurements we can obtain three physical quantities: the
luminosity, extent, and the SED of the halo emission.  With the deeper data sets,
we can map these quantities around the galaxy.

\subsubsection{Luminosity}

The luminosity $L_{\nu} = 4\pi d^2 F_{\nu}$ is computed from the integrated
halo fluxes (Table~\ref{table.halo_fluxes}).  As most of the flux comes from near
the galaxy, the true halo UV luminosity is sensitive to the choice of the
innermost bin height; our $L_{\nu}$ values are measured starting a few kpc from
the midplane (well above the optical disk).

Figure~\ref{figure.uv_lum} shows $L_{\nu}$ as a function of $M_K$ (galaxies
with $M_K > -22$\,mag are not shown).  The different symbols break the
galaxies up by type: blue squares represent late-type (Sab through Sd) galaxies,
black triangles are S0 galaxies, and red diamonds are elliptical galaxies.

\begin{figure}
\begin{center}
\hspace{-0.5cm}\includegraphics[width=0.5\textwidth]{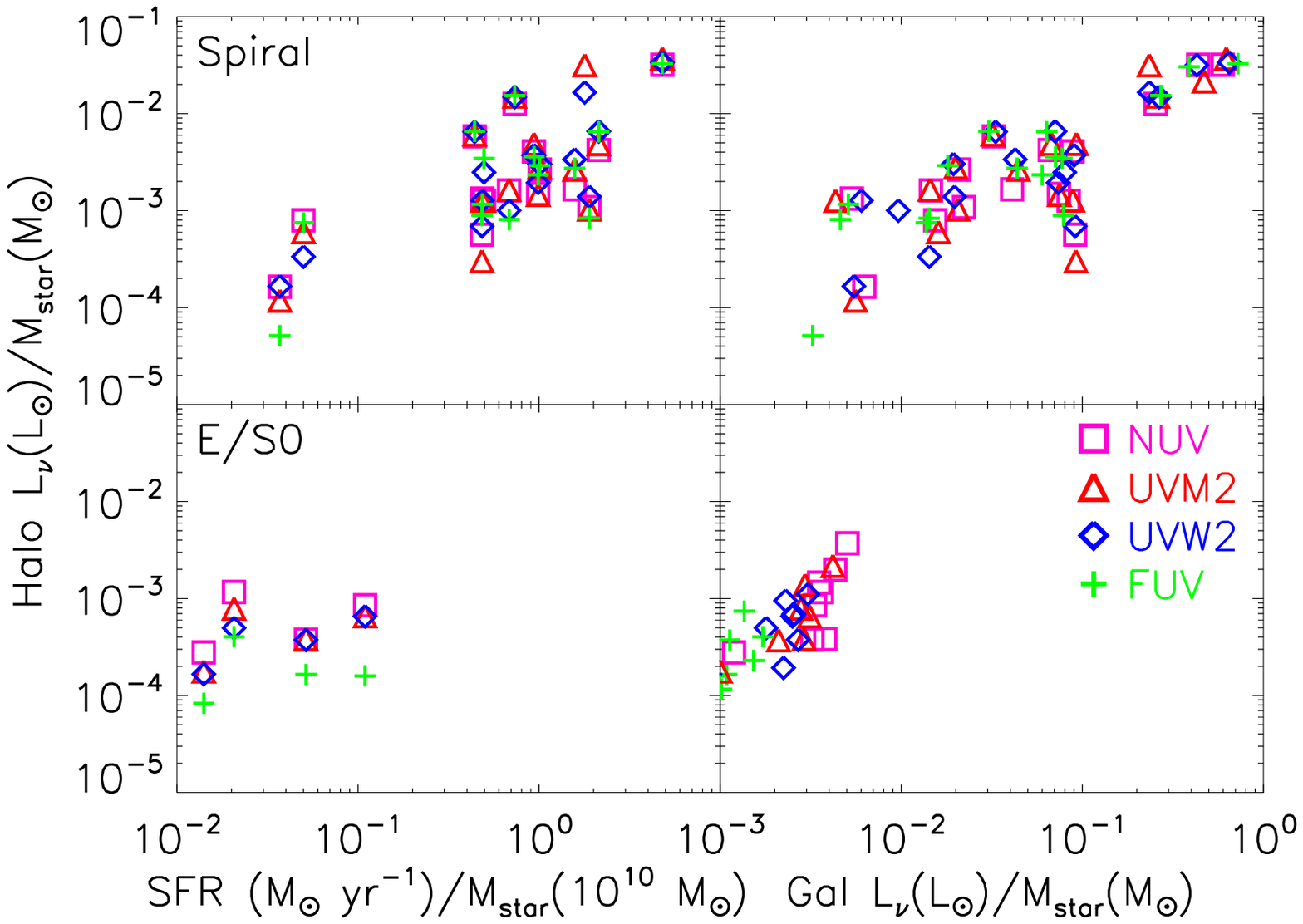}
\caption{\scriptsize \label{figure.uv_sfr}
Specific halo luminosity in the NUV, \uvmtwo{}, \uvwtwo{}, and FUV bands
plotted against specific SFR (left panels) and specific galaxy luminosity
(right panels) for spiral and early-type galaxies (top and bottom panels).
The values are divided by the stellar mass shown in Table~\ref{table.sample}.
See text for discussion.}
\end{center}
\end{figure}

\begin{figure}
\begin{center}
\includegraphics[width=0.5\textwidth]{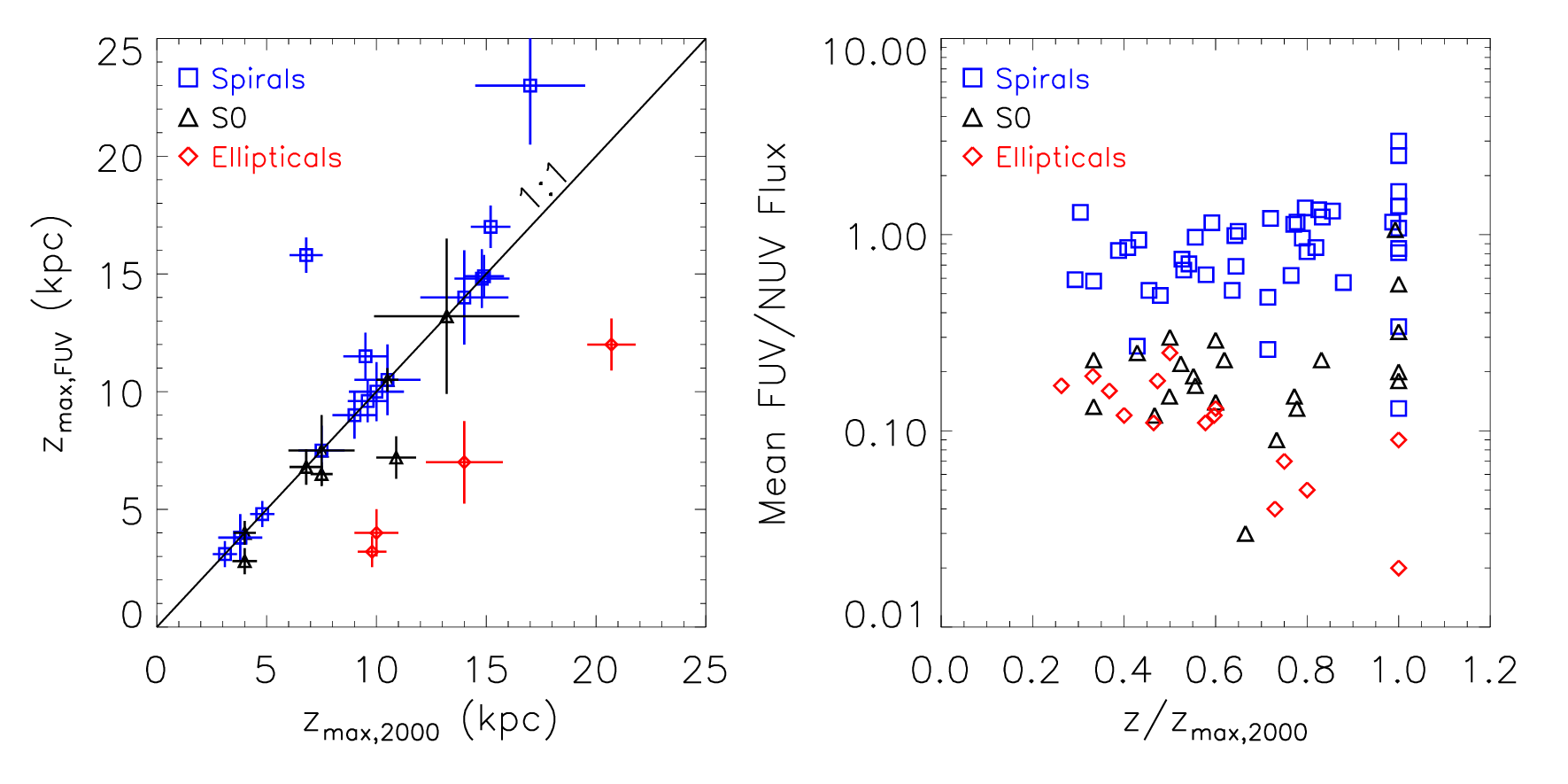}
\caption{\scriptsize \label{figure.zmax_fuv}
\textit{Left}: A plot of the maximum extent
  ($z_{\text{max}}$) of the halo as measured in the FUV filter against
  the average $z_{\text{max}}$ in the NUV, \uvmtwo{}, and \uvwtwo{} filters 
  $z_{\text{max,2000}}$ (see
  Section~\ref{section.results}).  The ``error bars'' are the width of the flux
  extraction bin representing $z_{\text{max}}$ for each galaxy.
  \textit{Right}: The ratio of the FUV to NUV fluxes as a function of
  height above the midplane for the galaxies in our sample, broken into
  spirals, S0s, and ellipticals.  Error bars are omitted for clarity.
  See Section~\ref{section.results}
  for discussion.
}
\end{center}
\end{figure}

Most of the galaxies in the sample have $M_K$ similar to the Milky Way 
($M_K \sim -24$\,mag) and are not cleanly separable in any of the bands. The
largest differences are seen in the FUV band, where the elliptical and S0 galaxies
tend to have lower FUV luminosities than the late-type galaxies; as we discuss
below, this is even more apparent when comparing the fluxes as a function of
height above the midplane.  
Since the galaxies have different sizes and the
sample is not objectively defined, the overlap is not surprising.

In Figure~\ref{figure.uv_sfr} we show the specific integrated halo luminosities in
the NUV (pink boxes), \uvmtwo{} (red triangles), \uvwtwo{} (blue diamonds), and
FUV (green crosses) as a function of the specific SFR and the specific galactic
luminosities, where ``specific'' is in reference to the stellar mass $M_{\text{star}}$ 
(galaxies without the information to compute the SFR or $M_{\text{star}}$ are omitted).
We have separated the galaxies by type into spiral and early-type (E/S0)
systems.  To measure the correlation between these quantities (in each band), we use the
Spearman ranked correlation coefficient $\rho$, which we prefer over the
Pearson coefficient because of the small number of data points. 

With regard to the SFR, the only significant correlation at better than
99\% is in the 
\uvwtwo{} band ($\rho = 0.67$, $p=0.004$).  The other bands do not even have
marginally significant correlations ($p=0.16$, 0.05, and 0.24 for the NUV, 
\uvmtwo{}, and FUV respectively).  If we consider only specific
SFR$>0.1 M_{\odot}$\,yr$^{-1}$, there is no significant correlation in 
any band, so among galaxies with SFR similar to or greater than the Milky
Way the specific SFR does not predict the halo luminosity.  For the early-type
galaxies, there are not enough data to measure a significant correlation.

On the other hand, the specific halo luminosity in each band except the NUV
is strongly and significantly correlated with the specific galaxy luminosity.
For the NUV filter $\rho = 0.57$ and $p=0.02$, for \uvmtwo{}
$\rho = 0.69$ and $p=0.002$, for \uvwtwo{} $\rho = 0.76$ and $p=0.0004$, and
for FUV $\rho = 0.80$ and $p = 0.0002$.  For the early-type galaxies, what
looks like a significant correlation in Figure~\ref{figure.uv_sfr} is due to
offsets between each band (which also appear in the SEDs; cf. Figure~\ref{figure.results_e}),
and the only significant correlation is in the FUV ($\rho = 0.86$ and $p=0.007$).
The others have $p > 0.05$.  

We also measured the correlation between the halo luminosity and $M_K$
and find no significant dependence in any band for the late-type galaxies,
whereas for the early-type galaxies there is a significant correlation in
the FUV ($\rho=0.86$, $p=0.0003$) and a marginally significant correlation
in the NUV ($\rho = 0.59$, $p=0.02$).  

The correlation between halo and galaxy UV luminosity hints at a physical connection
in the late-type galaxies, but the connection is evidently not mediated by 
rapid (present-day) star formation.  The actual correlations may be even
stronger, as we have made no attempt to correct for the effects of inclination
and extinction.

\subsubsection{Scale Heights and Maximum Extent of the Halo Emission}

Differences between galaxy types are more evident when considering the
UV luminosity as a function of height above the midplane $L_{\nu}(z)$.  We
cannot reliably measure scale heights for the halo emission in most galaxies
because of the coarseness of the flux measurement bins and the
differences in the bin widths between galaxies.  For many galaxies,
there is a large artificial minimum on the measurable scale height.

It is possible to measure a scale height in the galaxies with the best
data (where the $S/N$ is high in narrow bins).  These include the late-type
galaxies NGC~5775, NGC~5907, and NGC~3079, and the early-type galaxies NGC~4594,
NGC~5866, and NGC~3613.  We fit profiles of the form $I(z) = I_0 e^{-z/H}$ in 
each band for these galaxies, where we averaged the flux on each side of the
midplane at each height, and the results are given in Table~\ref{table.scaleheights}.
The early-type galaxies are fit well in each band except the FUV by the
exponential fit (we also tried Gaussian fits, which were never better for any
galaxy), with best-fit scale heights of $H \sim 2.75-3.5$\,kpc for NGC~4594,
$H \sim 2-3$\,kpc for NGC~5866, and $H \sim 6-8$\,kpc for NGC~3613.  The late-type
galaxies also had generally good fits, although the exponential fit is too
steep for most bands.  We find $H \sim 2.5-3.5$\,kpc for NGC~5775, 
$H \sim 2-3$\,kpc for NGC~5907, and $H \sim 3-5$\,kpc in NGC~3079.  

\begin{deluxetable*}{lccccccccc}
\tablenum{5}
\tabletypesize{\scriptsize}
\tablecaption{UV Scale Heights}
\tablewidth{0pt}
\tablehead{
\colhead{Galaxy} & \colhead{$H_{\text{NUV}}$} & \colhead{$\chi^2$} & \colhead{$H_{\text{UVM2}}$} & \colhead{$\chi^2$} &
\colhead{$H_{\text{UVW2}}$} & \colhead{$\chi^2$} & \colhead{$H_{\text{FUV}}$} & \colhead{$\chi^2$} & d.o.f. \\
 & \colhead{(kpc}) &  & \colhead{(kpc}) &  & \colhead{(kpc}) &  & \colhead{(kpc)} &  &  
}
\startdata  
\tableline
\multicolumn{9}{c}{Late-type Galaxies}\\
\tableline
NGC 3079 & $2.9\pm0.3$ & 15.6 & $3.5\pm0.5$ & 0.56 & $5.1\pm1.5$ & 0.80 & $4.5\pm0.1$ & 17.8 & 3 \\
NGC 5775 & $2.4\pm0.3$ & 3.5  & $3.0\pm0.2$ & 10.7 & $2.65\pm0.08$ & 4.6 & $3.4\pm0.1$ & 15.9 & 8 \\ 
NGC 5907 & $2.7\pm0.4$ & 1.3  & $2.2\pm0.3$ & 1.45 & $3.0\pm0.7$ & 0.53 & $2.32\pm0.03$ & 22.5 & 8 \\
\tableline
\multicolumn{9}{c}{Early-type Galaxies}\\
\tableline
NGC 3613 & $6.9\pm0.1$ & 5.8 & $7.8\pm0.2$ & 27.4 & $6.0\pm0.2$ & 16.2 & $6.1\pm0.1$ & 2.3 & 6 \\
NGC 4594 & $3.5\pm0.1$ & 4.2 & $2.8\pm0.3$ & 2.2  & $3.1\pm0.3$ & 4.0  & $3.07\pm0.05$ & 3.7 & 8 \\
NGC 5866 & $2.0\pm0.1$ & 1.5 & $2.5\pm0.4$ & 7.5  & $4.0\pm1.5$ & 0.11 & $2.0\pm0.1$ & 17 & 4 
\enddata
\tablecomments{\label{table.scaleheights}Scale heights are measured by fitting exponential profiles $I_0 e^{-z/H}$ to 
the data averaged over both sides of the midplane, where $z$ is the projected distance
above the midplane.}
\end{deluxetable*}

The UV scale heights for the late-type galaxies are larger than
the scale height of the thick disks, which range from 0.3--1.5\,kpc 
\citep[e.g.,][]{yoachim06}.  The scale height of the \ion{H}{1}, where it
has been measured \citep[e.g.,][]{fraternali02,oosterloo07,kamphuis08} is 
comparable but systematically lower (1--2\,kpc) than UV values,
probably because we start measuring at larger heights to avoid stars.  

For the galaxies where $H$ cannot be measured, the maximum extent of
the emission $z_{\text{max}}$ in each band is a useful proxy for the
observable halo size.  We define $z_{\text{max}}$ as the mean height
above the midplane for which we detect flux at 2$\sigma$ on both sides
of the galaxy.  Because of real angular variation in the sky
foreground (which dominates $\sigma_{\text{sky}}$, there is a natural
limit to the halo flux that can be detected; additional exposure time
may detect some emission beyond $z_{\text{max}}$, but not much.
Within $z_{\text{max}}$, additional exposure time improves the $S/N$
and allows smaller flux measurement box sizes.

The $z_{\text{max}}$ values are tabulated in Table~\ref{table.zmax} for the
non-\uvwone{} filters.  The final column is the average $z_{\text{max}}$ for
the NUV, \uvmtwo{}, and \uvwtwo{} filters, which gives a measure of the 
maximum detectable emission at or above 2000\,\AA.  The left panel of
Figure~\ref{figure.zmax_fuv} shows $z_{\text{max,FUV}}$ as a function of 
the average NUV $z_{\text{max,2000}}$.  Spiral galaxies are represented as blue squares,
S0 as black triangles, and elliptical galaxies by red diamonds.  The ``error
bars'' show the width of the flux measurement bin for each galaxy, and the
line shows where a galaxy would fall if the FUV $z_{\text{max}}$ were equal
to $z_{\text{max,2000}}$.  

All of the spiral galaxies except for NGC~4173 have $z_{\text{max,FUV}} \ge z_{\text{max,2000}}$,
whereas the ellipticals fall below this line.  The S0 galaxies also tend to
fall below or on the line.  Since spiral and elliptical galaxies overlap in
specific halo UV luminosity (see the right panels of Figure~\ref{figure.uv_sfr})
and scale height (Table~\ref{table.scaleheights}), the observation that the
FUV flux falls less rapidly than the other bands among the spirals and more
rapidly among the early-type galaxies hints at a different origin.  

\begin{deluxetable}{lcccccc}
\tablenum{6}
\tabletypesize{\scriptsize}
\tablecaption{Maximum Extent of UV Halo Emission}
\tablewidth{0pt}
\tablehead{
\colhead{Galaxy} & \colhead{Binsize} & \colhead{NUV} & \colhead{\uvmtwo{}} &
\colhead{\uvwtwo{}} & \colhead{FUV} & \colhead{$z_{\text{max,2000}}$} \\
\colhead{} & \colhead{(kpc)} & \colhead{(kpc)} & \colhead{(kpc)} & \colhead{(kpc)} &
\colhead{(kpc)} & \colhead{(kpc)} 
}
\startdata  
ESO 243-049 & 6.6 & 13.2 & 13.2 & 13.2 & 13.2 & 13.2 \\
IC 5249     & 1.7 & 4.8  & 3.1  & 3.1  & 4.8  & 3.7\\
NGC 24      & 1.1 & 5.9  & 3.7  & 3.7  & 4.8  & 4.4 \\
NGC 527     & 3.0 & 7.5  & 7.5  & 10.5 & 7.5  & 8.5 \\
NGC 891     & 2.1 & 5.4  & 7.5  & 9.7  & 7.5  & 7.5 \\
NGC 1426    & 1.3 & 16.2 & 9.8  & 5.7  & 3.2  & 10.6 \\
NGC 2738    & 3.3 & 5.0  & 5.0  & 5.0  & 5.0  & 5.0\\
NGC 2765    & 1.8 & 10.9 & 12.7 & 10.9 & 7.2  & 11.5 \\
NGC 2841    & 1.8 & 16.6 & 14.9 & 7.9  & 14.9 & 13.1 \\
NGC 2974    & 3.5 & 17.5 & 10.5 & 14.0 & 7.0  & 14.0 \\
NGC 3079    & 2.5 & 14.8 & 12.2 & 14.8 & 14.8 & 13.9 \\
NGC 3384\tablenotemark{b}    & 1.1 & $>$4.0  & $>$5.0  & $>$4.0  & 2.8  & $>$4.3 \\
NGC 3613    & 2.2 & 14.1 & 20.7 & 20.7 & 12.0 & 18.5 \\
NGC 3623\tablenotemark{a}    & 0.88 & 7.4 & 7.4 & 6.6 & 7.4 & 7.1\\
NGC 3628    & 2.0 & 7.5  & 9.5  & 7.5  & 11.5 & 8.2 \\
NGC 3818    & 2.0 & 12.0 & 10.0 & 8.0  & 4.0  & 10.0 \\
NGC 4036    & 1.5 & 5.3  & 6.8  & 8.2  & 6.8  & 6.8 \\
NGC 4088    & 2.2 & 12.4 & 12.4 & 10.1 & 12.4 & 11.6 \\
NGC 4173\tablenotemark{a}    & 1.1 & 3.1  & 3.1  & 6.5  & 3.1  & 4.2 \\
NGC 4388    & 2.5 & 8.0  & 10.0 & 10.0 & 10.0 & 9.3 \\
NGC 4594    & 1.0 & 12.5 & 8.5  & 7.5  & 10.5 & 9.5 \\
NGC 5301    & 1.8 & 9.6  & 9.6  & 6.1  & 9.6  & 8.4 \\
NGC 5775    & 1.8 & 13.4 & 15.2 & 15.2 & 17.0 & 14.6 \\
NGC 5866    & 1.0 & 7.5  & 7.5  & 7.5  & 6.5  & 7.5 \\
NGC 5907    & 1.5 & 6.8  & 5.2  & 6.8  & 15.8 & 6.3 \\
NGC 6503\tablenotemark{a}    & 0.5 & 3.3  & 2.3  & 5.8  & 3.8  & 3.8 \\
NGC 6925    & 2.0 & 12.0 & 10.0 & 12.0 & 12.0 & 11.3 \\
NGC 7090    & -   & -    & 6    & 6    & -    & 6\\
NGC 7582    & 3.0 & 7.5  & 7.5  & 16.5  & 10.5 & 10.5 \\
UGC 6697	& -   & 17   & -    & 17   & 23   & 17\\
UGC 11794\tablenotemark{a}   & 4.0 & 14.0 & 14.0 & 10.0 & 14.0 & 11.3
\enddata
\tablenotetext{a}{Flux was only measured on one side of the galaxy.}
\tablenotetext{b}{The galaxy was on the chip edge.}
\tablecomments{\label{table.zmax} We obtain a crude measurement of $z_{\text{max},i}$ of the
halo emission in each filter $i$ by finding the height above each galaxy $z$ where 
there is at least a 2$\sigma$ detection in both extraction bins at $z$.  The 
final column is the mean $z_{\text{max}}$ for the NUV, \uvmtwo{}, and \uvwtwo{}
filters.}
\end{deluxetable}

A difference can also be seen in the ratio of the FUV and NUV fluxes as a 
function of height above the midplane, normalized to $z_{\text{max}}$
(the right panel of Figure~\ref{figure.zmax_fuv}).  There is a clear 
separation between the spirals and E/S0 galaxies, but notably the S0 galaxies
differ from the elliptical galaxies in that the FUV/NUV ratio is roughly
constant with height, whereas in the ellipticals it drops (however, there
are only four elliptical galaxies in our sample).

\begin{figure*}
\begin{center}
\mbox{}
\vspace{-0.56cm} \includegraphics[width=0.85\textwidth]{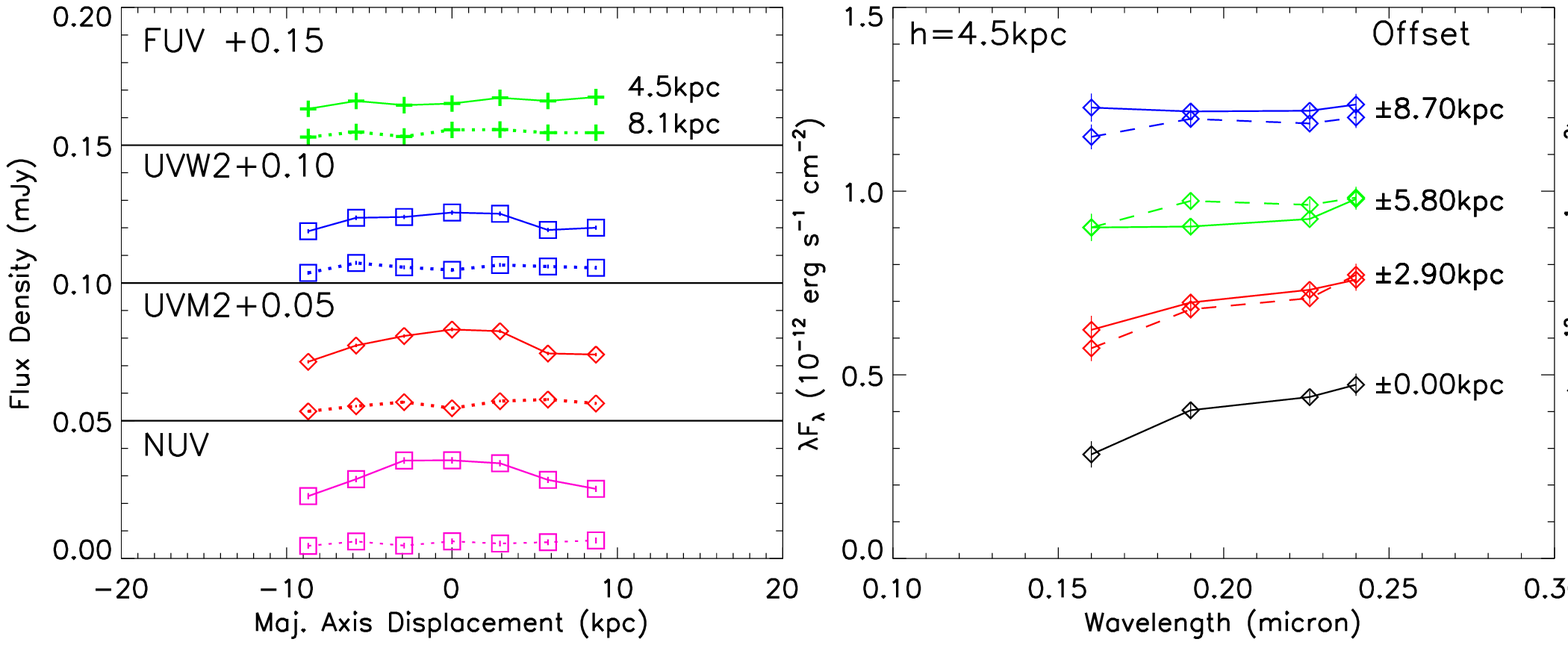}
\vspace{-0.56cm}\includegraphics[width=0.85\textwidth]{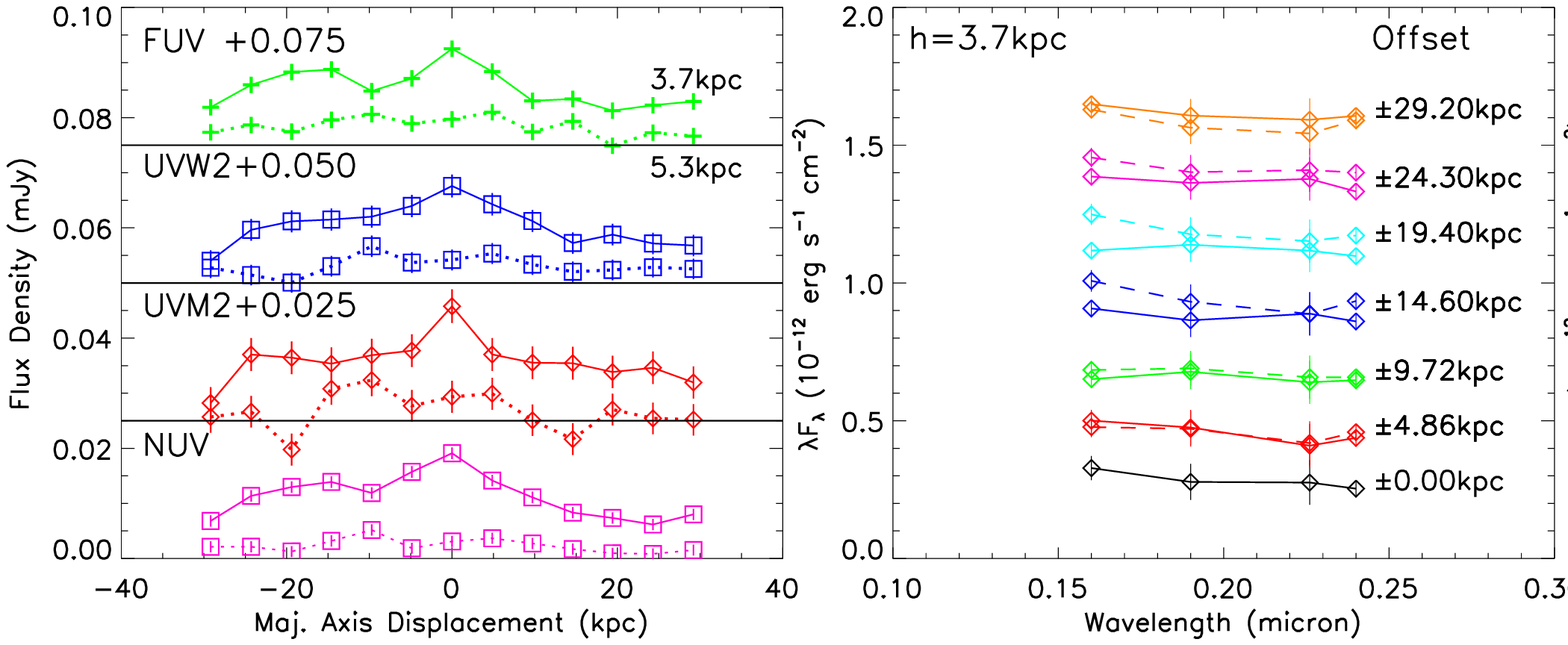}
\includegraphics[width=0.85\textwidth]{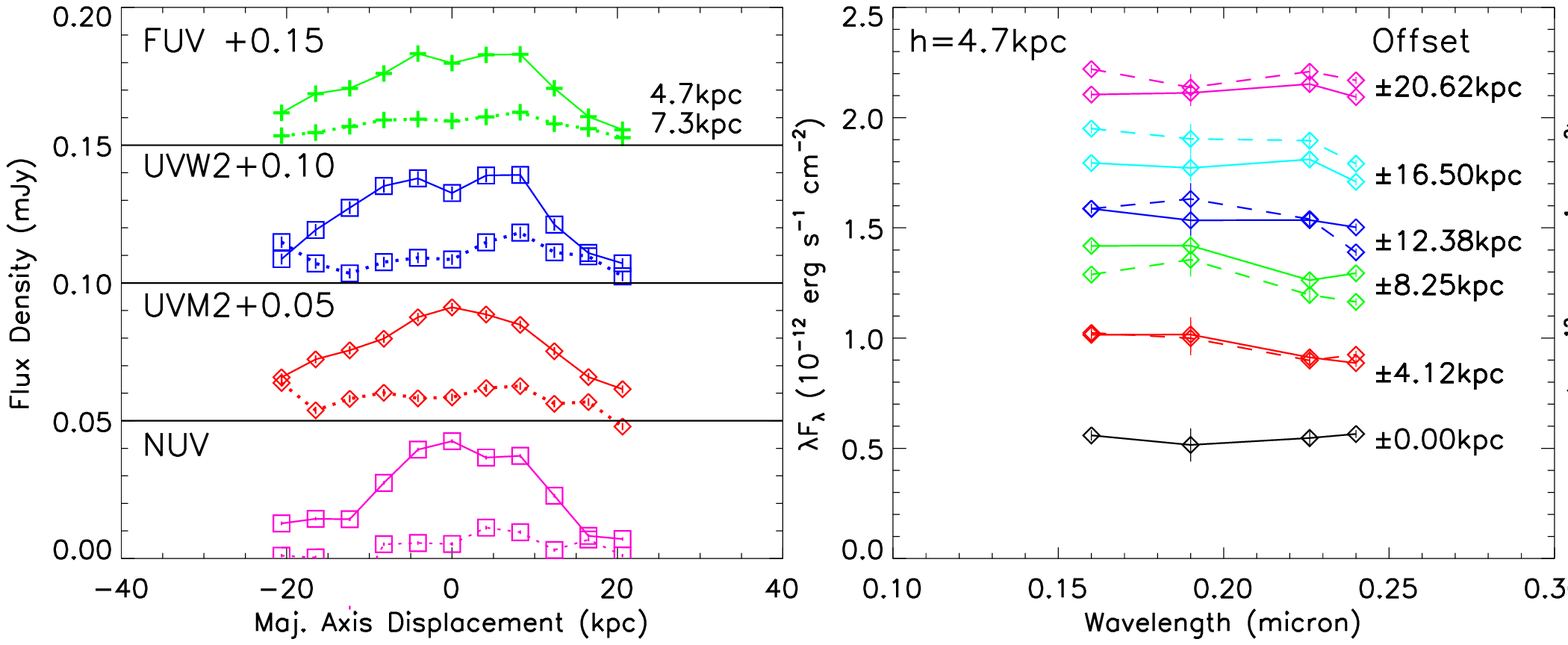}
\caption{\scriptsize \label{figure.radial}
Variation in halo emission as a function of displacement along the projected
major axis for NGC~5907,
NGC~5775, and NGC~3079.  The left panel shows the flux profile along the major axis at two heights
above the midplane (averaged over both sides along the minor axis).
The solid and dotted lines show the profile at two different offsets
along the minor axis.
The 
UV SEDs at each height above the minor axis are shown at center and right.  In all panels, we have
added artificial offsets for visual clarity and show the SED on both
sides of the midplane (solid and dashed lines) at each major axis
offset (different colors).  See text for discussion.
}
\end{center}
\end{figure*}

\subsubsection{Spectral Energy Distributions}

In the third column of Figures~\ref{figure.results_sc}, \ref{figure.results_sb},
\ref{figure.results_s0}, and \ref{figure.results_e}, we show the four-point
SED as a function of height above the midplane for each galaxy based on the
fluxes in the middle column of the same figures.  The SED of the galaxy itself
(scaled to fit on the plot) is shown as a dotted line at the top of each 
panel, and the \uvwtwo{} data points have been corrected using the correction
factor listed in Table~\ref{table.halo_fluxes}.  The galaxy SEDs have not been 
corrected for extinction within the galaxy. 

There are several remarkable features of the SEDs.  First, the halo SEDs
around most
late-type galaxies and some of the S0 galaxies differ markedly from the
galaxy SED, whereas the SEDs of the elliptical galaxies (Figure~\ref{figure.results_e})
appear similar near and far from the nucleus.  The
halo emission is different from the integrated starlight of the
galaxy.  Second, the halo SED appears to be connected to the galaxy
type..
Most Sc and Sb galaxies (Figures~\ref{figure.results_sc}
and \ref{figure.results_sb}) have halo SEDs that are either flat or
decline from the FUV to NUV,
while elliptical galaxies have ``halo'' SEDs
that rise from the FUV to NUV.  Third, we do not find any systematic differences
due to inclination in disk galaxies.  
This suggests that the SEDs measured in galaxies of lower inclination are
from the halo rather than an outer spiral arm (i.e., an XUV disk).
Finally, the halo SEDs of the
S0 galaxies (Figure~\ref{figure.results_s0}) do not have an archetypal
shape.  In some cases the SED rises continuously from the FUV through the NUV, whereas
in others the NUV flux is lower than the \uvmtwo{} flux.  This does not seem
to depend on the SED of the galaxy, and the SED may turn over with increasing
height.

The ratio of halo flux to galaxy flux varies substantially between galaxies,
but is typically a few percent or less for the late-type galaxies while it 
ranges from 10--100\% for the early-type galaxies (average 25\% for S0s and
45\% for ellipticals).
The FUV fluxes and SEDs of the elliptical galaxies indicate that 
none of the ellipticals in the sample is a ``UV upturn'' galaxy 
\citep[e.g.,][]{code79,burstein88,yi11} where the FUV luminosity is larger
than the NUV luminosity and the UV flux rises between 2500\AA\ and the Lyman
limit. 

\subsubsection{Variation along the Major Axis}

Some of the data allow us to measure the flux as a function of displacement
along the projected major axis as well as projected height above the galaxy.
Figure~\ref{figure.radial} shows
the flux profiles along the major axis at two different heights for three of
the galaxies with the best data.  The fluxes in each band are shown
with an artificial offset for clarity.  The bins closest to the galaxy
follow a typical trend where a central enhancement is visible and the flux
declines to larger displacements (except the FUV band in NGC~5775), but higher above the galaxy the profile is much
flatter.  In NGC~5907 and NGC~3079, we measure more flux on one side of the galactic center
than the other.  In the galaxies shown here (especially NGC~5907) most variations along
the major axis are correlated between filters; this may be due to where
star formation occurs in the disk or the structure of the halo or
cirrus substructure.

The SEDs at each height are shown in the right two panels of
Figure~\ref{figure.radial} (with artificial offsets) and show that
while there is general agreement on each side of the galactic center
there are real variations.  Again, it is not clear whether the source
is halo structure, variation in the disk, or cirrus.  Notably, the SED
of NGC~5775 flattens out at large displacement.  This may be due to
the rapid, but ``distributed'' star formation in the galaxy (producing
a bluer spectrum at larger offsets), or perhaps due to extinction
within the halo itself modifying the observed spectrum.  In the other
two galaxies, the increased flux near the disk does not have a strong
effect the SED as a function of displacement, except near $R = 0$ in
NGC~3079.

Similar trends are seen in other galaxies in our sample which have deep data in
fewer than three filters.  At any given height, differences as a 
function of displacement are usually visible in more than one band, so the background
is low enough to permit searches for halo substructure.  

\subsection{Comments on Individual Galaxies}

\paragraph{Cirrus around NGC~891, NGC~3623, and NGC~6925}

Galactic cirrus is in virtually every field, but for most of the sample the
contribution is small and the variation across the field is slight.  
The cirrus is faint and can only be seen in heavily smoothed, stretched images
(it cannot be seen in the galaxy images presented in this work), but it can be
brighter than the halo light.  Some
galaxies cannot be used at all because of the cirrus (e.g., NGC~7331), but there
are several borderline cases.  NGC~891 has more cirrus emission on the west
side of the galaxy (but the surface brightness is still very low relative to
the galaxy).  This was partially corrected by using different
background regions on each side of the galaxy.
NGC~3623 has spatially variable cirrus emission on the east side of the galaxy, but
no visible cirrus on the west side, so we measure fluxes only on this side.  
NGC~6925 is in a region of patchy cirrus with stronger emission on the east
side, which may explain the irregular appearance of the halo flux profile
(Figure~\ref{figure.results_sb}).  

\paragraph{NGC~2974}

NGC~2974 is classified as an E4 galaxy, but the \Galex{} data 
\citep[described in detail in][]{jeong07} reveal a blue star-forming ring in
the galaxy.  However, this ring does not seem to differentiate the halo emission
from NGC~2974 from the other elliptical galaxies (Figure~\ref{figure.results_e})
except in the innermost bin, where the relative FUV flux is higher.

\paragraph{IC~5249, NGC~24, NGC~6503, and NGC~4173}

These galaxies are significantly smaller than the other spiral galaxies in our
sample, and were included because they met our cutoff criteria and had good
\Swift{} data.  Their inclusion in the sample is useful since they demonstrate
that UV emission around smaller galaxies has the same general character as that from the
large ones. 

NGC~24 and NGC~6503 have large UV halos for their specific SFR
relative to the other galaxies in the sample (on absolute scales, they
are smaller than the halos of the more massive galaxies).  NGC~4173
also has a relatively large halo, assuming it is indeed at 8\,Mpc,
where its $M_K = -18$\,mag indicates a stellar mass of
$M_{\text{star}} \sim 3\times 10^8 M_{\odot}$.
We are only able to measure the halo emission to one side of NGC~4173 because
of the members of HCG~061 to the south; NGC~4173 was originally classified as
a member of HCG~061 (HCG~061b), but it is known to be a foreground object.  
IC~5249 has a smaller halo.

Since $z_{\text{max}}$ does not scale with the 
angular size of the galaxy for these or other galaxies in our sample, it seems
likely that this is a real effect, but the reason for the large UV halos is not
clear, since the small galaxies do not have unusually high specific $L_{\nu}$.
One possibility is that the galaxies have less extinction in the disk.  Another
is that the galaxies have large dusty halos despite their smaller stellar mass
relative to the rest of the sample.  Data for more small galaxies are required
to determine how $z_{\text{max}}$ is related to galaxy mass and other
parameters.

\begin{figure*}
\begin{center}
\includegraphics[width=0.5\textwidth]{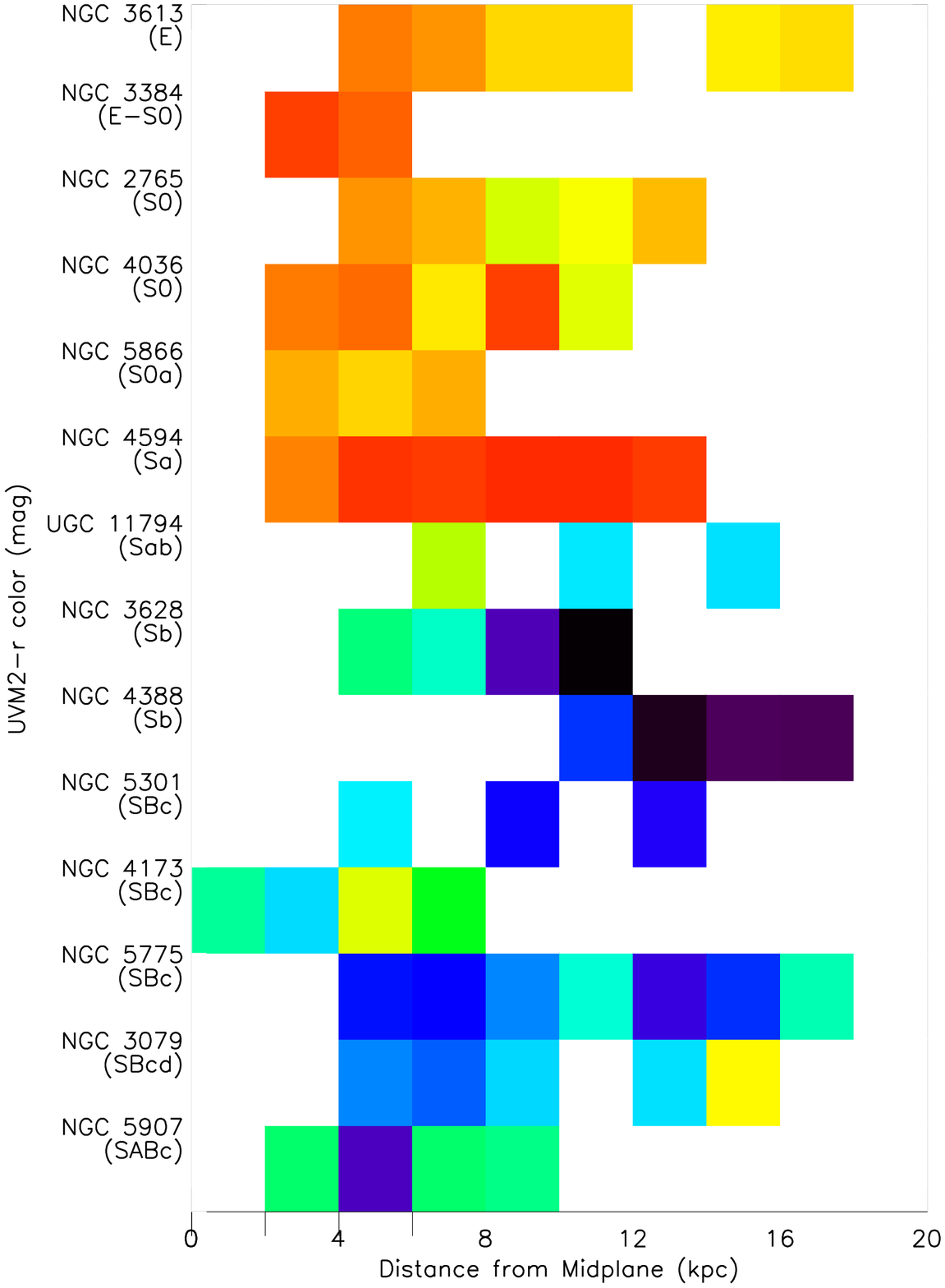}
\hspace{-1.05cm}\includegraphics[width=0.5\textwidth]{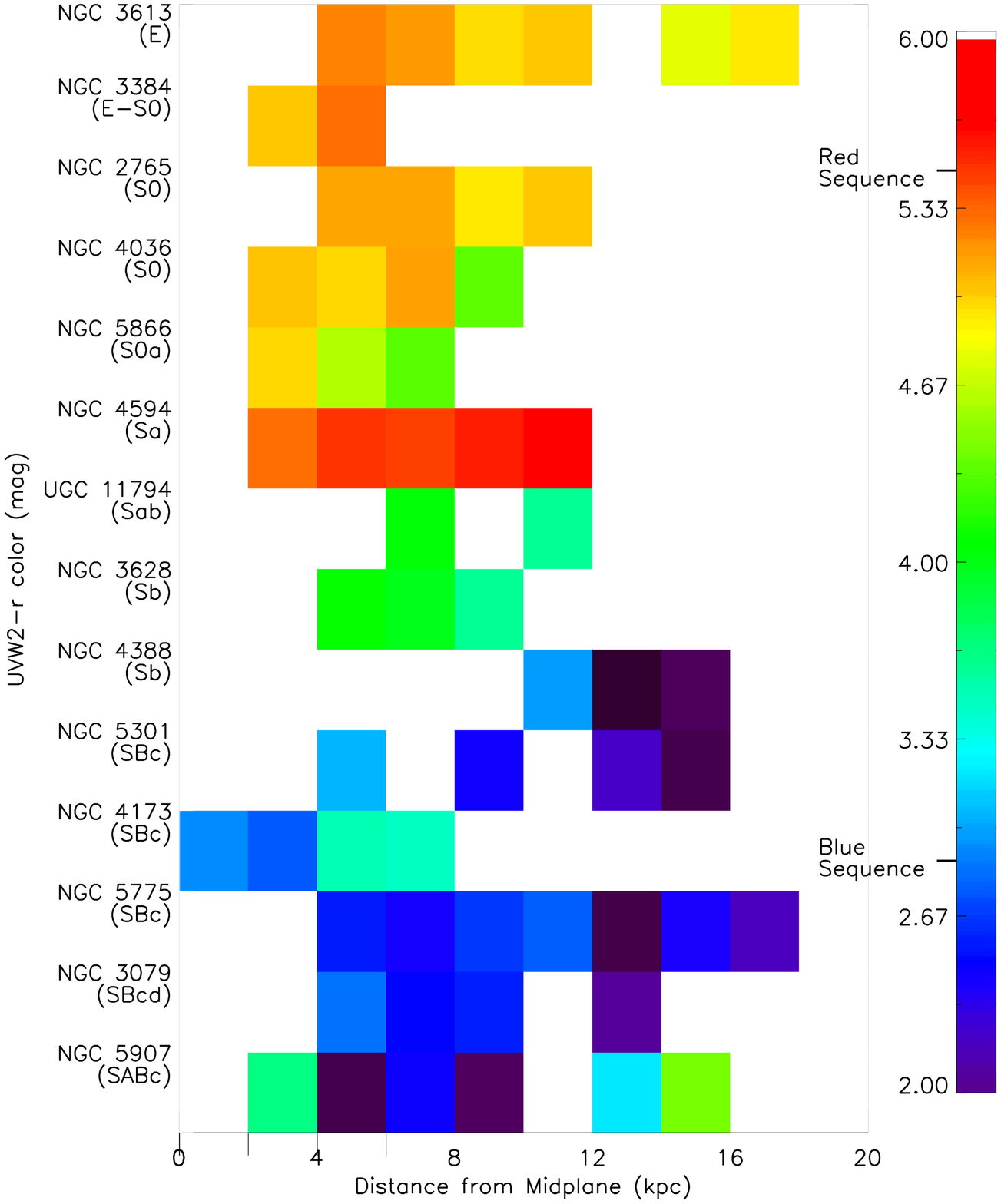}
\caption{\scriptsize UV$-r$ colors for the \uvmtwo{} and \uvwtwo{} filters for galaxies
with SDSS $r-$band data.  The galaxies are ordered from top to bottom in 
increasing morphological sequence.  In each row, we show the UV$-r$ color as a
function of height above the midplane in 2\,kpc bins, where the UV$-r$ color
is in magnitudes and colored according to the colorbars on the right, which
were generated based on the magnitudes in \citet{wyder07}.  See Section~\ref{section.sed_fits} for
additional details.}
\label{figure.tempmaps}
\end{center}
\end{figure*}

NGC~24 is remarkable in that it has very high quality \Swift{} data.  
Its flux height profile and SED 
(Figure~\ref{figure.results_sc}) look very similar to other spiral
galaxies with very deep data, 
suggesting that all star-forming late-type galaxies have similar UV halos.

\paragraph{UGC~6697, UGC~11794, NGC~527, and ESO~243-049}

Most of our sample is restricted to $d < 50$\,Mpc because the angular sizes of
the halos become too small to use multiple flux measurement boxes.
However, for galaxies with sufficiently deep data it may be possible
to measure a halo size.  

We detect halos out to about the same distances from the midplane, but as a
result of the larger physical sizes of the bins, we only see the halo emission
in the two innermost bins.  If these galaxies are representative of galaxies
at $50 \text{Mpc} < d < 100 \text{Mpc}$, halo detection should be straightforward
but the maximum extent, SED, and total flux will depend strongly on the 
bin sizes used and are more sensitive to undetected point sources or background
variations.  Indeed, the profiles and SEDs for UGC~11794 and NGC~527 look more
confused than the other galaxies in the sample, and for UGC~11794 it is 
questionable whether we see any halo emission.

\section{Halo SEDs and the Nature of the UV Light}
\label{section.sed_fits}

We now address the nature of the halo emission, which could either be 
intrinsic to the halo or a reflection nebula produced by scattering of UV photons that 
escape the disk.  In the former case, starlight from the stellar halo is the
most likely candidate (gas tends to be cold or very hot), so we expect SEDs 
and fluxes consistent with a stellar
halo population.  In the latter case, we expect the SED to be consistent with
predictions using dust models and galaxy SED templates.  

In this section, we examine which scenario is more consistent with the
data from the perspective of UV$-r$ color in the halo, SED fitting
using dust models and galaxy template spectra, and leveraging the UVOT
red leaks to predict a contribution from the stellar halo.  We find
that the UV light is more consistent with a reflection nebula for
late-type galaxies, but that extended emission in elliptical galaxies
is probably from stars in the outskirts.

\subsection{UV$-r$ Color}

If the UV light is consistent with a single-population stellar halo,
we expect it to have a UV$-$optical color comparable to an early-type
galaxy (perhaps bluer if the halo consists of metal-poor stars, but
not as blue as a late-type galaxy).

We measured the UV$-r$ colors using the \uvmtwo{}, \uvwtwo{}, and
Sloan Digital Sky Survey (SDSS) DR7 \citep{abazajian09} $r$ bands for
the diffuse emission.  We use the UVOT rather than \Galex{} filters
because of the better spatial resolution and the difficulty of
removing the large-scale artifacts from the \Galex{} images
(Section~\ref{section.scattered_light}).  We use the
\textit{uncorrected} \uvwtwo{} fluxes to eliminate potential bias, but
note that this means the filter has an effectively longer central
wavelength.  The $r$-band flux was measured from source-free regions
above the galaxy corresponding to the UV flux extraction boxes, and
both the UV and $r$-band fluxes are converted to surface brightnesses
(mag\,arcsec$^{-2}$). 

Figure~\ref{figure.tempmaps} shows the UV$-r$ colors for the two bands
as a function of height above the midplane along the $x-$axis and as a
function of increasing morphological type along the $y$-axis.  Only
galaxies with suitable SDSS data are included.  The UV$-r$ color (in
magnitudes) is encoded as an RGB color based on the UV$-r$
color-magnitude diagram (CMD) from \citet{wyder07}, i.e., a blue color
corresponds to a galaxy on the blue sequence and a red color to a
galaxy on the red sequence in their CMD.  \citet{wyder07} based their
CMD on \Galex{} data, so we computed a small small magnitude shift due
to the difference in filters.  Figure~\ref{figure.tempmaps} is
\textit{not} itself a CMD, but one can see a clear dichotomy among the
early- and late-type galaxies.  The blueness of the emission in the
late-type galaxies indicates that the UV emission from late-type
galaxy halos is not consistent with an older stellar population.  In
contrast, the redness of the emission above the early-type galaxies
and the small extent of the FUV emission relative to the spirals
(Figure~\ref{figure.zmax_fuv}) suggests that we are seeing the
outskirts of the galaxy.

\begin{figure}
\begin{center}
\hspace{-0.5cm}\includegraphics[width=0.5\textwidth]{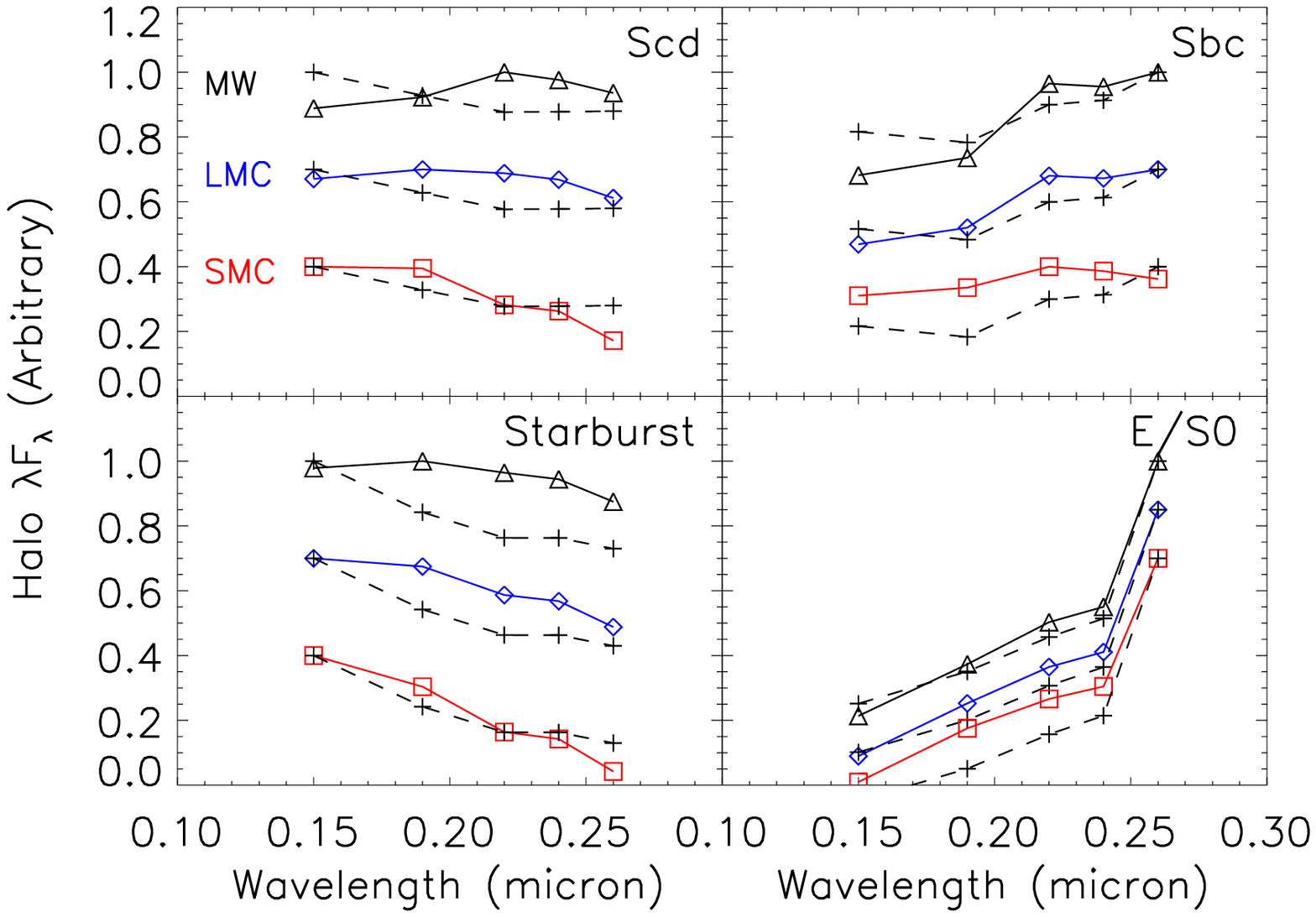}
\caption{\scriptsize \label{figure.template_halo_seds} Model SEDs for four representative
galaxy templates and MW, LMC, and SMC dust models, folded through the 
FUV, \uvwtwo{}, \uvmtwo{}, NUV, and \uvwone{} filters.  The NUV filter is
plotted with an offset to 0.24\,micron.  The fluxes are a prediction
of the halo emission in the reflection nebula scenario for each
template, but note that the UV bump in the MW-type dust may not be
real (see text).
}
\vspace{-0.0cm}
\end{center}
\end{figure}

Figure~\ref{figure.tempmaps} is imprecise.  The coloring is
based on the \citet{wyder07} CMDs, which are measured for \textit{galaxies}
rather than halos.  There are also issues with measuring diffuse emission
in the SDSS that add uncertainty to the $r$-band magnitudes, and the colors
in the outer bins are unreliable.  Still, the difference between
early- and late-type galaxies is larger than the expected magnitude
uncertainties (in total, less than 0.3\,mag near the galaxy).  

\subsection{Fitting Reflection Nebula Models}

If the halo light is a reflection nebula, then its spectrum will be the
emergent galaxy spectrum modified by the dust in the halo.  Here we
test this scenario by fitting reflection nebula models (based on
galaxy template spectra and Local Group dust models from \citetalias{weingartner01}) to the
measured halo SEDs (from fluxes in Table~\ref{table.halo_fluxes}). 

Even if the UV halos are reflection nebulas, we do not know the dust
type or the galaxy spectrum as seen by the halo (the galaxies are
highly inclined), so we do not expect our models to produce formally
good fits in many or most cases.  To provide context, we fit each
halo SED with a suite of reflection nebula models produced for several
galaxy types and dust models.  A good or marginal fit for a model
constructed from the ``right'' galaxy template and dust model that is
significantly better than fits with other models would support the
reflection nebula hypothesis (but not rule out other, non-reflection
nebula models).

\subsubsection{Model SEDs}

For an optically thin halo with a single type of dust, the scattered
spectrum is 
\begin{equation}
\label{equation.scattering}
L_{\text{halo}}(\lambda) = L_{\text{gal}}(\lambda)(1-e^{-\tau(\lambda)\varpi(\lambda)})
\end{equation}
where $\tau(\lambda) = N_H \delta_{\text{DGR}}
\sigma_{\text{ex}}(\lambda)$ is the optical depth and $\varpi(\lambda) =
\sigma_{\text{scat}}(\lambda)/\sigma_{\text{ex}}(\lambda)$ is the scattering albedo.
$N_H$ is the column density of hydrogen, $\delta_{\text{DGR}}$ is the
dust-to-gas ratio, and $\sigma_{\text{ex}}(\lambda)$ is the extinction
cross-section for a given dust mixture.  $\sigma_{\text{ex}}(\lambda)$
and $\varpi(\lambda)$ come from a dust model.  $L_{\text{halo}}$ and
$L_{\text{gal}}$ are measured quantities, so if one knows two of the
constituents of $\tau(\lambda)$, the UV measurements yield the third.
In our case, we know neither $N_H$ nor $\delta_{\text{DGR}}$, but
these do not affect the spectrum (assuming an optically thin halo).
Thus, we can fit each model SED to the halo SED with the dust 
column $N_d = N_H \delta_{\text{DGR}}$ as a free parameter.

\begin{figure*}
\begin{center}
\includegraphics[width=0.52\textwidth]{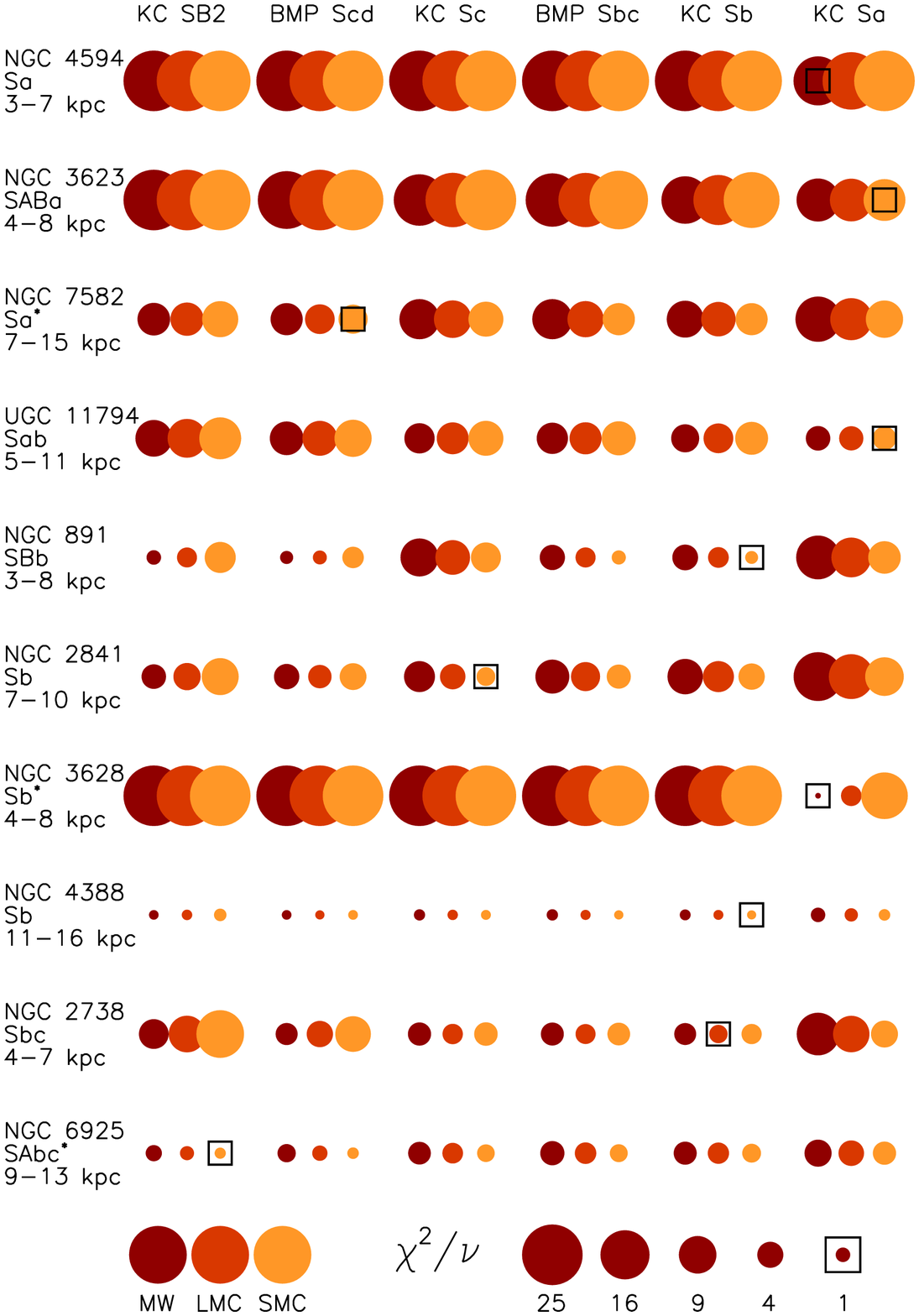}
\hspace{-1cm}\includegraphics[width=0.52\textwidth]{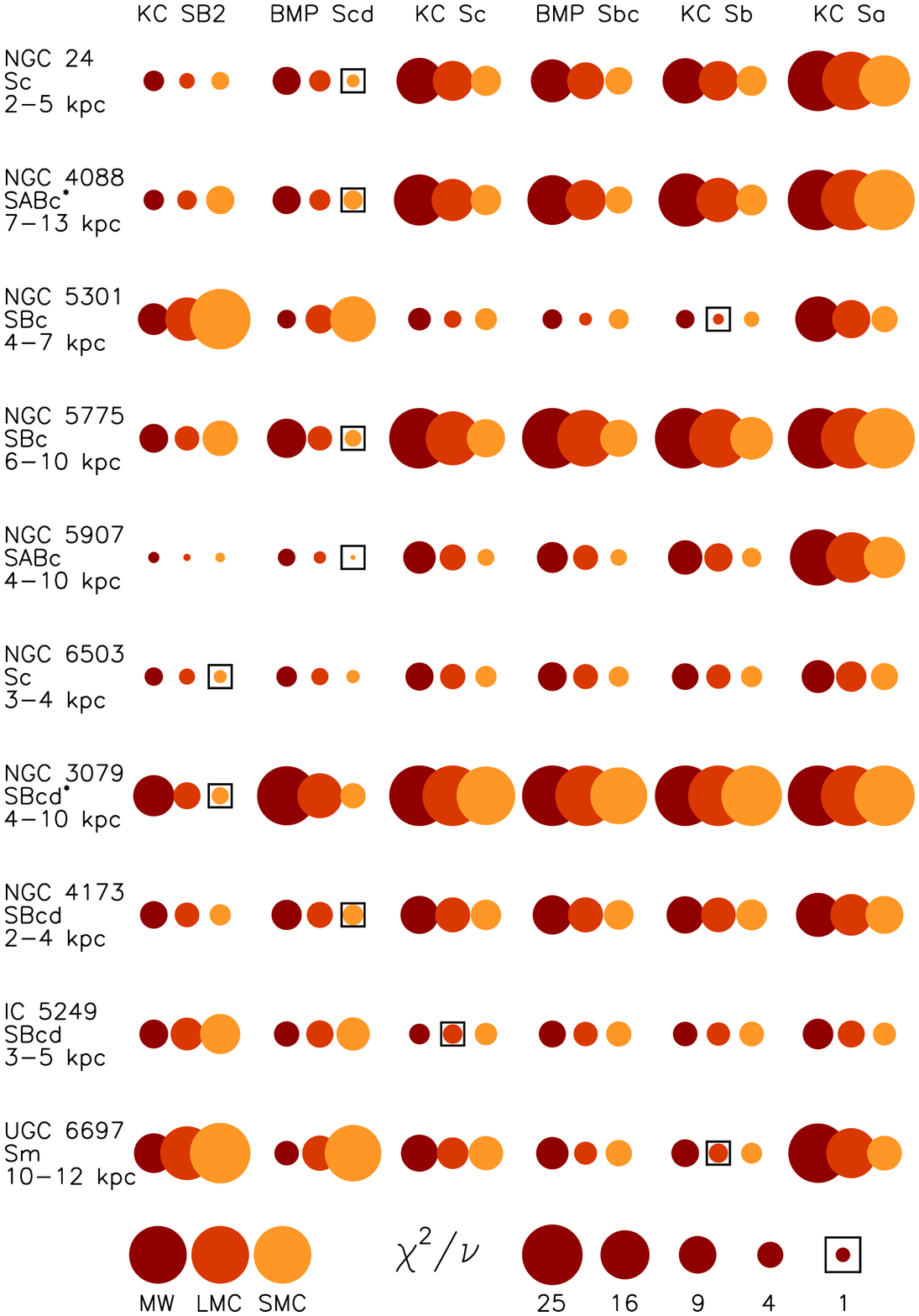}
\caption{\scriptsize Reflection nebula model fits for the spiral galaxies as a
  function of galaxy template and dust model used.  MW-, LMC-, and
  SMC-type dust are shown as dark red, orange, and light orange
  respectively.  The reduced $\chi^2$ values (clipped at 25) are shown
  as different sizes of circles, and the best-fit model for each
  galaxy is enclosed in a box.  Asterisks denote starburst galaxies,
  and we give the height where the SED was measured.  The templates
  are the \citet{kinney96} starburst2, Sc, Sb, and Sa (KC SB2, KC Sc,
  KC Sb, and KC Sa) and \citet{bolzonella00} Scd and Sbc (BMP Scd and
  BMP Sbc) models.  See text for discussion.}
\label{figure.chisq_tempmaps}
\end{center}
\end{figure*}

If $\tau \ll 1$, then $L_{\text{halo}} \approx
L_{\text{gal}}\tau \varpi$.  This depends on $N_H$ and
$\delta_{\text{DGR}}$, since $\sigma_{\text{ex}} \delta_{\text{DGR}}
\sim 10^{-21}$\,cm$^{2}$ and $\varpi \lesssim 0.5$ between
1500--3000\AA\ for dust models based on Local Group dust.  It is
difficult to separate the truly extraplanar gas from the disk gas,
which has a finite height, without a fully 3D view of the galaxy, but
here we loosely define the halo as a region above several optical
scale heights (as seen around edge-on systems), which is where we
measure the UV light.  In practice, this means a few kpc.  At this
height, $N_H \sim 0.1-1\times 10^{20}$\,cm$^{-2}$ around many Milky
Way-sized edge-on galaxies \citep[e.g.,][and references
therein]{sancisi08}.  Since the extraplanar \ion{H}{1} at a few kpc
from the midplane or above is typically consistent with an exponential
atmosphere, the mean column from the ``bottom'' of the halo to
infinity is perhaps $10^{20}$\,cm$^{-2}$ ($L_{\text{gal}}$ must then
also be defined as the emergent galactic light at this height).  For
the $\delta_{\text{DGR}} \sim 1/100$ in the Milky Way \citep[and
galaxy halos;][]{menard12}, this means $\tau \lesssim 0.1$ and the
optically thin approximation is valid. 

This implies a $L_{\text{halo}}/L_{\text{gal}}$ ratio of a few
percent for $\varpi \lesssim 0.5$.  We discuss this in detail in
Section~\ref{section.msfr10}, but here the relevant point is that the
attenuation of the reflection nebula spectrum by dust in the halo is
less than 10\%, and often much less because by the time a typical
photon has scattered it will have passed through the densest region of
the halo and scattering is highly forward-throwing \citep{draine03}.
In other words, the scattered light we see at any point in the halo
was largely traveling outwards and thus sees a smaller $\tau$ on the
way out of the halo.  A self-consistent Monte Carlo radiative transfer
scheme is required to rigorously compute the reflection nebula
spectrum (including a disk emission model, halo model, and dust
model), and this will be the subject of a future paper.  However, we
expect the extinction of the reflection nebula spectrum in the halo is
between 1--5\%, which is smaller than the error on the measured
fluxes.  Thus, the effect will not be discernable in our fitting.

Our model spectra consist of galaxy templates and Local Group dust
models from \citetalias{weingartner01}.  We used galaxy templates for the input spectra
because the sample galaxies are highly inclined and a detailed model
is required to de-redden them.  For the spiral galaxies we used six
templates, 
including \citet{bolzonella00} templates for the Sbc and Scd galaxies
(denoted as BMP~Sbc and BMP~Scd respectively) and the \citet{kinney96}
templates (starburst2, Sc, Sb, and Sa), which we denote as KC~SB2,
KC~Sc, KC~Sb, and KC~Sa respectively.  The different starburst
templates are based on different levels of internal extinction, and
for our sample a low level is appropriate (the starbursts in the
sample are not luminous infrared galaxies).  For the early-type
galaxies we used the \citet{kinney96} and \citet{bolzonella00}
templates (denoted KC~E and BMP~E).

Figure~\ref{figure.template_halo_seds}
shows model SEDs for four galaxy templates with the MW ($R_V = 3.1$),
LMC average, and SMC bar dust models from \citetalias{weingartner01}.  In each panel, we show the
models and a raw template spectrum normalized to their maxima in order
to show the differences.  The LMC and SMC models have offsets for
visual clarity.  While models with MW and LMC dust appear redder than
the input spectrum in this wavelength range, the SMC dust model is
bluer.  The region of the spectrum near the 2175\AA\ UV bump is
unreliable.  The bump is thought to be purely absorptive
\citep[][also A.~Witt 2014, private
communication]{andriesse77,calzetti95} because of the size of the
grains that produce it.  Although the \citetalias{weingartner01} model does show the bump is
primarily absorptive \citep[cf.][]{draine03}, $\varpi$ is a derived
quantity in their models and there is a small residual bump that may
indicate a small inaccuracy.  Thus, the peak seen in the \uvmtwo{}
filter for the MW model (and, to a lesser extent, the LMC model) in
the Scd and Sbc models may be inaccurate.  However, the LMC and SMC
dust are adequate representations of a scattered spectrum for testing
whether the halo light is consistent with a reflection nebula.  Thus,
in the reflection nebula scenario we expect the best fits with LMC- or
SMC-type dust and a galaxy template that matches the galaxy type.

\subsubsection{Fitting Results}

We fit the measured halo SED using least-squares fitting with the
$\chi^2$ goodness-of-fit statistic.  The measured SEDs come from the
fluxes in Table~\ref{table.halo_fluxes}, and the \uvwone{} filter is
not used because of the red leak (we return to this later).  It is
also formally inappropriate to use the \uvwtwo{} correction factor,
which depends on the true underlying spectrum, but for these fits the
difference in correction factor between galaxy templates is only a few
percent after the dust model has been folded in because extinction is
much more efficient in the UV.

Figure~\ref{figure.chisq_tempmaps} shows
$\chi^2/\nu$ for the best-fit scaling factor in each model for the
spiral galaxies.  The left-hand panel shows the Sa--Sbc galaxies, and
the right-hand panel the Sc--Scd galaxies.  In each row, we show the
$\chi^2/\nu$ values for the best fit for each model, where there are
three circles per template representing (from left to right) the
$\chi^2/\nu$ value for MW, LMC, and SMC-type (dark red, orange, and
light orange) dust models respectively.  The best fit overall for each
galaxy is marked by a box, and the values are encoded as circles whose
sizes correspond to $\chi^2/\nu$, which is clipped at 25.  Smaller
circles are better fits.  We omit the early-type templates, but these
are never preferred.

The best 2--3 fits
typically occur for a galaxy template close to the host galaxy type
(Figure~\ref{figure.chisq_tempmaps}, note that an asterisk after the
galaxy type denotes a starburst), and the best-fit model uses the
SMC or LMC dust in 18/20 cases.  If we consider only those cases where
$\chi^2/\nu \le 3$, 9/11 use the SMC model, which we think has the
most accurate $\varpi$ of the \citetalias{weingartner01} models.  For most 
galaxies the fit gets significantly worse as the template becomes a
worse match to the galaxy type if we use the MW dust model.  An
obvious exception is NGC~3628 (Figure~\ref{figure.chisq_tempmaps}),
whose halo SED is closer to an Sa type galaxy
(Figure~\ref{figure.results_sb}).  

Only a few of the early-type galaxies in the sample have ``halo''
emission that is fit well by any of the reflection nebula models (not
shown).  In most cases, the observed SED rises continously towards
longer wavelengths and the FUV emission is too weak to obtain
$\chi^2/\nu \lesssim 25$.  

\begin{figure}
\begin{center}
\hspace{-0.5cm}\includegraphics[width=0.5\textwidth]{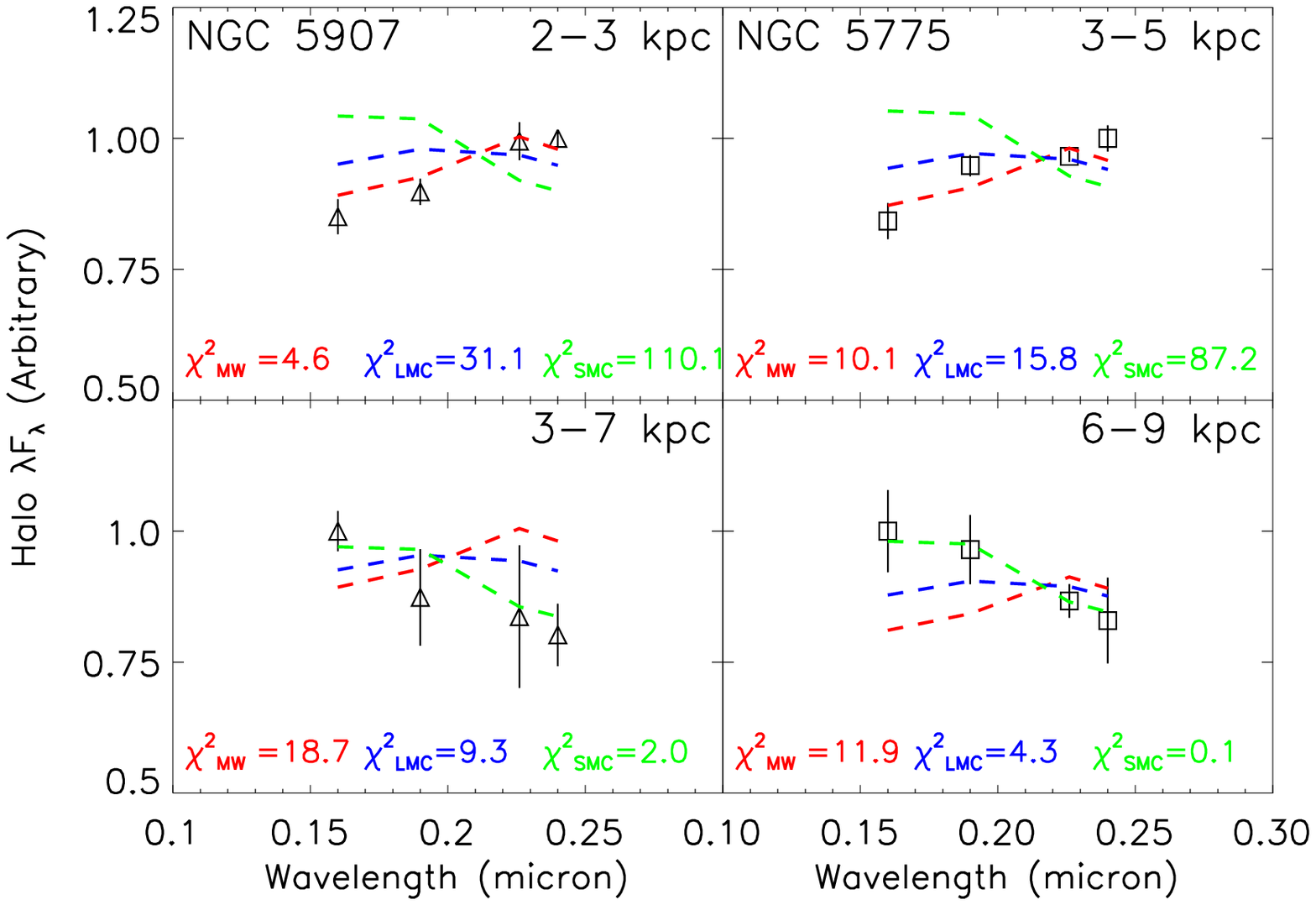}
\caption{\scriptsize \label{figure.height_sedfit} Measured SEDs and the best-fit models
(using the BMP~Scd template) for two zones around NGC~5907 and NGC~5775.  
Closer to the galaxy, the SED is more consistent with MW or LMC-type dust,
whereas the slope reverses farther out and SMC-type dust is a better fit.  
This is true of other late-type galaxies in the sample, but the data 
are not as good.
}
\end{center}
\end{figure}

Some halo SEDs get bluer farther from the galaxy, and a few 
galaxies have data of sufficient quality to measure a change in
the SED slope.  Figure~\ref{figure.height_sedfit} shows the SED
measured near and farther from the galaxy for NGC~5907 and NGC~5775,
where the sign of the slope reverses from closer to the disk (though
still several kpc above the midplane) to farther out.  We have also
plotted the best-fit models for each dust model using the BMP~Scd
template.  The SEDs closer to the disk cannot be fit well with a
reflection nebula model, whereas 
the SEDs higher up are
fit very well by the SMC model.  Thus, near the disk
there may be an additional component or the reflection nebula model
is invalid.  Extinction of a reflection nebula \textit{by dust in the
halo} cannot explain the change in slope, as the required $N_H$ is
several$\times 10^{22}$\,cm$^{-2}$ (for $\delta_{\text{DGR}} \sim
1/100$).  A similar transition is also seen in NGC~891 and NGC~5301,
but not in NGC~3079 (or NGC~24 and NGC~4088, which are less inclined,
so the halo extraction boxes slice through multiple heights).

We note that the SMC $\delta_{\text{DGR}}$ is 5--10\,times smaller
than in the Milky Way \citep[depending on the
region;][]{gordon03,leroy07}, so for SMC dust the implied $N_H$ to
produce a given $L_{\text{halo}}$ is likewise higher.  However,
\citet{menard12} found $\delta_{\text{DGR}} = 1/108$ for the halo
dust with an SMC-like extinction curve (i.e., a similar dust
composition to the SMC but a different $\delta_{\text{DGR}}$).  
Using their value, we find a mean $N_H \sim 5\times 10^{19} - 3\times
10^{20}$\,cm$^{-2}$ based on Equation~3.  For truly SMC dust, the
value is commensurately higher.  The column density through the halo
is discussed in more detail in Section~\ref{section.msfr10}.

\subsection{The UVOT Red Leaks}

The UVOT red leak means that the \uvwone{} and \uvwtwo{} filters have
redder central wavelengths than the nominal values.  The effect is stronger in
the \uvwone{}, which we excluded from our SED fitting.  
The best-fit models for the
4-point SEDs all underpredict the measured \uvwone{} flux for the
spiral galaxies, so the proportion of excess flux tells us about the
luminosity of another component.  If the second component is a
(classical) stellar halo and the reflection nebula hypothesis is
correct, we expect the model to fit well when the stellar component
has about the same luminosity (and perhaps color) as observed halos.

We attempted to determine the stellar halo luminosity in this way for
NGC~5907 and NGC~5775, which have some of the best data.  For the
reflection nebula component, we used the best-fit models from above,
and for the stellar halo we tried the elliptical galaxy templates
from above as well as GALEV \citep[a population synthesis and evolution
  code;][]{kotulla09} E/S0 and Sa models with metal-poor populations,
which may better represent the stellar halo.  We fit the halo model to
measured SEDs including the \uvwone{} flux, where the free parameter
is the proportional flux in the stellar halo.

\begin{figure}
\begin{center}
\hspace{-0.5cm}\includegraphics[width=0.5\textwidth]{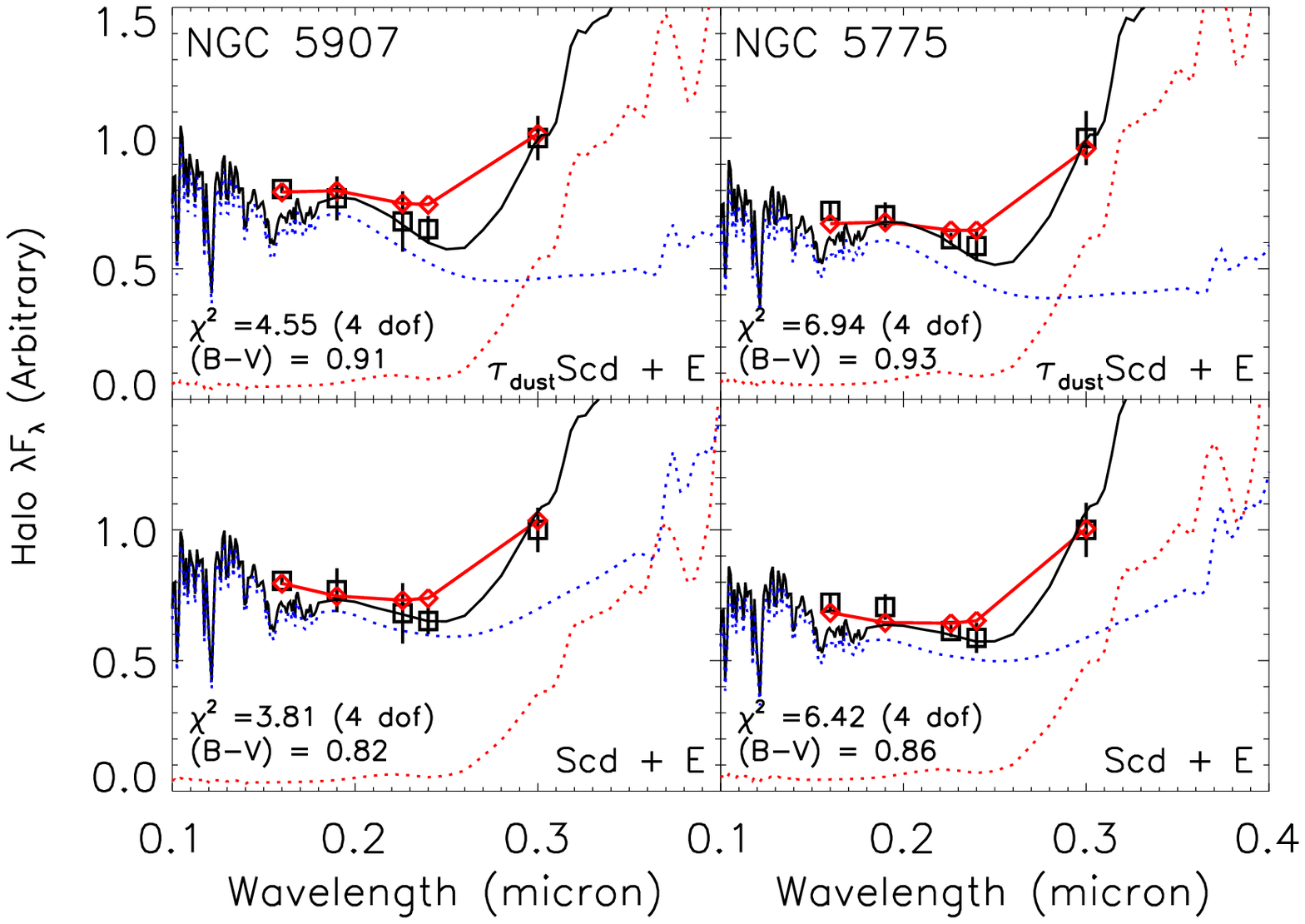}
\caption{\scriptsize \label{figure.uvw1corr} 
SEDs including the \uvwone{} filter (black) with best-fit reflection 
nebula$+$BMP~elliptical template model (top panels) and
BMP~Scd$+$elliptical template model (bottom panels) for NGC~5907 and NGC~5775 
(solid red line).  The model $\lambda F_{\lambda}$ is overplotted in arbitary
units (not a fit) with the proportional components shown as dotted blue
(reflection nebula or Scd template) and red (elliptical template) lines.  
The \uvwone{} flux is plotted at the effective wavelength computed for that
model, not its nominal wavelength of 2600\AA.  See
text for discussion.
}
\end{center}
\end{figure}

The best fits are shown in the top panels of
Figure~\ref{figure.uvw1corr}.  The measured SEDs are shown as black
boxes, and the best-fit model SED is shown as red diamonds connected
by a red line.  We have also overplotted the model $\lambda
F_{\lambda}$ and its components (the dotted blue and red lines
represent the reflection nebula and stellar halo respectively).
The \uvwone{} point is shown at an effective wavelength computed from the 
model and filter response.  This wavelength is strongly model dependent.
In both galaxies, the
best stellar halo model is an early-type template rather than a
metal-poor template, but the $\chi^2$ values are not much different.
For NGC~5775, $\chi^2/\nu = 1.7$, whereas for NGC~5907 $\chi^2/\nu =
1.1$ (NGC~5775 also has a worse $\chi^2$ for the best fit in the prior
subsection).

For NGC~5907, the model predicts a bolometric stellar halo luminosity
of $2\times 10^9 L_{\odot}$, and for NGC~5775 about $5\times 10^9
L_{\odot}$.  This is consistent with stellar halo measurements for
these galaxies and with the Milky Way \citep{carney89}, and stellar
halos typically tend to have luminosities of a few percent of the host
galaxy.  We also computed the model $B-V$ colors by folding the
  spectrum through the $B$ and $V$ filters and adding a constant
  factor for comparison with observations, following \citet{fukugita95}.
The model $B-V$ color is 0.91\,mag for NGC~5907 and 0.93 for
NGC~5775.  \citet{lequeux96} measured $B-V = 0.90$ for NGC~5907
\citep[which is remarkable for being a ``red'' halo;][]{sackett94}, so
for this galaxy the model is consistent with the observations.  

We also tried fitting a dual stellar halo model, where the first
component is simply a late-type galaxy template instead of the
reflection nebula spectrum.  This is motivated by the possibility that
the UV light comes from halo stars of a separate population.  
Stars might form in some of the gaseous material expelled
into the halo.  Molecular gas has been seen up to a few kpc around
some edge-on galaxies \citep[e.g.,][]{garcia-burillo92,lee02}, and
only a small amount of halo star formation is needed to explain the UV
fluxes.  For example, in NGC~891, the \Galex{} FUV halo flux
integrated above 1\,kpc reqires a halo SFR of $\sim 0.028
M_{\odot}$\,yr$^{-1}$ \citep[using the formulas
  in][]{kennicutt98,parnovsky13}.  We would expect some associated
H$\alpha$ emission, but NGC~891 has a bright H$\alpha$ halo; using the
scale height of 520\,pc and total $L_{\text{H}\alpha} \sim 9\times
10^{40}$\,erg\,s$^{-1}$ from \citet{howk00}, the diffuse H$\alpha$
above 1\,kpc is over 7 times higher than that expected from halo star
formation.

The dual halo model provides equally good fits (bottom panels of
Figure~\ref{figure.uvw1corr}), but the luminosity of the redder halo
is several times smaller than for the reflection nebula (plus halo)
model, and the $B-V$ colors are bluer.  The reflection nebula and
stellar halo model predicts more realistic stellar halos.

There are several other galaxies where we can do similar fits: NGC~24,
NGC~891, NGC~2841, and NGC~5301.  In each case, 80--85\% of the flux
in the best-fit models comes from the reflection nebula component in
that model, with $B-V \sim 0.84-0.93$.  The early-type galaxies, by
comparison, have \uvwone{} fluxes consistent with the outskirts of the
galaxy.

\subsection{Maximum Extent of the FUV Emission}

In Section~\ref{section.results}, we noted that around spiral galaxies the
FUV emission is visible at least as far or farther than emission in the other
bands, whereas in elliptical galaxies the FUV emission drops below 2$\sigma$
before the other bands (Figure~\ref{figure.zmax_fuv}).  This is probably not due to a difference in scale height
between bands (e.g., Table~\ref{table.scaleheights}) but rather a difference in
SED and emission mechanism.

The extent of the FUV emission is simply explained in the reflection
nebula scenario by the relatively large FUV flux emerging from
star-forming galaxies as well as the higher scattering cross-section
at shorter wavelenths.  A traditional stellar halo (even a metal-poor
one) is not likely to be more detectable in the FUV than other bands.
If the light is from younger stars, the scale heights of 3--5\,kpc for
NGC~3079, NGC~5775, and NGC~5907 indicate that there are some young stars
over 10\,kpc from the disk.

\subsection{Summary and Caveats}

In late-type galaxies, the reflection nebula model is a better
explanation than a stellar halo for the UV halo SED.  We have
considered the UV$-r$ colors in the halo and the halo SEDs.  
The early-type galaxies, on the other hand, have UV$-r$ colors and SEDs in
their ``halos'' that are consistent with being the faint galactic outskirts
rather than a reflection nebula.  The S0 galaxies have some
distinct halo emission (the halo SEDs differ from the galaxy and 
the FUV emission extends far from the galaxy), but additional
data are required to separate the light from the stellar halo/galaxy
outskirts.

We caution that we have not tried models other than the reflection
nebula, classical stellar halo, and two-component stellar halo, so our
results do not prove a reflection nebula origin.  Even if the UV halo
is a reflection nebula, the dust model we used may not be accurate.  A
better dust model \citep[perhaps patterned after][]{nozawa13},
including Monte Carlo radiative transfer, is required to make stronger
statements about the dust content, and will be the subject of a future
paper.

\begin{figure*}
\begin{center}
\includegraphics[width=0.4\textwidth]{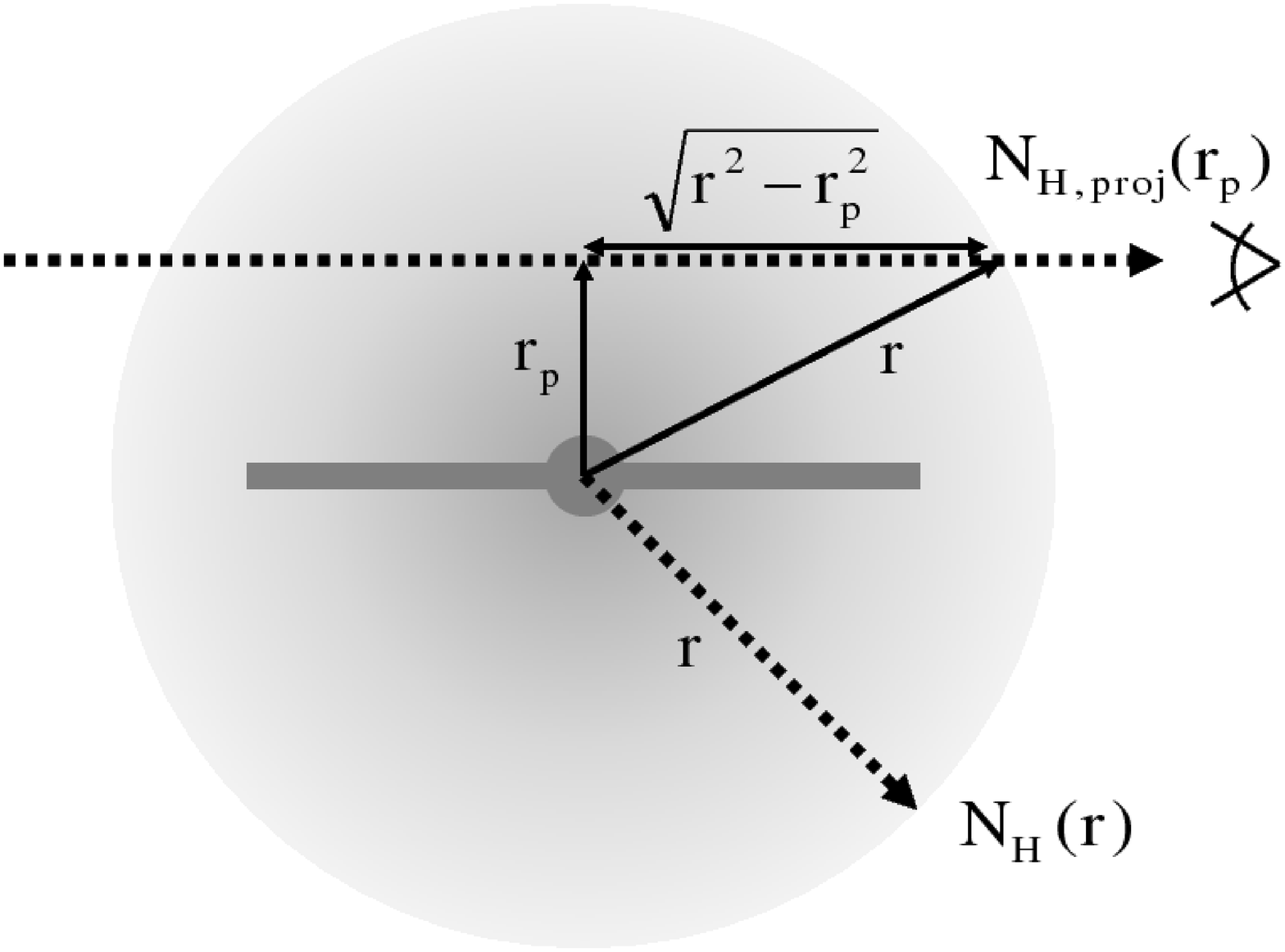}
\includegraphics[width=0.5\textwidth]{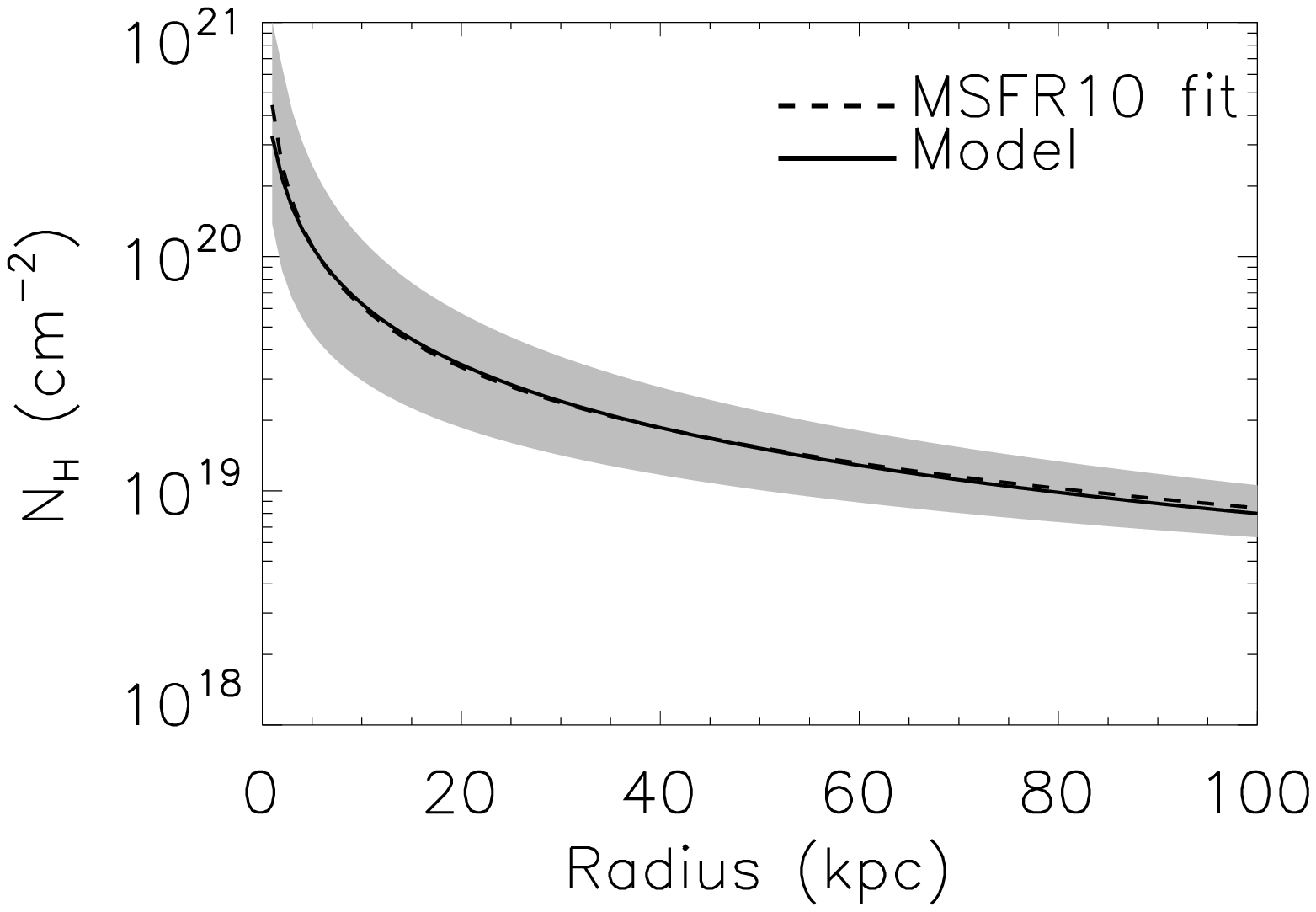}
\caption{\scriptsize \label{figure.colden} 
\textit{Left}: Cartoon of a spherical halo showing the relationship
between $N_H$ measured by an observer and $N_H$ from the origin
through the halo. \textit{Right}: \citetalias{menard10} fit a curve to their mean
$A_V(r_p) \propto N_H(r_p)$ measurements (dashed line, 1$\sigma$ error
shown in grey) that is consistent with a
volume density $n(r) \propto r^{-1.96}$ (solid line).  See text for discussion.
}
\end{center}
\end{figure*}
	
\section{Comparison with Optical Reddening from MSFR10}
\label{section.msfr10}

In the reflection nebula picture, the ratio of the halo and galaxy
luminosity depends on the amount of dust
(Equation~\ref{equation.scattering}).  Since the halo dust also
reddens background sources, we can determine whether the UV data are
consistent with the dust that produces the extinction measured by \citetalias{menard10} at
larger radii by comparing the halo luminosity and light profile we
expect if the \citetalias{menard10} results are extrapolated to smaller radii.

\citetalias{menard10} fit a curve to the mean extinction as a function of projected
galactocentric radius, $r_p$.  
In this section, we first derive the volume density profile implied by the
\citetalias{menard10} fit, which is required to measure extinction within the halo.  We
then compute a plausible range of $L_{\text{halo}}/L_{\text{gal}}$ values
for such a halo, using the \citetalias{weingartner01} SMC dust model that we think best approximates
the scattered spectrum with the $\delta_{\text{DGR}}$
from \citet{menard12}.  Comparing this range to the 
observed data requires several additional considerations, such as the distributed
nature of the galaxy emission, the anisotropy of scattering, and accounting
for the extinction in the disk for estimating $L_{\text{gal}}$.  We adopt a 
simplistic model (a more realistic model is reserved for a future paper) to
show that the data are consistent with the halo luminosity expected from the
\citetalias{menard10} profile at small radii.  Finally, we compare the height profile of the
measured UV fluxes to that expected from the \citetalias{menard10} curve and compute the
gas mass.

We focus on the \uvmtwo{} data, where there is no contamination
from the stellar halo and where we think the background is the most
uniform.  However, the analysis could be extended to the other bands.

\subsection{A Spherical MSFR10 Halo}

\citetalias{menard10} fit a curve to the mean optical extinction in the $V$ band, $A_V(r_p)$, towards background sources
with $r_p = 15-1000$\,kpc (Equation~30 in their paper).  $A_V(r_p)$,
depends on the projected column density $N_{H,\text{proj}}(r_p)$ through the halo and the dust model:
\begin{equation}
\begin{split}
A_V(r_p) &= 0.23\pm0.06 \biggl(\frac{r_p}{1\text{kpc}}\biggr)^{-0.86\pm0.19} \\
	 &= 1.086\sigma_{\text{ex}}(V) \delta_{\text{DGR}} N_{H,\text{proj}}(r_p)
\end{split}
\end{equation}
where $\sigma_{\text{ex}}(V)$ is the extinction cross-section in the
$V$ band.

If the halo of dust-bearing gas is spherically symmetric, we can use 
$N_{H,\text{proj}}(r_p)$ to obtain the volume density as a function of galactocentric
radius $r$ (Figure~\ref{figure.colden}).  From $n(r)$, we can then find
the column density between the center of the halo and a point within,
$N_H(r) = \int n(r) dr$.  $N_{H,\text{proj}}(r_p)$ is integrated
through the halo on a path perpendicular to $r_p$ (see
Figure~\ref{figure.colden}), i.e., $N_{H,\text{proj}}(r_p) = \int n(s)
ds$.  The integral is effectively over all volume densities $n(r>r_p)$
and the path length from $r_p$ to some higher $r$ along the path is $s
= \sqrt{r^2 - r_p^2}$.
Thus,
\begin{equation}
\label{equation.nh_proj}
\begin{split}
N_H(r_p) &= \int n(s) ds = 2\int_{r_p}^{\infty} n(r)\frac{r}{\sqrt{r^2 -
         r_p^2}} dr\\
         &= \frac{A_V(r_p)}{1.086\sigma_{\text{ex}}
         \delta_{\text{DGR}}}
\end{split}
\end{equation}
It is possible that $n(r)$ does not have an easily integrable form,
but based on the \citetalias{menard10} fit a reasonable choice is $n(r) = n_0
r_{\text{kpc}}^{-x}$, where $x$ is determined by choosing a value, evaluating the
integral, and comparing the resulting $N_{H,\text{proj}}(r_p)$ to the \citetalias{menard10} curve.  
The best-fit exponent is $x = 1.96 \pm 0.06$, which is consistent with
$x=2$ within the \citetalias{menard10} error bars (Figure~\ref{figure.colden}).
Thus, we adopt $n(r) = n_0 r_{\text{kpc}}^{-2}$.  Since $n(r)$
diverges at $n=0$, $N_H(r)$ is effectively normalized by some minimum radius,
\begin{equation}
N_H(r) = n_0\text{(1 kpc)}\biggl(\frac{1}{r_{\text{min}}} - \frac{1}{r}\biggr) 
\end{equation}
where the radii are in kpc.
We do not know $r_{\text{min}}$, and in real galaxies the disk--halo interface
is nebulous.  Considering that thick (stellar) disks in disk galaxies have
scale heights of $0.2-1.5$\,kpc \citep{yoachim06}, a reasonable
choice of $r_{\text{min}}$ is 2\,kpc for Milky Way-sized disk
galaxies, which is consistent with where we measure UV fluxes.  
Of course, a spherical halo model breaks down near the disk
\citep[extraplanar gas tends to look like an exponential atmosphere near the
disk; e.g.,][]{sancisi08}, so $r_{\text{min}}$ could be thought of as a normalization.
For $r_{\text{min}} = 2$\,kpc
and $\delta_{\text{DGR}} = 1/100$, the total $N_H$ through the halo is 
a few$\times 10^{20}$\,cm$^{-2}$.  As $N_H$ depends on $\delta_{\text{DGR}}$
but $\tau$ depends only on the dust column, we defer a discussion of $N_H$ and
gas mass to the end of this section.

The spherical halo described here (hereafter the \citetalias{menard10} halo) 
will be used through the rest of this
section to estimate $L_{\text{halo}}$ in the \citetalias{menard10} model, its light profile,
and the total mass.  We also assume that the halo is optically thin and that
a scattered photon scatters only once before escaping, based on the
$N_H$ and $\delta_{\text{DGR}} \sim 1/100$.
While the spherical approximation must break
down near the disk, we will see below that it is useful.  

\begin{figure}
\begin{center}
\includegraphics[width=0.5\textwidth]{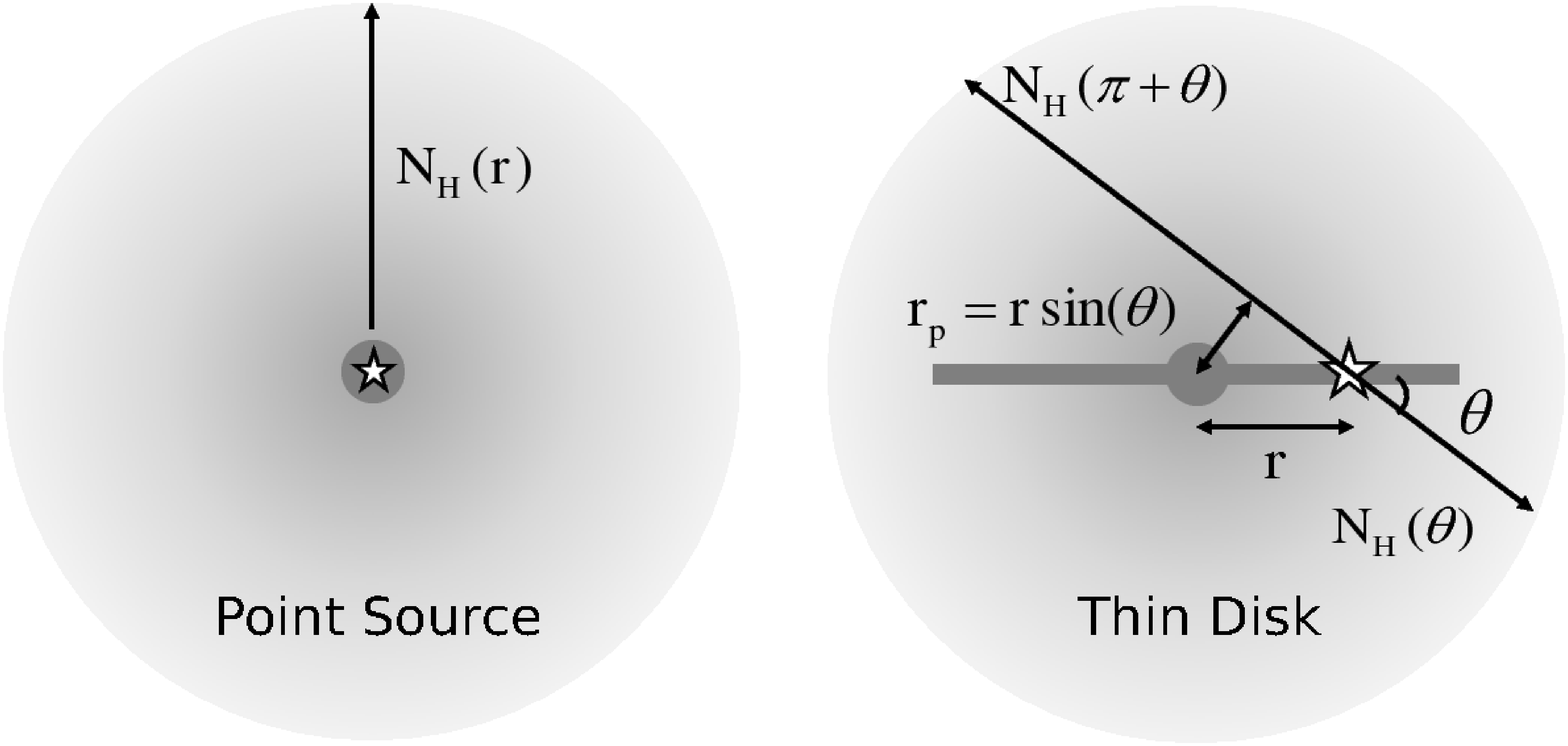}
\caption{\scriptsize \label{figure.ps_disk} 
A ``point source'' galaxy and a uniformly bright disk in a spherical
\citetalias{menard10} halo provide upper and lower bounds on the $L_{\text{halo}}$ we
expect because of the different mean column density.  In the disk
case, one can see that $N_H(\theta)+N_H(\pi+\theta) =
N_{H,\text{proj}}(r\sin\theta)$  for a ray emerging at $\theta$ from a point source in
the disk (right).  
}
\end{center}
\end{figure}

\subsection{$L_{\text{halo}}/L_{\text{gal}}$ in the MSFR10 Halo}

$L_{\text{halo}}/L_{\text{gal}}$ can be derived in the \citetalias{menard10} halo 
from a dust model and a source/halo geometry.  In this subsection, we determine
the total luminosity and ignore the directional dependence
in the scattering cross-section (which affects how much of the light
is seen at a given viewing angle).

A ray traveling from the origin a distance $r$ is attenuated by $e^{-\tau}$,
where $\tau(r,\lambda) =
N_H(r)\sigma_{\text{ex}}(\lambda)\delta_{\text{DGR}}$
(Equation~\ref{equation.scattering}).  If $\tau \ll 1$ and we
integrate over all rays to $r = \infty$,
Equation~\ref{equation.scattering} becomes
\begin{equation}
L_{\text{halo}}/L_{\text{gal}} \sim \tau(\lambda) = \overline{N_H} \sigma_{\text{ex}}(\lambda)\varpi(\lambda)\delta_{\text{DGR}}
\end{equation}
where $\overline{N_H}$ is the mean $N_H$ over rays in all directions (and thus contains
all geometric considerations).  

We now consider two limiting cases to bracket the plausible range of
$L_{\text{halo}}/L_{\text{gal}}$ for the \citetalias{menard10} halo: a ``point source'' galaxy and a thin galaxy disk
in a completely spherical halo.  In the case of the point source, $N_H
= \overline{N_H} 
= n_0/r_{\text{min}}$, so 
\begin{equation}
(L_{\text{halo}}/L_{\text{gal}})_{\text{ps}} \sim 0.2/r_{\text{min,kpc}} \sim 0.1
\end{equation}
for $r_{\text{min}} = 2$\,kpc.
In a real galaxy, $L_{\text{gal}}$ is distributed over a disk much larger
than $r_{\text{min}}$.  Even if the halo becomes like an exponential
atmosphere near the disk, the average $N_H$ seen by light emitted from farther
out in the disk will be smaller than near the center, so 
$(L_{\text{halo}}/L_{\text{gal}})_{\text{ps}}$ is higher than we expect in
a real galaxy.

On the other hand, if $L_{\text{gal}}$ is distributed over a thin, uniformly
bright disk in the \citetalias{menard10} halo (i.e., extraplanar gas does not look like an
exponential atmosphere near the disk), the $\overline{N_H}$ farther out in the disk
will be lower than we expect from a real galaxy, so this situation provides
a lower bound on $L_{\text{halo}}/L_{\text{gal}}$.  To obtain $L_{\text{halo}}/L_{\text{gal}}$,
we first find the $\overline{N_H}$ as a function of radius for a source emitting in
the disk.  We then integrate over the disk.

Consider a point source at a radius $r$ in a disk of radius $R$
(Figure~\ref{figure.ps_disk}).  The total column density
through the halo for each ray then depends on the angle $\theta$, but
one can see that the sum of $N_H(\theta)$ and $N_H(\pi+\theta)$ is 
equal to the projected column density $N_{H,\text{proj}}(r_p)$ at $r_p
= r\sin\theta$.  
The mean $N_{H,\text{proj}}(r_p)$ occurs at $r_p =
2r/\pi$, since the mean of $\sin\theta = 2/\pi$ between 0 and $\pi$,
so for 
$n(r) = n_0 r^{-2}$,
\begin{equation}
\overline{N_H}(r) = \frac{n_0}{2} \int_{r_p}^{\infty} dr \frac{1}{r^2}
\frac{r}{\sqrt{r^2-r_p^2}} = \frac{n_0 \pi^2}{8}\frac{1}{r}
\end{equation}
For a uniformly bright, thin disk of total
luminosity $L_{\text{gal}}$ and inner and outer radii $r_{\text{min}}$
(for consistency) and $R$, the total area is
$\pi(R^2-r_{\text{min}}^2)$ and the area element is $dA = 2\pi r dr$.  Thus,
\begin{equation}
\begin{split}
\biggl(\frac{L_{\text{halo}}}{L_{\text{gal}}}\biggr)_{\text{disk}} &=
\frac{\sigma_{\text{ex}}\delta_{\text{DGR}}\varpi}{\pi(R^2-r_{\text{min}}^2)} \int_{r_{\text{min}}}^{R}
2 \pi r
\biggl(\frac{\pi^2}{8}\frac{n_0}{r}\biggr) dr\\
&=
r_{\text{min}}\frac{\pi^2}{4}\frac{(R-r_{\text{min}})}{(R^2-r_{\text{min}}^2)}
\biggl(\frac{L_{\text{halo}}}{L_{\text{gal}}}\biggr)_{\text{ps}}
\end{split}
\end{equation}
The Milky Way-sized disk galaxies in our sample have UV luminous disks
with projected semimajor axes
between 10--20\,kpc.  Taking this range for $R$, the scattered halo luminosity from our
idealized disk is between 1/3 and 2/11 of that from a point source of
the same $L_{\text{gal}}$ at the center of the halo, or
\begin{equation}
(L_{\text{halo}}/L_{\text{gal}})_{\text{disk}} \sim 0.02-0.03
\end{equation}
The value can be scaled up or down if the disk is not uniformly
bright, based on the concentration of light at each radius.

For the parameters above, we expect $0.02 \lesssim L_{\text{halo}}/L_{\text{gal}} \lesssim 0.1$.
Considering differences in $N_H$, morphological type, disk size between
galaxies of a similar absolute magnitude, and differences between
$\sigma_{\text{ex}}$ model curves, we expect the late-type galaxies to
have 
\begin{equation}
L_{\text{halo}}/L_{\text{gal}} \sim 0.04-0.1
\end{equation}
This ratio does not depend on the galaxy color.

\subsection{Comparison to Observations}

The total $L_{\text{halo}}/L_{\text{gal}}$ is not directly comparable to the
measured values.  First, we must determine the fraction of $L_{\text{halo}}$ that we
would see from a region of interest around a galaxy of a given inclination
and disk thickness.  This is important because scattering is forward-throwing
\citep{draine03} and extinction in galaxy disks makes $L_{\text{gal}}$ 
anisotropic.  Second, we must correct the measured $L_{\text{gal}}$ for
extinction within the disk, as the galaxies in our sample are highly inclined.

\begin{figure}
\begin{center}
\includegraphics[width=0.5\textwidth]{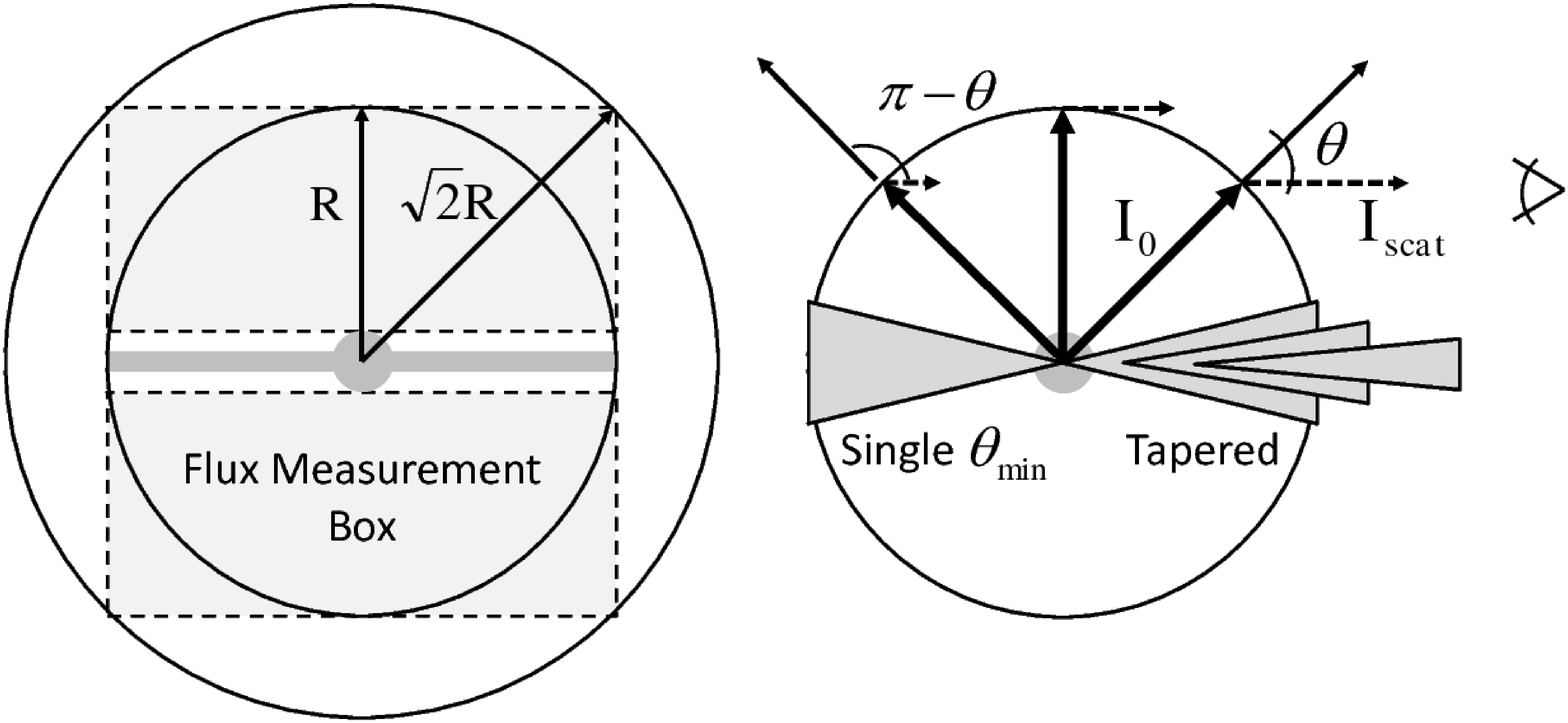}
\caption{\scriptsize \label{figure.obscat} 
\textit{Left}: The column density from the origin to the edge of the
flux measurement box varies with angle, but the mean $N_H$ occurs for
a sphere with a radius in between the smallest circle to bound the box
and the largest that fits inside.  \textit{Right}: The anisotropy in
$L_{\text{halo}}$ occurs because most light is scattered close to the
incident angle.  For an edge-on galaxy, light emerging from the galaxy
close to the line of sight is strongly attenuated by the disk.
}
\end{center}
\end{figure}

\subsubsection{Observable Fraction of $L_{\text{halo}}$}

If we compute an isotropic luminosity from the measured UV fluxes 
($4\pi d^2 F_{\text{halo}}$), it will underestimate the true $L_{\text{halo}}$
for several reasons.  The most obvious is that the flux measurement
boxes (e.g., Figure~\ref{figure.results_sc}) are smaller than the halo,
but as most of the scattering occurs near the disk, inclination effects
are more important.  Due to extinction in the disk, more
light emerges along the minor axis than near the major axis.  
A scattered photon is most likely to have a new trajectory at a small angle
from the incident path \citep{draine03}, so more of the total $L_{\text{halo}}$
is visible from above the disk than from the side (if it could be separated
from the non-scattered light).

For the \citetalias{menard10} halo we can estimate the
amount of light that would be scattered into the line of sight for a given
inclination, measurement region, and disk emission model.  The ``isotropic'' halo
luminosity derived from the visible light we call $L_{\text{halo,vis}}$.
$L_{\text{halo,vis}}$ underestimates the true $L_{\text{halo}}$ for an 
edge-on galaxy and overestimates it for a face-on galaxy.  For the remainder of
this subsection, we consider a perfectly edge-on galaxy for the two limiting
cases of a point source and a uniformly bright disk of radius $R$ described
above (Figure~\ref{figure.ps_disk}).  $L_{\text{halo,vis}}$ is derived for a
given flux measurement region by finding $\overline{N_H}$ for rays passing through that
region and integrating over $d\sigma_{\text{scat}}/d\Omega$ from \citet{draine03}.

The boxes where we measure fluxes have finite width and height, but measure
light from an infinite depth.  For the \citetalias{menard10} halo, we can define a box and
integrate numerically to evaluate $N_H$ and $d\sigma_{\text{scat}}/d\Omega$,
but with a few geometric simplifications we can make an instructive estimate.
The box width is chosen to match the projected major axis of each galaxy, which
corresponds to a radius $R$.  The maximum height of the boxes in our sample is typically less
than $R$, but as the halo fluxes drop below our sensitivity at lower heights,
we make only a small error by assuming the distance from the top of the
measurement boxes to the midplane of the galaxy is $R$.  Since most
scattering occurs at small radii, we can likewise truncate the depth and
consider a cube with each side having a length $2R$ (the extension and 
truncation cause small errors in opposite directions).  This geometry is shown
in the left panel of Figure~\ref{figure.obscat}.

\begin{deluxetable*}{lccccccc}
\tablenum{7}
\tabletypesize{\scriptsize}
\tablecaption{Halo/Galaxy \uvmtwo{} Luminosity Ratio}
\tablewidth{0pt}
\tablehead{
\colhead{Galaxy} & \colhead{\uvmtwo{} $F_{\text{gal}}$} & \colhead{$F_{\text{IR}}$} & \colhead{$A_{\text{UV}}$} & \colhead{$A_B$} & \colhead{$L_{\text{halo,vis}}$} & \colhead{$L_{\text{gal,corr}}$} & \colhead{$L_{\text{halo,vis}}/$}\\
                 & \colhead{($10^{-11}$ erg\,s$^{-1}$} & \colhead{($10^{-9}$ erg} & \colhead{(mag)} & \colhead{(mag)} & \colhead{($10^{8} L_{\odot}$)} & \colhead{($10^{8} L_{\odot}$)} & \colhead{$L_{\text{gal,corr}}$} \\
				 & \colhead{cm$^{-2}$\,\AA $^{-1}$)} & \colhead{s$^{-1}$\,cm$^{-2}$)}
}
\startdata 
\tableline
NGC 24    & 6.72$\pm$0.02 & 0.17 & 0.95 & 0.5 & 0.23 & 4.8 & 0.05 \\
NGC 891   & 3.09$\pm$0.07 & 6.16 & 5.06 & 0.5 & 0.71 & 75. & 0.01 \\
NGC 2841  & 10.0$\pm$0.03 & 0.54 & 1.37 & 0.4 & 0.58 & 22. & 0.03 \\
NGC 3079  & 9.82$\pm$0.01 & 4.81 & 3.39 & 0.6 & 1.4  & 113.& 0.01 \\
NGC 3628  & 8.32$\pm$0.04 & 5.51 & 3.78 & 0.6 & 1.2  & 81. & 0.01 \\
NGC 4088  & 1.13$\pm$0.01 & 2.41 & 2.56 & 0.5 & 0.41 & 73. & 0.006 \\
NGC 4388  & 3.00$\pm$0.01 & 1.08 & 3.07 & 0.4 & 0.37 & 43. & 0.009 \\
NGC 5301  & 2.53$\pm$0.02 & 0.35 & 2.17 & 0.5 & 0.54 & 24. & 0.02 \\
NGC 5775  & 2.69$\pm$0.01 & 2.13 & 3.93 & 0.5 & 1.9  & 165.& 0.01 \\
NGC 5907  & 7.23$\pm$0.03 & 1.55 & 2.57 & 0.4 & 2.7  & 61. & 0.04 \\
NGC 6503  & 14.6$\pm$0.01 & 1.12 & 1.69 & 0.5 & 0.009& 4.5 & 0.002\\
NGC 6925  & 7.43$\pm$0.01 & 0.62 & 1.76 & 0.4 & 0.72 & 60. & 0.01 \\
NGC 7090  & 6.57$\pm$0.02 & 0.80 & 2.06 & 0.5 & 0.44 & 8.  & 0.06 \\
NGC 7582  & 3.68$\pm$0.03 & 4.58 & 4.46 & 0.6 & 0.69 & 220.& 0.003 \\
UGC 6697  & 2.61$\pm$0.04 & 0.17 & 1.57 & 0.6 & 9.3  & 190. & 0.05\\
UGC 11794 & 0.20$\pm$0.01 & 0.08 & 3.22 & 0.5 & 0.78 & 50.  & 0.02 
\enddata
\tablecomments{\label{table.gal_lum} Cols. (1) Galaxy name (2) \uvmtwo{} galaxy flux (3) 
FIR flux (IRAS $F_{60\mu\text{m}} + F_{100\mu\text{m}}$) (4) $A_{\text{UV}}$
estimated from the FIR flux using the \citet{buat99} relation, which
we use as a correction from the edge-on perspective (5) $A_B$ 
through the disk along the minor axis estimated from
\citet{calzetti01}, which we use as a correction perpendicular to the disk. 
$A_{\text{UV}} = 2.4A_B$ for /citetalias{weingartner01} MW-type dust or $2.8A_B$ for SMC-type dust.  
We use MW-type dust here. (6) \uvmtwo{} $L_{\text{halo,vis}}$ (7) \uvmtwo{} $L_{\text{gal,corr}}$ (8) Halo-to-galaxy
flux ratio.}
\end{deluxetable*}

We can then identify a sphere that is similar enough to the cube that the
$\overline{N_H}$ is approximately correct.  This has the advantage of making
$N_H$ independent of angle and simplifying the integral over 
$d\sigma_{\text{scat}}/d\Omega$.  The radius of the sphere can be obtained
from the largest sphere that fits within the cube and the smallest one that
bounds it (Figure~\ref{figure.obscat}).  
In the former case, the radius is $R$ and in the latter it is $\sqrt{2}R$, so
we take the mean $N_H$ from the galactic center to the
cube edge to be $N_H \approx n_0(1/r_{\text{min}} - 1/1.2R)$.  

In this framework, the anisotropy in flux scattered into a given line of sight
can be described in terms of an effective minimum polar angle $\theta_{\text{min}}$ for light emitted 
from the galaxy to escape its disk (Figure~\ref{figure.obscat}).  For a chosen viewing angle, the 
amount of scattered light can be obtained by integrating $d\sigma_{\text{scat}}/d\Omega$
between $\theta_{\text{min}}$ and $\pi - \theta_{\text{min}}$ 
and multiplying by $N_H$ (the integral over $\phi$ gives a factor of $2\pi$). 
If the disk is uniformly thick at
all radii and a gaseous disk extends beyond the UV-emitting region, 
$\theta_{\text{min}}$ is essentially constant, but one could also imagine a
tapered disk (left and right halves of the circle shown on the right side of
Figure~\ref{figure.obscat}).  Based on typical $N_H$ values in galaxy
disks, $\theta_{\text{min}}$ in our sample is near 
$\theta_{\text{min}} \sim \pi/9$ (20$^{\circ}$), but the extent of the range
is not known.  A real disk model is required and will be presented in a future
work.

Using the \citetalias{weingartner01} SMC bar model and the \citet{draine03}
curve for the \uvmtwo{} wavelength, $R=10$\,kpc and $r_{\text{min}} = 2$\,kpc,
and $\theta_{\text{min}} = \pi/9$, we 
find $L_{\text{halo,vis}}/L_{\text{gal}} \sim 0.001$ and 0.03 for
the uniform disk and point source cases respectively.  The lower bound is
small because the flux measurement boxes are truncated at the disk width.  

Considering the range in disk radii and uncertainty in $\theta_{\text{min}}$, 
the geometric simplifications we made, and the different dust models, a plausible
range for $L_{\text{halo,vis}}/L_{\text{gal}}$ for edge-on galaxies in the
\citetalias{menard10} halo is
\begin{equation}
L_{\text{halo,vis}}/L_{\text{gal}} \sim 0.01 - 0.05
\end{equation}
When we numerically evaluate $L_{\text{halo,vis}}$ for extraction boxes of
width $2R$ in the two limiting cases, we find values that fall within this range.
The integral over $d\sigma/d\Omega$ increases for galaxies at
lower inclination, but we actually expect a lower $L_{\text{halo,vis}}/L_{\text{gal}}$
for these systems because most scattering occurs at small radii, and
the lower regions of the halo are seen in projection against the
galaxy (where we do not measure halo flux).
 
\subsubsection{Correcting for Extinction in the Measured $L_{\text{gal}}$}

$L_{\text{gal}}$ must be corrected because the observed flux is
measured at a viewing angle with maximal extinction in the disk,
whereas light escaping to the halo sees much less extinction.
We correct for this by estimating the intrinsic UV luminosity and then
estimating the extinction along the minor axis.

The UV flux absorbed by dust in the disk is reprocessed into the FIR, so one
can use the FIR-to-UV flux ratio to estimate the UV extinction.  As in 
\citet{buat96}, we use the sum of the IRAS 60\,$\mu$m and 100\,$\mu$m fluxes for the FIR flux.
We adopt the fit in \citet{buat99} to star-forming galaxies (measured at 2000\AA),
\begin{equation}
\begin{split}
A_{\text{UV}} = & 0.466(\pm0.024)+ \\ & 1.00(\pm0.06)\log(F_{\text{FIR}}/F_{\text{UV}})+ \\ & 
0.433(\pm0.051)\log(F_{\text{FIR}}/F_{\text{UV}})^2
\end{split}
\end{equation}
and apply the extinction to the \uvmtwo{} fluxes to obtain the intrinsic
UV luminosity.   

The light is attenuated as it exits the galaxy along the minor axis.
We can estimate this factor from extinction corrections towards
face-on galaxies (which also include any halo extinction, and are thus
slightly too large).  The extinction towards face-on disks depends on the
galaxy type, luminosity, and the radius where it is measured.  Based
on \citet{calzetti01} and references therein, the effective $B$-band
extinction along the minor axis for late-type galaxies is perhaps $A_B
= 0.4-0.6$\,mag for an Sc or Sd galaxy and $A_B = 0.3-0.4$\,mag for Sa
and Sb galaxies.  
For Milky Way-type dust with $R_V = 3.1$, $A_{\text{UV}}/A_B = 2.4$,
so $L_{\text{gal}}$ as seen by the halo is about $0.7-1.7$\,mag
smaller than the intrinsic UV luminosity obtained from the
\citet{buat99} correction.  For SMC-type dust, $A_{\text{UV}}/A_B =
2.8$.  Even within galaxies, it is not clear what dust model to use,
as some have SMC-type dust \citep{calzetti94} and others MW-type
dust.  The location of the dust relative to star forming regions may
also play a role \citep{panuzzo07}, and starbursts tend to have
higher $A_B$.  As the difference is rather small, we adopt the MW-type
dust value for the dust within the disk.

Thus, 
\begin{equation}
L_{\text{gal,corr}} = 4\pi d^2 F_{\text{gal}} 10^{(A_{\text{UV,maj}}-A_{\text{UV,min}})/2.5}
\end{equation}
where the ``maj'' and ``min'' subscripts refer to extinction through
the disk along the major and minor axes respectively.  Considering the scatter in the
\citet{buat99} FIR-to-UV relation, reported differences in the
extinction towards face-on galaxies, and the dependence of the
extinction on the galaxy model \citep{calzetti01,marcum01}, we suppose
that $L_{\text{gal,corr}}$ is accurate to within 50\% in most systems.

\subsection{Comparison to Observed Fluxes}

We can now compare the \citetalias{menard10} $L_{\text{halo,vis}}/L_{\text{gal}}$ to that derived
from the UV data.  Since we rely on
the FIR data, we quote values for the 15 highly inclined Sa--Sc
galaxies with FIR and \uvmtwo{} data in
Table~\ref{table.gal_lum}.  The other quantities in the table go into
computing $L_{\text{gal}}$.  For these galaxies, we find 
\begin{equation}
(L_{\text{halo,vis}}/L_{\text{gal,corr}})_{(\text{obs})} = 0.002-0.06
\end{equation}
This is remarkably similar to the range predicted by our \citetalias{menard10} extrapolation,
\begin{equation}
(L_{\text{halo,vis}}/L_{\text{gal}})_{\text{MSFR10}} = 0.01-0.05
\end{equation}
Galaxies with smaller inclination angles tend to have smaller values, as expected.

There are several comments worth making at this point.  First, we expect 
$L_{\text{halo,vis}}/L_{\text{gal}}$ values of a few percent for a range of
geometries and parameters because the \citetalias{menard10} $A_V$ curve has a maximum
column density of a few$\times 10^{20}$\,cm$^{-2}$ for a realistic $r_{\text{min}}$,
and integrating the \citet{draine03} expression gives a cross-section of
a few$\times 10^{-22}$\,cm$^{2}$.  The errors we make through the various
simplifying assumptions are smaller than the range from our upper/lower bound
analysis.  

Second, the observed $L_{\text{halo,i}}/L_{\text{gal}}$ values are 
consistent with the \citetalias{menard10} fit, but also any halo model that produces
$N_H \sim 10^{20}$\,cm$^{-2}$.  The ratio by itself does not prove the
\citetalias{menard10} model.

Third, there is intrinsic variation in the sample and $L_{\text{halo}}/L_{\text{gal}}$
does not necessarily follow \ion{H}{1} halo mass (for example, NGC~891 has an
unusually massive \ion{H}{1} halo but a relatively low $L_{\text{halo}}/L_{\text{gal}}$).
This could be due to variation in extinction through the disk, the location of
star formation in the disk, and the difference in ``real'' $r_{\text{min}}$
for each galaxy.  A larger sample is also required to rigorously
determine whether the \citetalias{menard10} halo is valid within 15\,kpc of the disk.

Finally, the results are shown for the \uvmtwo{} data only, but we can also
use the \Galex{} NUV and FUV with appropriate corrections and using the /citetalias{weingartner01}
and \citet{draine03} data for the right wavelengths.  For example, for the
FUV the \citet{draine03} model and \citetalias{menard10} curve predict 
$L_{\text{halo,vis}}/L_{\text{gal}} \sim 0.02-0.09$.  The \citet{buat99} 
correction is measured at longer wavelengths and so may not be valid, but
blindly applying an extrapolation to the FUV band, we find
$L_{\text{halo,vis}}/L_{\text{gal}} \sim 0.01-0.04$ for several galaxies where
the FUV fluxes are well constrained.  

\begin{figure*}
\begin{center}
\includegraphics[width=0.65\textwidth]{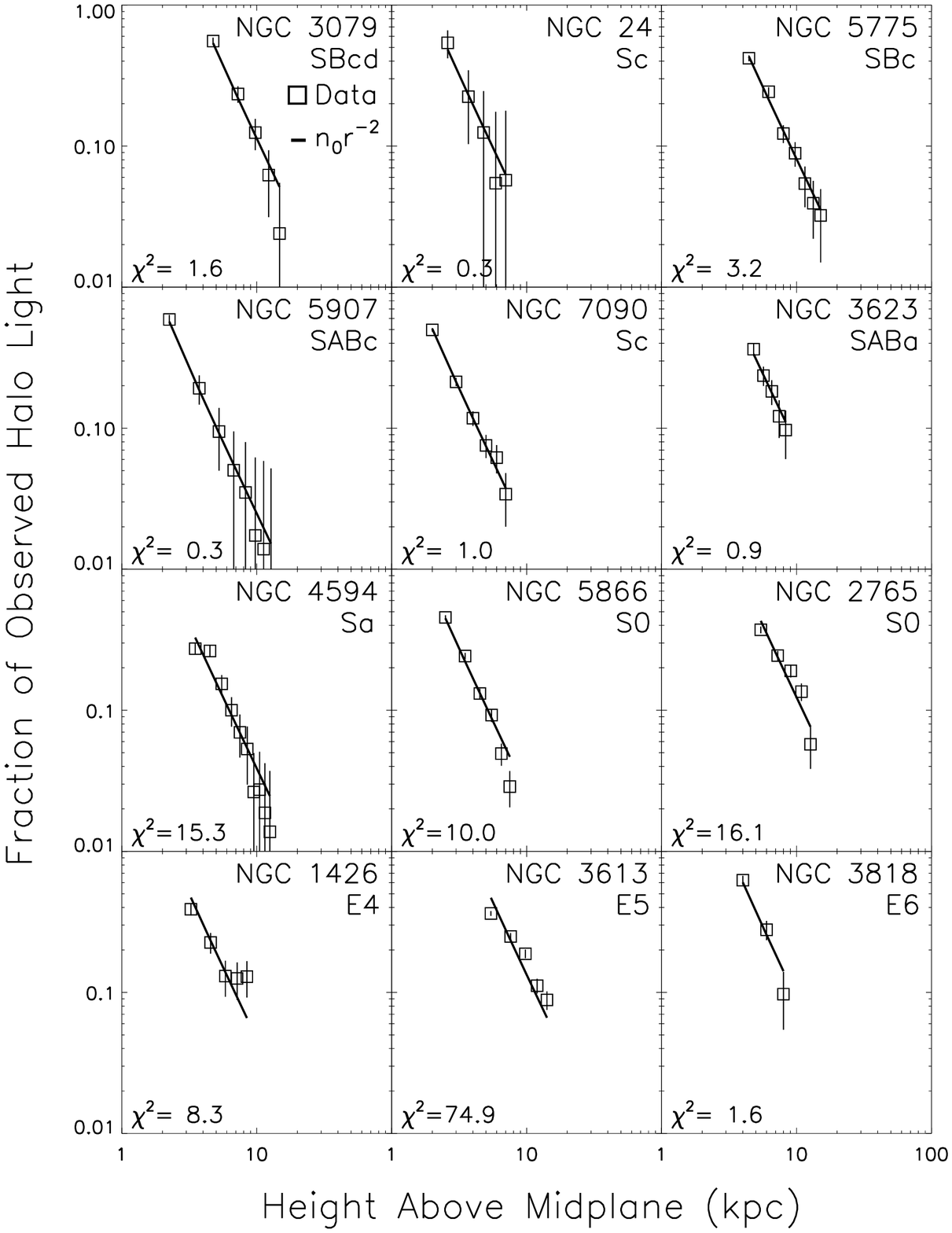}
\vspace{-1.25cm}
\caption{\scriptsize \label{figure.flux_profiles} Measured halo light profiles for some of
our sample (data points) and the prediction from the \citetalias{menard10} $A_V$ fit (line).  
The lines are not fits to the data, but we compare them to the data with the
$\chi^2$ statistic shown in each panel.  
}
\end{center}
\end{figure*}

\subsection{Light Profile}
We can also compare the expected light profile from \citetalias{menard10} to the 
observed fluxes.  In Figure~\ref{figure.flux_profiles} we show the observed
flux profiles for several of the galaxies in our sample with excellent data (boxes),
where the fluxes have been normalized so that the $y$-axis shows the percentage 
of the total \textit{observed} halo light seen in each bin.  We have also overplotted
a model based on the \citetalias{menard10} $n(r)$ using the same heights and 
projected bin sizes (albeit making a spherical approximation as for 
$L_{\text{halo,vis}}$ above).  There is excellent agreement between
the late-type galaxies in Figure~\ref{figure.flux_profiles}, consistent with
$n(r) = n_0 r^{-2}$ even at small radii.  

It is worth noting that the profiles in Figure~\ref{figure.flux_profiles} do not depend on the galaxy
luminosity, and the corresponding curves do not depend on the dust model
(which affects total extinction).  The profiles are thus a clean measure of
how the density varies with radius, assuming a single type of dust and $\delta_{\text{DGR}}$.
A few of the early-type galaxies also show good agreement in 
Figure~\ref{figure.flux_profiles}.  The reason is unclear (most of the early-type
galaxies in the sample prefer a shallower radial dependence), but in these
galaxies the observed $L_{\text{halo}}/L_{\text{gal}} \sim 0.3-1.5$
before any correction for galactic extinction \citep[the][fit is calibrated to
star-forming galaxies]{buat99}, which is badly inconsistent with a 
scattered light origin.

The flux profiles and $L_{\text{halo,vis}}/L_{\text{gal}}$ derived from 
extrapolating the \citetalias{menard10} fit inwards are consistent with the observations.

\subsection{Total $N_H$ and Mass}

If the \citetalias{menard10} halo extends to small radii, we can estimate the mass of 
dust-bearing gas within 20\,kpc.  

The $N_H$ and mass implied by the \citetalias{menard10} curve depend on
$\delta_{\text{DGR}}$, which varies by location in the galaxies of the
Local Group.  For the Milky Way, $\delta_{\text{DGR}} \sim 1/100$,
whereas in the Small Magellanic Cloud it is 5--10 times smaller
\citep{gordon03,leroy07}.  \citetalias{menard10} prefer SMC-type dust for the (outer)
halo, but \citet{menard12} find that $\delta_{\text{DGR}} = 1/108$ for
the halo dust outside 20\,kpc.  Thus, it seems likely that the true
$\delta_{\text{DGR}}$ between 5--20\,kpc is also about $1/100$.  This
gives a maximum $N_H(r) \sim 3\times 10^{20}/r_{\text{min,kpc}}$.  For
$r_{\text{min}} = 2$\,kpc, this is similar to typical sight lines out
of the Milky Way at high latitudes, but these include disk material.
It is about three times higher than $N_H$ viewed out of the Lockman
hole.

Column densities measured in projection towards edge-on extraplanar
galaxies \citep[e.g.,][and references therein]{sancisi08} indicate
maximum extraplanar $N_{H,\text{proj}}(r_p)$ of a few$\times 10^{20}$\,cm$^{-2}$
(NGC~891 has a particularly massive halo with a maximum $N_H(r_p)$
about ten times higher).  In the spherical halo model, at $r_p =
2$\,kpc, $N_H(r_p) \approx 2.4\times 10^{20}$\,cm$^{-2}$, which is in
agreement with these galaxies.

From the density profile we can measure the enclosed mass within a
given radius (one could also integrate the projected column density implied
by the \citetalias{menard10} extinction curve over a visible area),
\begin{equation}
\begin{split}
M(<r) &= 4\pi \mu m_p \int r^2 n(r) dr \\
      &\sim 3.1\times 10^{7} (r_{\text{kpc}}-r_{\text{min,kpc}})
\end{split}
\end{equation}
where $\mu$ is the mean atomic weight per particle and $n \sim 0.1
r_{\text{kpc}}^{-2}$\,cm$^{-3}$.  Within 20\,kpc, this yields
$5\times 10^8 M_{\odot}$ of gas, or several percent of the gas in the
disk for a typical late-type galaxy.  

\subsection{Summary and Caveats}

This section started from the proposition that if the extraplanar UV
emission is a reflection nebula, we can use the fluxes to determine
whether the UV emission is consistent with the same dust seen in
extinction at larger radii.  We found that the halo luminosities and
flux profiles are indeed consistent with the \citetalias{menard10} fit within 20\,kpc
down to the edge of the thick disk, with
a total dust-bearing gas mass of around $5\times 10^8 M_{\odot}$ in this volume.  

There are several potential weaknesses in this analysis.  First, the
results depend on the \citetalias{weingartner01} models and the \citet{menard12}
$\delta_{\text{DGR}}$ for their halo dust, which determines $n_0$ (and thus
$N_H$).  $N_H$ also depends on $r_{\text{min}}^{-1}$, so both $L_{\text{halo}}$ and
$M_{\text{gas}}(<r)$ in the halo depend on this value.  Our choice of
$r_{\text{min}} = 2$\,kpc is based on physical considerations (albeit
in an unphysical halo approximation), and perhaps validated by the
extraplanar $N_H$ measured in other galaxies.  Still, a real halo
model where the halo connects smoothly to the disk gas is required to
explore this issue.  Third, the $L_{\text{gal}}$ that emerges from the
disk depends on the disk thickness and density.  We incorporated this
as a $\theta_{\text{min}}$ above which light escapes into the halo,
but a disk model is required to do a more careful computation.
Finally, our corrections to the observed
$L_{\text{gal}}$ are based on the \citet{buat99} work and measurements
of $A_B$ through galaxy disks.  There is a large scatter in both
between galaxies, so our range of ``measured''
$L_{\text{halo}}/L_{\text{gal}}$ may not be conservative (i.e., wide)
enough.

It is worth noting that most of the geometric simplifications we make
introduce errors in $N_H$ that are small compared to the range between
the limiting cases we considered.  A real model is required to go
beyond this simple analysis.  Likewise, the choice of SMC- or
MW-type dust from the /citetalias{weingartner01} models does not make a large difference if
we adopt the \citet{menard12} $\delta_{\text{DGR}}$ for SMC-type halo
dust.  

\section{Discussion and Conclusions}
\label{section.discussion}

The main findings in this work are
\begin{enumerate}
\item Diffuse UV emission is ubiquitous around highly inclined
  late-type galaxies and in the outskirts of early-type galaxies.
\item In disk galaxies, the extent of this emission is 5--20\,kpc
  (projected, no $\sin i$ correction)
  above the disk midplane.  Halo UV luminosities ($L_{\nu}$) range from $10^7$ to
  a few$\times 10^8 L_{\odot}$ in the \Swift{} and \Galex{} filters.
\item Close to the galaxy, the UV flux becomes higher near the
  projected disk center, but at larger minor axis offsets (Figure~\ref{figure.radial}) the emission is
  more uniform.  There are differences in diffuse luminosity that do
  not correspond to differences in the SED.
\item The halo emission is not consistent with a classic stellar halo,
  but it is consistent with a reflection nebula produced by dust in
  the halo.
\item The distribution of this dust around the galaxy is consistent
  with extrapolating the \citetalias{menard10} fit to the mean extinction outside of
  15--20\,kpc to within a few kpc of the disk.  This implies a typical
  dust-bearing gas mass of $\sim5\times 10^8 M_{\odot}$ for
  $\delta_{\text{DGR}} \sim 1/100$ within 20\,kpc.  
\item Unlike the \citetalias{menard10} detection of halo dust, which required
  averaging over extinction towards sources behind many galaxies, dust
  is detectable around individual galaxies in the UV, and possibly
  even substructure.
\end{enumerate}
A definitive test of the reflection nebula hypothesis may be possible
because the scattering cross-section is anisotropic \citep{draine03}
and scattered light is highly polarized.  Thus, in deep observations
of highly inclined, but not edge-on, galaxies we expect a higher flux
on one side of the galaxy than the other.  The existing data for this
sample are insufficient to make this measurement, and a larger sample
with very little cirrus contamination is required.  NGC~24 is a good
candidate for inclusion in this sample, but the data are not deep
enough.  A UV polarimeter would determine whether the halo emission is
highly polarized, but none is currently available, and the \uvwone{}
red leak indicates that the reflection nebula emission becomes dominated by
a classical stellar halo somewhere between the NUV and $U$ band.

Based on the existing evidence, we conclude that star-forming spiral
galaxies produce reflection nebulas in their halos.  There are a few
important consequences:

First, the scattered-light spectrum could be used to constrain the
metallicity with data in a few more filters.  The fractional
composition of the dust (e.g., the proportion of graphitic or silicate
grains) determines the slope of
the extinction curve in the FUV \citep[e.g.,][]{nozawa13}.  With
additional measurements shortward of the 2175\AA bump, we could
determine the slope (fits would be particularly sensitive to any data
shortward of the \Galex{} FUV).  Since the grain abundance depends on the
gas metallicity, for galaxies with $N_H$ measurements we could put a
lower bound on the metal mass through $\delta_{\text{DGR}}$.

Changes in the SEDs as a function of height would also determine how
(or whether) dust evolves in the disk--halo interface.  Extraplanar
dust filaments are seen in NGC~891 \citep{rossa04} and NGC~5775
\citep{howk09} that extend out to several kpc from the disk, but
beyond this they find no dust structures despite having the
sensitivity to do so.  On the other hand, some dust filaments do
extend to greater heights \citep{irwin06,mccormick13}, which may be
related to galactic superwinds.  In either case, at some point
the dust becomes more diffuse, which may suggest that the dust is
cosmologically old, that galaxies are inefficient at expelling dust
to arbitrary radii, or that the disk--halo interface is good at mixing
small-scale structures into large ones.

Second, given that the \citet{menard12} $\delta_{\text{DGR}}$ beyond
20\,kpc from the galaxy is close to the value within the Milky Way, if
the dust is primarily in Mg~II absorbers, the inferred mass suggests a
comparable amount of hot and cold gas in galaxy halos within 20\,kpc
\citep[for examples of hot mass, see][]{bregman94,strickland04,li13}.
This may be true at larger radii as well: \citetalias{menard10} estimate a total dust
mass within the virial radius of $5\times 10^7 M_{\odot}$ (measured
outside 20\,kpc), corresponding to a gas mass of $M_{\text{gas}} \sim
6\times 10^9 M_{\odot}$.  \citet{anderson13} estimate a total hot mass
within 50\,kpc of late-type galaxies of about $5\times 10^9
M_{\odot}$.  This balance is consistent with hydrodynamic cosmological
simulations in \citet{cen13}.

Another possibility is that some of the dust is hosted in 
the hot gas itself, which may help the hot gas cool.
The discovery of diffuse intergalactic dust in galaxy clusters
\citep{chelouche07}, which are filled with comparatively dense, hot gas, 
indicates that dust must be accreted regularly into these systems because
of the sputtering times \citep{mcgee10}.  However, in the more tenuous hot halos of
galaxies, grains can survive for considerably longer.  
 
Third, variations in flux along the projected major axis with no
corresponding change in SED (Figure~\ref{figure.radial} and seen in
several other galaxies) beyond several kpc from the disk may trace
denser clouds and thus gaseous structures in external galaxies.  For
nearby galaxies, deep UV exposures could probe the gas structure of
the halo on spatial scales of a few kpc.  However, it is also possible
that the differences arise from anisotropic illumination of the halo
by the disk, so a real galaxy--halo model is required for further
study.  Another intriguing possibility is to look for satellite
galaxies through variations in the SED.  Galaxies like NGC~5907 have
such uniform UV halo SEDs (Figure~\ref{figure.radial}) that bins with
markedly different SEDs may identify compact halo structures.
However, high resolution optical follow-up would be necessary to rule
out point sources or background galaxies.

If the SEDs can be connected to dust properties, UV photometry of
galaxy halos will be a powerful way to study individual galaxies with
relatively short exposures compared to those required for true
spectroscopy.

Fourth, if the UV emission is a proxy for halo gas then it may be much
easier to detect cool extraplanar gas within $\sim$20\,kpc of the disk
using UV imaging rather than radio interferometry, which provides more
information at the cost of much longer integration times in the large
arrays needed to achieve arcsecond resolution.

Finally, the detection of the diffuse UV halos indicates that studies
of diffuse UV emission (besides the Galactic cirrus) are possible with
\Swift{}.  We have examined potential contamination sources in detail,
and the agreement between the \Swift{} and \Galex{} fluxes strongly
indicates that we see intrinsic emission from the galaxy halo.  It is
worth notin gthat the correction needed to make the
\uvwone{} a true ``UV'' flux (in the halo) appears to be similar for all of 
the late-type galaxies in our sample, and the $UBV$ filters aboard \Swift{}
are also sensitive enough to constrain the stellar halo luminosity.  Thus, one can in 
principle use \Swift{} observations without \Galex{} data, and there
are a number of galaxies with deep \Swift{} data with shallow or
missing \Galex{} data.

\acknowledgments

The authors thank Eric Bell and Alice Breeveld for helpful input, and
Adolf Witt for pointing out that the ``UV bump'' is a purely
absorptive feature and cannot be used to distinguish dust models via 
scattered spectra.  The authors thank the referee for substantial and detailed
comments that greatly improved the paper.  EHK was supported by
NASA ADAP grant \#061951.

This research has made use of data and/or software provided by the High Energy
Astrophysics Science Archive Research Center (HEASARC), which is a service of
the Astrophysics Science Division at NASA/GSFC and the High Energy Astrophysics 
Division of the Smithsonian Astrophysical Observatory.

{\it Facilities:} \facility{GALEX}, \facility{Swift}


\end{document}